\documentclass[12pt,fleqn]{article}
\usepackage{a4,epsfig,epic}

\newcommand{\sibe}{\ifmmode\sigma_{\beta}
                   \else$\sigma_{\beta}$\fi}
\newcommand{\sicha}{\ifmmode\sigma_{\theta}
                   \else$\sigma_{\theta}$\fi}

\newcommand{\AmS}{{\protect\the\textfont2
   A\kern-.1667em\lower.5ex\hbox{M}\kern-.125emS}}

\begin{document}
\begin{titlepage}
\ 
\vskip 2cm
\centerline{\bf \Large Pattern Recognition and Event Reconstruction}
\vskip 0.8cm
\centerline{\bf \Large in Particle Physics Experiments}
\noindent

\vskip 2cm
\normalsize
\centerline{\large R.~Mankel\footnote{Email: Rainer.Mankel@desy.de}}
\vskip 5mm
\centerline{Deutsches Elektronen-Synchrotron DESY, Hamburg}

\vskip 5cm

\centerline {\bf Abstract}

This report reviews methods of pattern recognition and event
reconstruction used in modern high energy physics experiments. After a
brief introduction into general concepts of particle detectors and
statistical evaluation, different approaches in global and local
methods of track pattern recognition are reviewed with their typical
strengths and shortcomings. The emphasis is then moved to methods
which estimate the particle properties from the signals which pattern
recognition has associated. Finally, the global reconstruction of the
event is briefly addressed.

\end{titlepage}
\thispagestyle{empty}

\newpage

\tableofcontents
\newpage

\section{Introduction} 
Scientific discovery in elementary particle physics is largely driven
by the quest for higher and higher energies, which allow delving ever
more deeply into the fine structure of the microscopic
universe. Higher energies lead in general to an increased multiplicity
of particles. Since the acceleration of electrons is limited either by
synchrotron radiation in case of storage rings, or by field gradients
in case of linear colliders, multi-TeV energies are in the near future
only accessible by accelerating hadrons, the collision of which
generates even more particles.

\begin{figure}
\begin{center}
\unitlength1cm
\begin{picture}(11,19)
\put(1.5,0){\epsfig{file=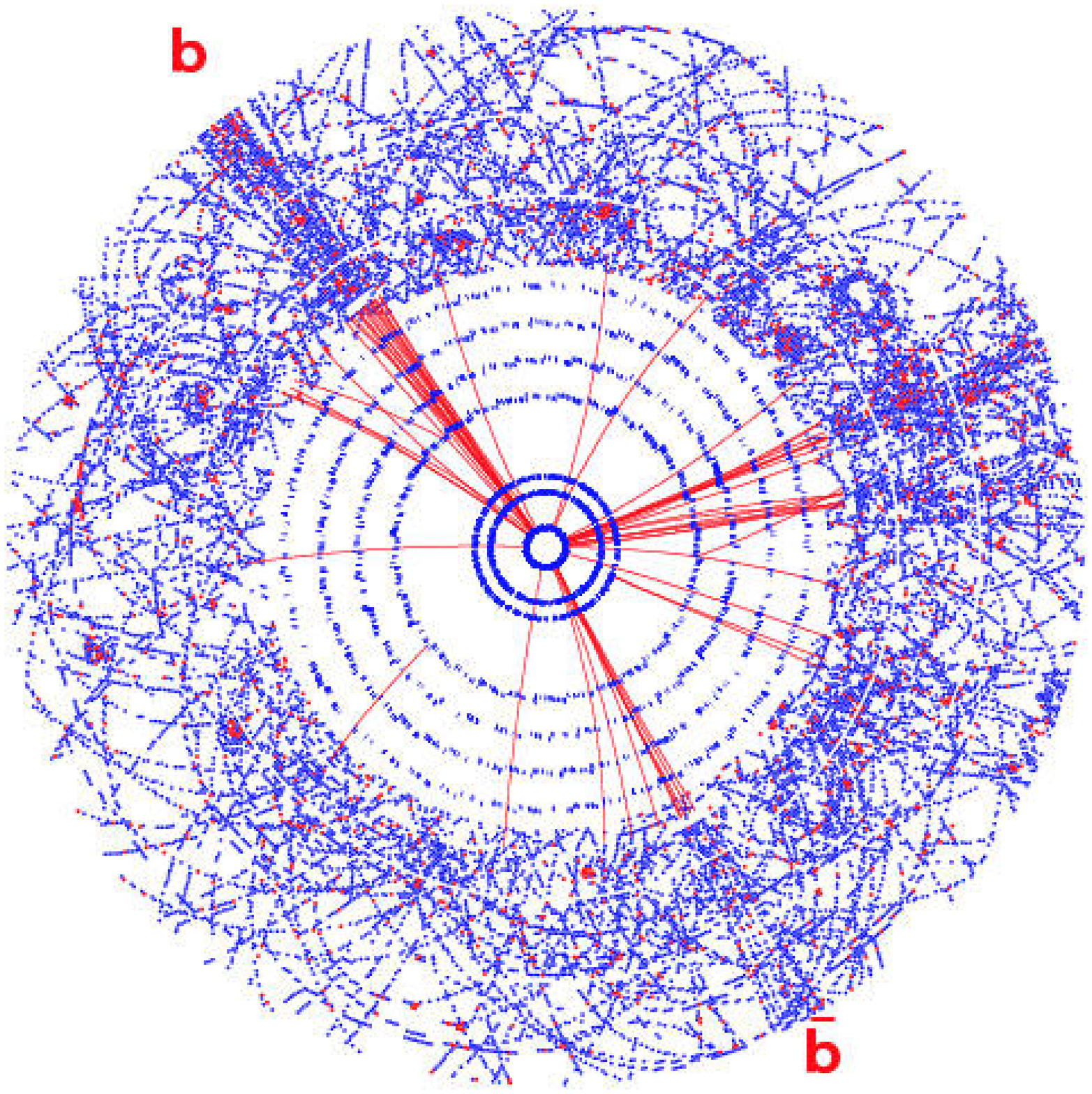,width=8.8cm}}
\put(0.0,10){\epsfig{file=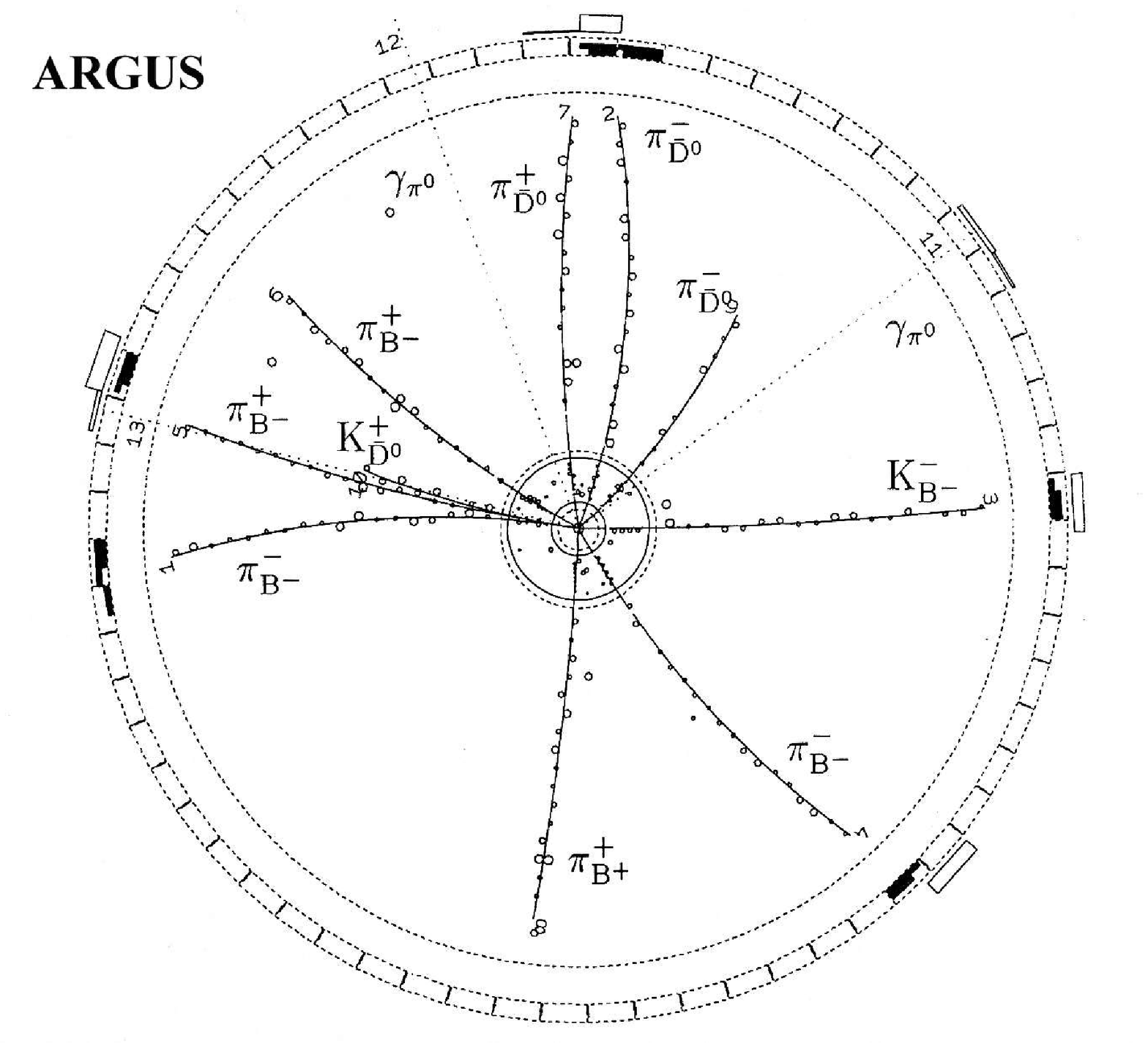,width=10.4cm}}
\put(1.8,9.3){\makebox(0,0)[t]{\bf ATLAS}}
\end{picture}
\end{center}
\caption{Comparison of event complexity in the experiments ARGUS and
  ATLAS. The ARGUS event (top) consists of two reconstructed $B$
  mesons, one of them being a candidate for the charmless decay $B^-
  \rightarrow K^-4\pi^{\pm}$ (from ~\cite{argusRareB}). The ATLAS
  display (bottom) shows a simulation of an event in the inner
  detector with a Higgs boson in the decay mode $H^0 \rightarrow
  b\bar{b}$, including the pileup at full LHC luminosity
  (from~\cite{atlasDesign}).}
\label{fig:argusAtlas}
\end{figure}
Reconstruction of charged particles from signals of tracking detectors
in spectrometers has always shown aspects of a discipline of art,
since the variety of experimental setups lead to development of very
diverse pattern recognition methods, which could not easily be ranked
among each other.  An general overview has been given in an earlier
review~\cite{grotereview}. It is remarkable that even today, no
generally accepted standard software package exists which performs
track finding in a variety of detector setups, a situation which is in
marked contrast e.g. to detector simulation. A new generation of
experiments is now emerging in which the track density is so high that
success will crucially depend on the power of the reconstruction
methods. One example for the development in tracking demands over 15
years is illustrated in fig.~\ref{fig:argusAtlas}, which shows in
direct comparison an event from the experiment
ARGUS~\cite{argusRareB}, which took data of $e^+e^-$ collisions in the
$\Upsilon$ range in the period 1982--1992, and the ATLAS
experiment~\cite{atlasDesign} which is currently under construction
and will operate from 2007 on with proton collisions at the LHC. The
new experiments also require huge computing resources for
reconstruction of their data.  Since track finding is usually the most
time consuming part in reconstruction, the sophistication and economy
of pattern recognition methods has considerable impact on the
computing effort.

Pattern recognition plays an important r\^ole also in other detector
components, for example cluster reconstruction in calorimeters, or
ring finding in ring imaging \v{C}erenkov detectors (RICH). It is
however in track reconstruction where the new generations of
experiments pose the most crucial challenges. This article will
therefore focus on track reconstruction as well as to related aspects
of event reconstruction.

The first of the following chapters will provide an introduction into
basic detector concepts and tracking devices and summarize
mathematical tools for estimating parameters and performance that
will be used later on. The two following chapters focus on track
pattern recognition with various methods, including applications in
several experiments. The next chapter then concentrates on parameter
estimation from particle trajectories, which is -- in contrast to
track finding -- in principle a straight-forward mathematical problem,
but contains several detailed issues worth mentioning. The last
chapter briefly discusses some track-related aspects of event
reconstruction.

\section{Basics} 
This section provides a brief introduction into the basic elements
influencing event reconstruction. It is not intended to cover the
subject of particle detectors in full detail, instead the detector
literature (see for example~\cite{detectorsGrupen,
  detectorsKleinknecht, detectorsGreen}) is referred to.

\subsection{Detector Layouts}
Modern detectors in high energy physics are usually sampling
detectors. The detector volume is filled with devices which the
particles traverse and in which they leave elementary pieces of
information, as e.g. an excitation in a solid-state detector, a primary
ionization in a gaseous chamber or an energy deposition in a sensitive
volume of a calorimeter. The event record of an experiment consists of
the amassed volume of the signals from all particles of an interaction
-- or possibly even several interactions -- joined together. After
sorting out which bits of information are related to the same particle
-- this process is called {\it pattern recognition} -- the kinematical
properties of each particle have to be reconstructed, to reveal the
physical nature of the whole event.

In general, experiments nowadays strive to record the interaction as a
whole, with all (significant) particles produced in the process. This
has lead to the development of $4\pi$ detectors, where almost the
whole solid angle region, as seen from the interaction, is covered.

In general, two main concepts have to be distinguished, which will be
discussed in the following.
\begin{figure}[htbp]
  \begin{center}
    \epsfig{file=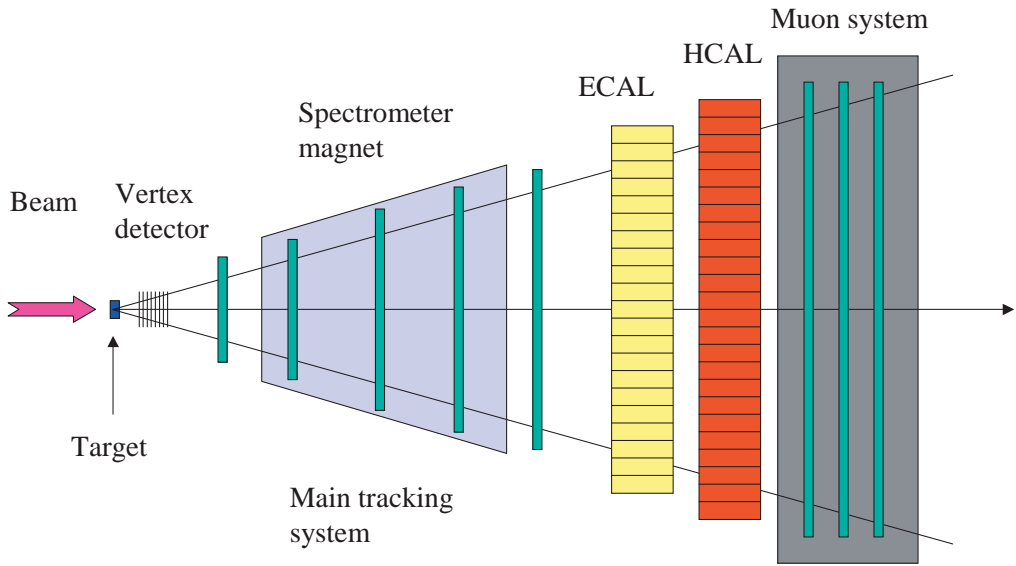,width=7.5cm,angle=270}
    \caption{Typical geometry of a forward spectrometer, as used
      e.g. in fixed-target setups.}
    \label{fig:fixedTarget}
  \end{center}
\end{figure}
\subsubsection{Forward or fixed target geometry}
When the interaction is generated by an incident beam hitting a fixed
target, the centre-of-mass system of the participating particles is
seen under a strong Lorentz boost, and the emerging particles are
moving within a cone into the forward direction. In this case, the
detector setup must cover this forward cone with instrumentation,
while the more backward part of the solid angle is generally
neglected. This scenario is called a {\it forward detector geometry}.
Similar situations exist where the dynamics of the interaction result
in all relevant particles to be produced under a huge Lorentz boost,
like heavy flavour production at large hadron colliders.

Figure~\ref{fig:fixedTarget} schematically shows a forward detector
geometry as it is used in fixed target experiments. The event is
generated through collision of a beam particle with a nucleus in the
target. Because of the momentum of the incident beam particles, the
whole event is seen under a Lorentz boost in the beam direction, so
that the emerging particles are confined to a cone whose opening angle
depends on the typical transverse momenta generated in the
interaction, and the size of the Lorentz boost.

The main components of a typical forward spectrometer are:
\begin{itemize}
\item the vertex detector, which is a precision tracking system very
  close to the interaction point. Its main purpose is the improvement
  of track resolution near the interaction point which allows
  reconstruction of secondary vertices or distinction of detached
  tracks which is used e.g. for the tagging of heavy flavour decays.
\item the spectrometer magnet with the main tracking system, which
  measures trajectories of charged particles and determines their
  momentum and charge sign from the curvature.
\item the calorimeter system, which is often split into an
  electromagnetic and a hadronic part. The calorimeter allows
  identification of electrons and hadrons by their deposited shower
  energy, and very often provide essential signals for the trigger
  system.  The calorimeter can also measure energies of individual
  neutral particles, in particular photons, though the actual
  capability in this task depends strongly on the particle density in
  the event.
\item the muon detector, which consists of tracking devices in
  combination with absorbers. Only muons are able to traverse the
  intermediate material, and are then measured in the dedicated
  tracking layers.
\end{itemize}

The design of a forward spectrometer is influenced by several factors.
The sheer size of the tracking volume depends on the leverage required
for the momentum resolution, since at sufficiently high momentum the
resolution is inversely proportional to the integral of the magnetic
field along the trajectory~\cite{gluckstern}, as will be discussed in
more detail in sec.~\ref{sec:fitting}.  Depending on the scope
of the experiment, further detector components may be introduced to
provide particle identification, for example ring-imaging \v{C}erenkov
counters (RICH) or transition radiation detectors (TRD).

\subsubsection{Collider detector geometry}
\begin{figure}[htbp]
  \begin{center}
    \unitlength1cm
    \begin{picture}(13,9.5)
      \put(-2,-1){\makebox{
          \epsfig{file=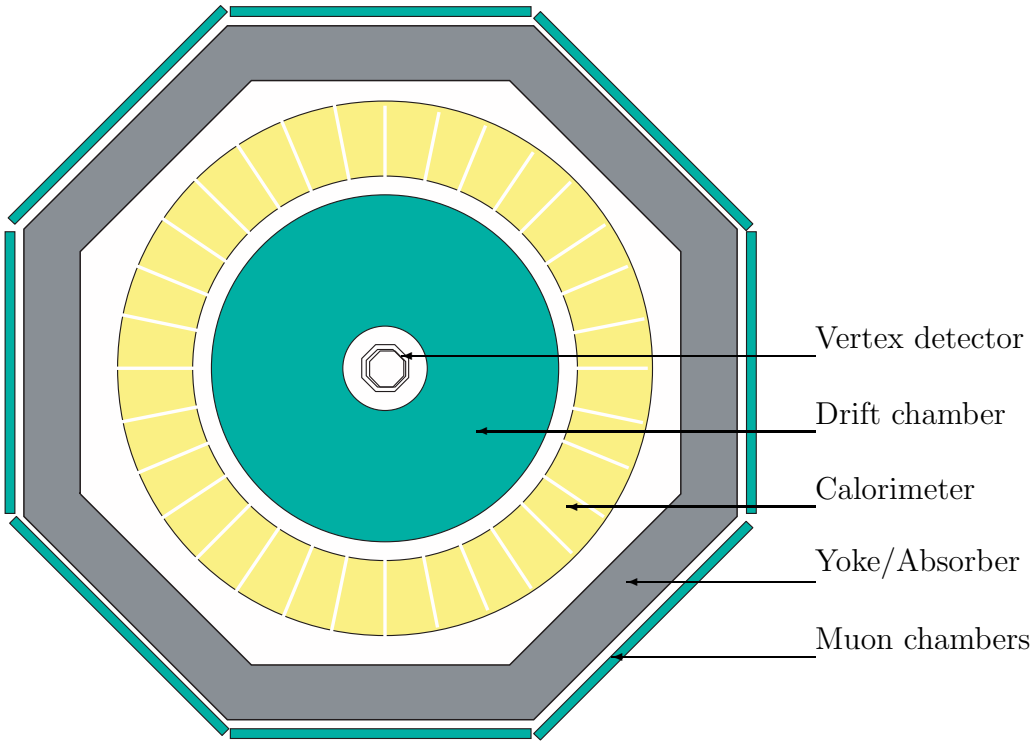,width=12cm,angle=0,
            bbllx=23,bblly=143,bburx=600,bbury=708}
          }}
      \put(10,5.0){\makebox(5,0.5)[l]{Vertex detector}}
      \put(10,4.0){\makebox(5,0.5)[l]{Drift chamber}}
      \put(10,3.0){\makebox(5,0.5)[l]{Calorimeter}}
      \put(10,2.0){\makebox(5,0.5)[l]{Yoke/Absorber}}
      \put(10,1.0){\makebox(5,0.5)[l]{Muon chambers}}
      \put(10,5.0){\vector(-4,0){5.5}}
      \put(10,4.0){\vector(-4,0){4.5}}
      \put(10,3.0){\vector(-4,0){3.3}}
      \put(10,2.0){\vector(-4,0){2.5}}
      \put(10,1.0){\vector(-4,0){2.7}}
    \end{picture}
    \caption{Typical setup of a collider detector.}
    \label{fig:collisionDet}
  \end{center}
\end{figure}

When two beams collide head-on, the centre-of-mass system of the
interactions is either at rest or moving moderately. In this case, the
detector should try to cover the full solid angle. This beam setup
usually leads to cylindrical detector layouts with a solenoid field
parallel to the beam axis (fig.~\ref{fig:collisionDet}).
In comparison to the forward geometry detector, the cylindrical
geometry differs in several details:
\begin{itemize}
\item the vertex detector requires modules parallel to the beam, at
  least in the central part of the angular acceptance, often referred
  to as the {\it barrel part}.
\item the main tracking system is generally contained in the magnetic
  field. Coil and yoke of the magnet usually have to be within the
  detector volume, where the general choice is to have the coil
  between drift chamber and calorimeter, where particles traverse it
  before their energy being measured in the calorimeter, or to make it
  large enough to enclose the calorimeter, which may be more costly to
  build and operate and where the field may have adverse effects on
  the calorimeter itself.
\item the calorimeter system now requires barrel and end cap parts to
  cover the solid angle. A main functionality at high energy colliders
  is the measurement of jets.
\item for the muon detector, the yoke of the solenoid lends itself
  readily as absorber.
\end{itemize}

\newpage
\subsection{Typical Tracking Devices}

\subsubsection{Linear single-coordinate measurements}
\label{sec:singleCoord}
\begin{figure}[htbp]
  \begin{center}
    \epsfig{file=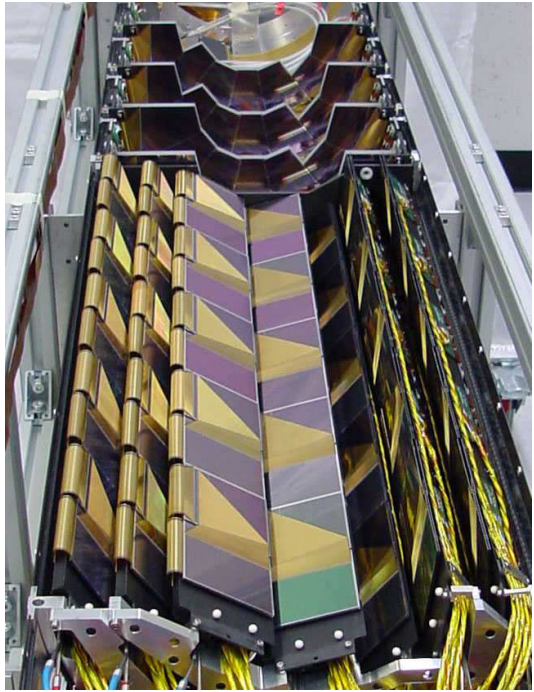,width=7cm}
  \end{center}
  \caption{Lower half barrel part of the Zeus micro-vertex detector}
  \label{fig:zeusmvd}
\end{figure}
A widespread type of tracking device measures one coordinate of the
particle whose trajectory intersects the device. A good example for
this type represent silicon strip detectors, which are
semiconductor-based devices structured in strips typically down to
widths of 25~$\mu$m.  Each strip works like a small diode, with a
voltage applied such that the border area is depleted and the
resistance is high. A traversing charged particle will then create
pairs of electrons and corresponding holes which drift apart under the
voltage and can be registered as a pulse.  In general several strips
will register a signal under traversal of a particle, and the pulse
heights of the participating channels can be evaluated with suitable
clustering algorithms, for example centre-of-gravity based, and
determine the location at which the particle has passed. Solid-state
detectors are presently the tracking devices with the highest spatial
resolution, and they are often installed very close to the
interaction region as {\it vertex detectors} where they allow or
improve the reconstruction of primary and secondary vertices. Another
favourable property of solid-state detectors is their resilience
against radiation damage.  The current limitation is in the size of
individual detector modules, which makes them expensive for coverage
of large volumes.  Figure~\ref{fig:zeusmvd} shows the micro-vertex
detector of the ZEUS experiment~\cite{zeusmvd}, prior to its
installation in 2001.

\subsubsection{Radial single-coordinate measurements}
The size of the tracking volumes is important, since momentum
measurement requires the particle to traverse a magnetic field, where
the length of the path provides the leverage that determines the
precision of the momentum reconstruction. This is one of the reasons
why gaseous chambers, in particular drift chambers are very commonly
employed when large areas have to be covered.

\begin{figure}[htbp]
  \begin{center}
    \epsfig{file=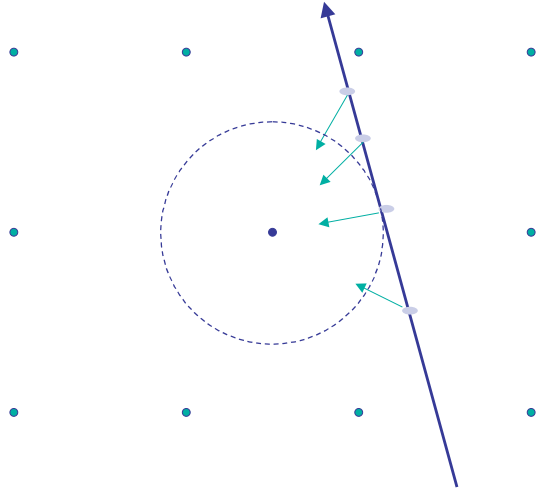,width=7cm}
  \end{center}
  \caption{Schematic view of a drift chamber cell. The filled
    circles indicate wires, with the sense wire in the middle of the
    cell and the field wires on the outside. The black arrow shows the
    trajectory of a particle, the grey arrows denote primary
    ionization charges drifting towards the sense wire.}
  \label{fig:driftcell}
\end{figure}
The basic principle of the drift chamber is displayed in
fig.~\ref{fig:driftcell}. A drift cell consists of an anode wire in
the centre and an arrangement of field wires. The geometry shown is
very similar to that in the ARGUS drift chamber~\cite{argusDrift} (see
also fig.~\ref{fig:argusClose} in section~\ref{sec:local}). The drift
cell need not be of rectangular shape, in the drift chamber of the
BaBar experiment, for example, it is hexagonal~\cite{babarDrift}.
Along the path of the particle, primary ionization occurs.  The
charges drift to the anode wire, where they create a locally confined
avalanche of particles within the large electrical field close to the
wire. This effect results in a multiplication of the ionization which
is called {\it gas amplification}. The rising edge of the signal
picked up by the anode wire triggers a time-to-digital converter (TDC)
which then measures the time until a common stop signal. This allows
measuring of the drift time for those charges that are the first to
arrive. In the simplest case, the drift field will be shaped such that
the drift velocity is uniform, and the time resolution can be directly
transformed into a uniform resolution of the drift distance. In
practice, numerous effects can lead to a non-linear {\it
  drift-time/space} relation, and the spatial resolution will depend
on the precise location of the traversal of the particle.

\begin{figure}[htbp]
 \begin{center}
   \unitlength1cm
   \begin{picture}(13,8)
     \put(0,0){\makebox{\epsfig{file=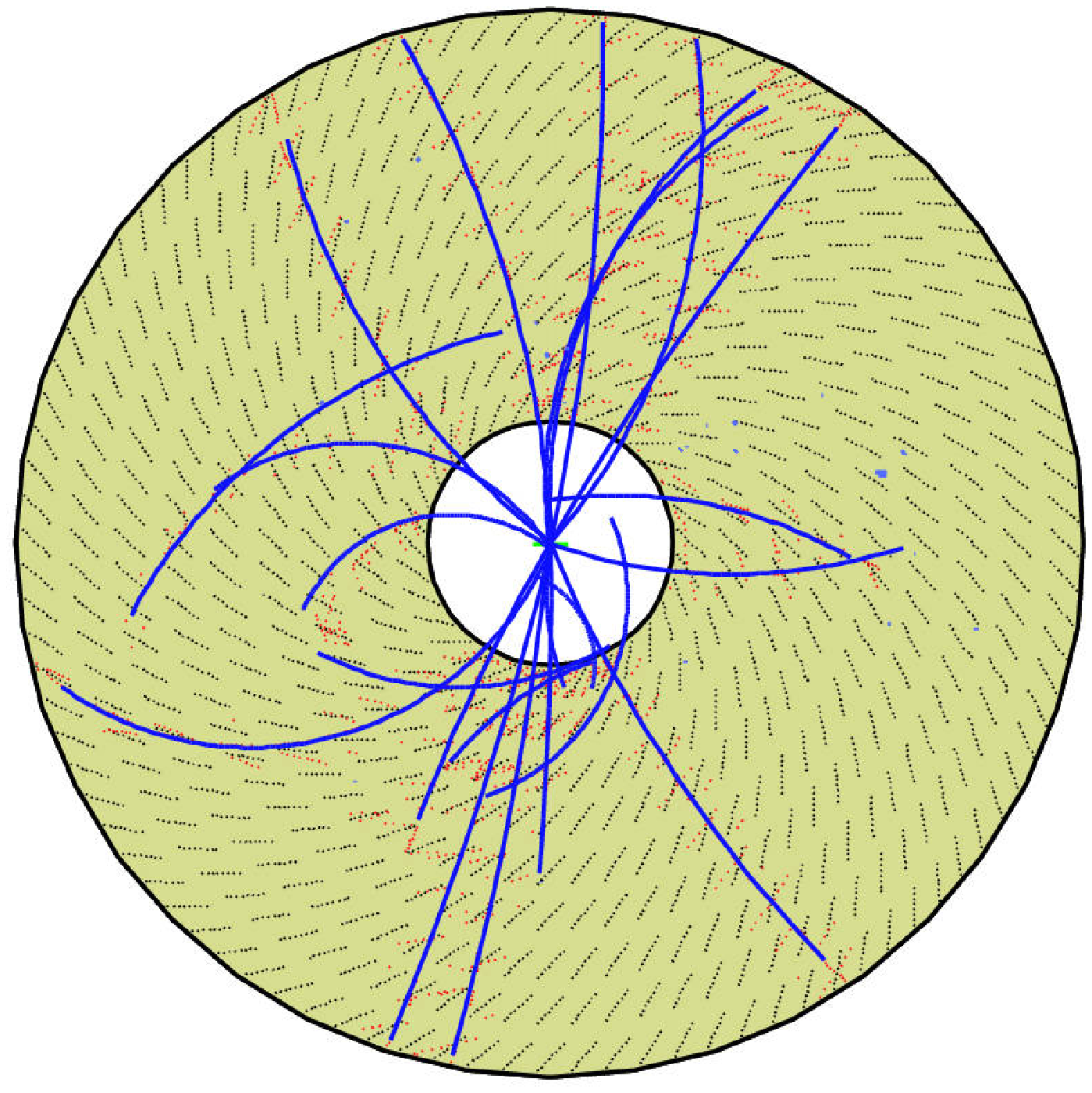,width=7cm,
           bbllx=70pt,bblly=39pt,bburx=500pt,bbury=468pt}}}
     \put(4,1){\framebox(1.2,1.2){}}
     \put(8,1){\makebox{\epsfig{file=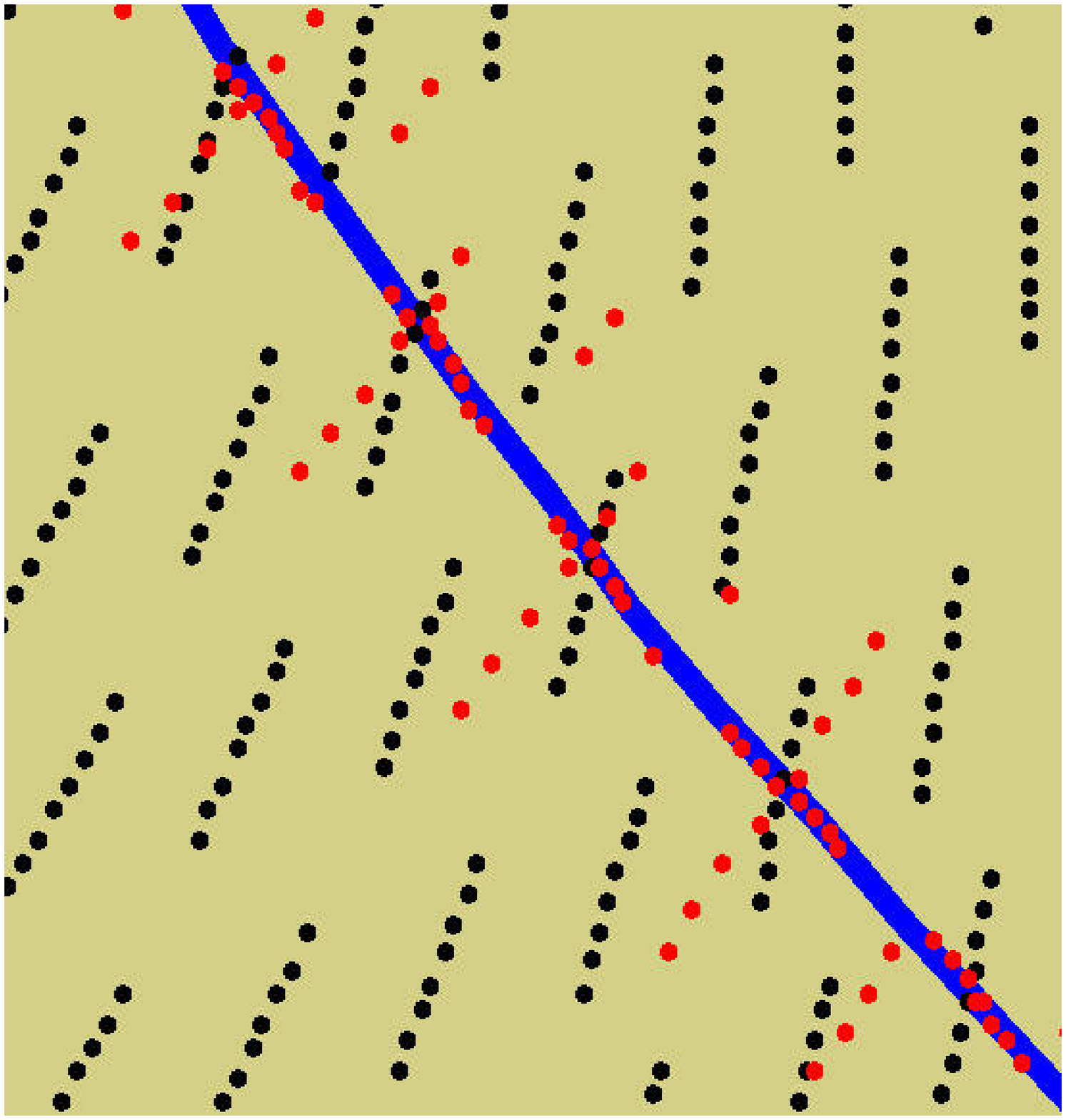,
           width=5cm,angle=0}}}
   \end{picture}
 \end{center}
 \caption{Left: event display from the ZEUS central tracking detector (CTD),
   showing sense wires and reconstructed tracks. Right: closeup around
   the track in the lower left area. The black dots represent the
   sense wires, the grey dots indicate the drift distance end points
   on both sides of the corresponding wire.}
\label{fig:zeusctd}
\end{figure}

Since the time measured by the TDC corresponds to the arrival of the
first charges, usually those with the smallest distance to the
wire, the drift chamber measures the distance of closest approach of
the particle to the wire. In cases where more than one particle
traverses the same drift cell within the same interaction window, in
general only the particle closest to the wire is registered. This
effect may cause complications for pattern recognition which depend on
the degree of occupancy. Another typical property of drift
chambers is that the single measurement cannot distinguish on which
side of the wire the particle has traversed; this uncertainty is
called {\it left-right ambiguity}. In the worst case, left-right
ambiguity may lead to a {\it mirror track} that cannot be
distinguished from the real one.  Concepts have therefore been
developed how to design drift chambers such that left-right ambiguity
can be resolved in all cases, e.g. the {\it butterfly
  geometry}~\cite{butterfly}.

Drift in gases is influenced also by magnetic fields. The deviation of
the gas drift direction from the vector of the electric field is
described by the {\it Lorentz angle}. Figure~\ref{fig:zeusctd} shows
an event display of the central tracking detector (CTD) of the ZEUS
experiment, in the view along the beam axis, which has been created
using the tool described in~\cite{zevis}. The Lorentz angle in this
case is $45^\circ$, and it is reflected in the design of the cell
structure.

\subsubsection{Stereo angles}
\label{sec:stereo}
\begin{figure}[htbp]
  \begin{center}
    \epsfig{file=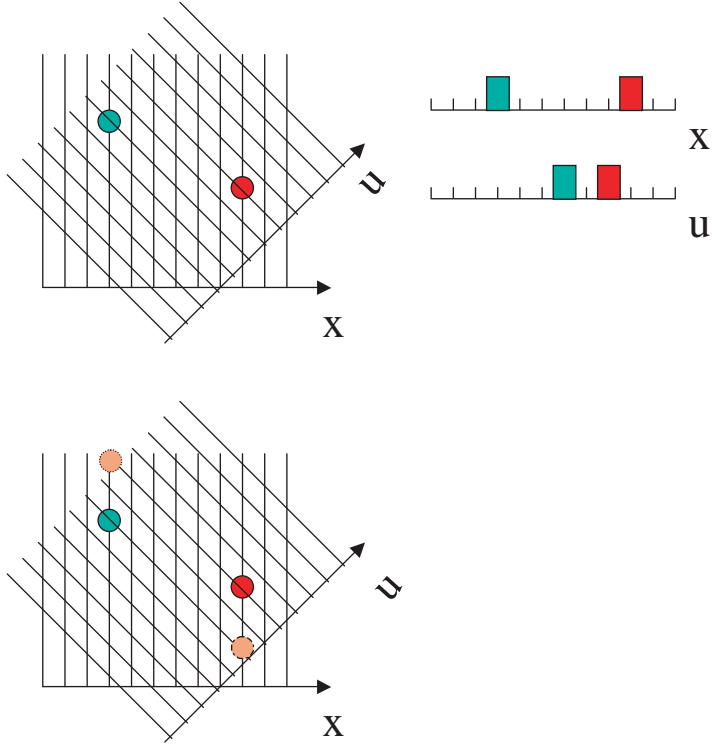,width=12cm}
  \end{center}
  \caption{Hit ambiguities with two stereo views}
  \label{fig:stereo}
\end{figure}
Devices measuring single coordinates do not provide
three-dimensional\footnote{The shorthands 2D (two-dimensional) and 3D
  (three-dimensional) will frequently be used in the following.}
points on a trajectory, but measure only in a projected space. While
such devices can be very economic in the sense that a relatively small
number of channels is needed to cover a region at good resolution, 3D
information can only be obtained by combining several projections,
usually named {\it stereo views}. While two views are in principle
sufficient to reconstruct spatial information, the presence of more
than one track leads in general to ambiguities regarding the
assignment of projected information. This is illustrated in
fig.~\ref{fig:stereo}, where two particles are measured in two strip
detector views of $0^\circ$ ($x$) and $45^\circ$ ($u$).  Ambiguity in
the assignment of the measured hits in the $x$ and $u$ views to each
other leads to the reconstruction of two ghost points. This
illustrates that in general at least three views are necessary to
avoid this kind of ambiguities. On the other hand, in special cases of
limited track density, the use of only two views may be justified,
since in this case the majority of ghosts may be discarded for
geometrical reasons. This can already be guessed from
fig.~\ref{fig:stereo}: since the true tracks are well separated, the
uppermost ghost combination is already just outside the chamber
acceptance of the $u$ view. Such concepts are called {\it all-stereo}
designs.

\setlength{\unitlength}{1.15cm}
\begin{figure}
\begin{center}
\begin{picture}(13,7)
  \put(0,0){\epsfig{file=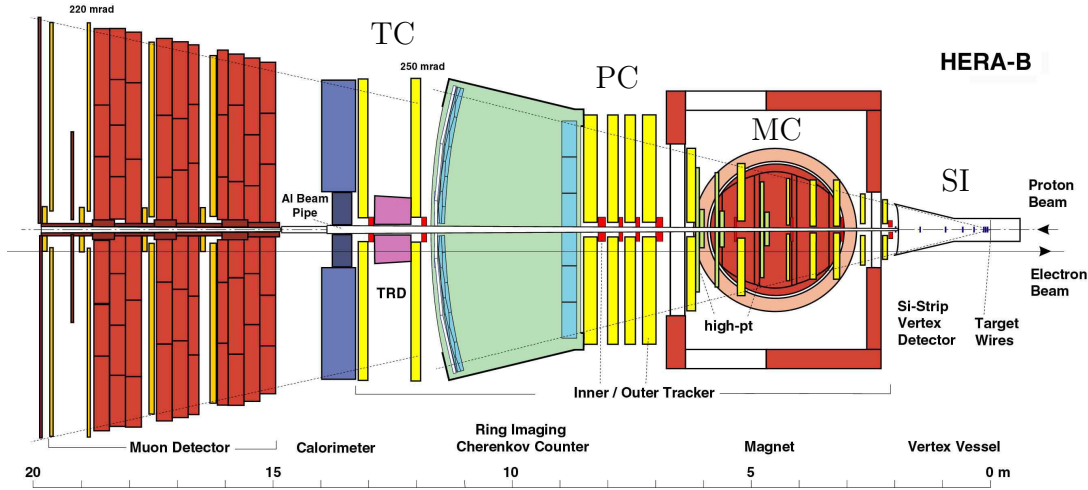,width=15cm}}
  \put(4.2,5){\makebox(1,0.5)[l]{TC}}
  \put(6.8,4.5){\makebox(1,0.5)[l]{PC}}
  \put(8.6,3.9){\makebox(1,0.5)[l]{MC}}
  \put(10.8,3.3){\makebox(1,0.5)[l]{SI}}
\end{picture}
\end{center}
\caption{Layout of the HERA-B spectrometer. The labels
  TC, PC, MC and SI indicate groups of tracking stations that comprise
  the vertex and main tracking system.}
\label{fig:herabDetector}
\end{figure}
An example for a spectrometer that combines several types of
single-coordinate measurements is the HERA-B
detector~\cite{herabBeauty94,herabDesign,herabVancouver} which is
shown in fig.~\ref{fig:herabDetector}. The vertex detector (labelled
{\it SI}) consists of eight superlayers of silicon strip detectors
with four different stereo angles. The design of the main tracker is
structured into the three areas within the magnet ({\it MC}), between
magnet and RICH ({\it PC}) and between RICH and calorimeter ({\it
  TC}), it contains 13 superlayers of honeycomb drift chamber modules
for the outer area and 10 superlayers of micro-strip gaseous chambers
(MSGC) for the region close to the beam\footnote{The layout of
  tracking stations has been modified later with the shift of emphasis
  away from $B$ physics.}.

\subsubsection{Three-dimensional measurements}
\begin{figure}[htbp]
  \begin{center}
    \epsfig{file=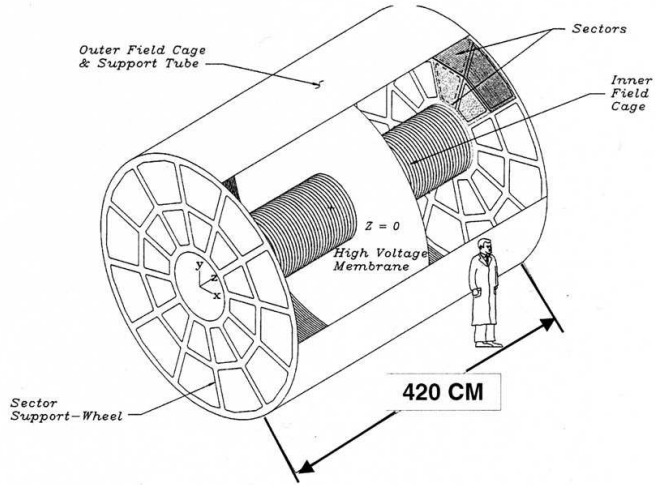,width=9cm}
  \end{center}
  \caption{TPC of the STAR experiment (from~\cite{starTPC}).}
  \label{fig:starTPC}
\end{figure}
In general pattern recognition will benefit considerably if the
tracking device itself is able to measure 3D space points. A modern
example is solid-state pixel detectors, as for example the CCD-based
vertex detector of the SLD experiment~\cite{sldVertex}, where the
pixels have a size of $20\times 20\ \mu {\rm m}^2$. A gaseous detector
capable of covering large tracking volumes with 3D measurement is the
{\it time projection chamber} (TPC).  Figure~\ref{fig:starTPC} shows
the TPC of the STAR experiment~\cite{starTPC}. The gas volume itself
is free of wires; instead, an axial electrical field, produced with
the help of a membrane electrode in the middle plane, lets the primary
charges drift to the anodes at the end caps, where they are
registered, for example with multi-wire proportional chambers with pad
readout. While this provides a direct measurement of the $x$ and $y$
coordinates, the $z$ coordinate is inferred from the time measurement.
The magnetic field is also axial, and plays an important r\^ole in
limiting diffusion effects during the drift.

\subsection{Track Models and Parameter Representations}
\subsubsection{Forward geometry}
\label{sec:forwardGeometry}
In the forward geometry, the interaction region lies very often in an
area without magnetic field, since the spectrometer magnet is located
further downstream. The natural choice of parameters, assuming that
the $z$ coordinate points down the spectrometer axis and $x$ and $y$
are the transverse coordinates, is then
\begin{description}
  \item[{\bf $x_0$}] the $x$ coordinate at the reference $z_0$
  \item[{\bf $y_0$}] the $y$ coordinate at the reference $z_0$
  \item[{\bf $t_x=\tan \theta_x$}] the track slope in the $xz$ plane
  \item[{\bf $t_y=\tan \theta_y$}] the track slope in the $yz$ plane
  \item[{\bf $Q/p$}] the inverse particle momentum, signed according to
    charge
\end{description}
where $z_0$ denotes the location of a suitable reference plane
transverse to the beam, for example at the position of the target, or
at the nominal interaction point. The slope parameters allow for a
convenient transformation of the parameters to a different reference
$z$ value, as is needed during vertex reconstruction. In cases of a
very homogeneous magnetic field, it may be advantageous to substitute
the parameter $Q/p$ by $Q/p_{\perp}$, where $p_{\perp}$ is the
momentum in the plane transverse to the magnetic field, or by
$\kappa=Q/R$, the signed inverse radius of curvature.

\subsubsection{Cylindrical geometry}
\label{sec:cylindrical}
In collider detectors with cylindrical geometry, the magnetic field
normally encompasses the whole tracking volume, including the
interaction region where the particles are produced. In a homogeneous
solenoid field, the particle trajectory will be a helix curling around
an axis parallel to the magnetic field. Assuming the $z$ coordinate is
oriented along the detector axis, and the radius is given by
$r=\sqrt{x^2+y^2}$, typical track parameters given at a reference
value $r=r_0$ may be
\begin{description}
\item[{\bf $\phi_0$}] the azimuth angle where the trajectory intersects the
    reference radius
\item[{\bf $z_0$}] the $z$ value where the trajectory intersects the
    reference radius
\item[{\bf $\psi_0$}] the phase angle of the helix at the
  reference radius intersection, which corresponds to the angle of the
  tangent at this point
\item [{\bf $Q/R$}] the signed inverse curvature radius of the helix
\item [{\bf $\tan \lambda$}] where $\lambda = \arctan
  p_z / p_{\perp}$ is the dip angle of the helix
\end{description}

\subsection{Parameter Estimation}
The estimation of the kinematical parameters of a particle, as
position (or impact parameter), direction of flight and momentum at
its point of origin from spatial measurements along its trajectory is
generally referred to as track fitting. We assume at this point that
the measurements related to a particle have been correctly identified
in the pattern recognition step (which will be discussed in more
detail in sections~\ref{sec:globalPattern} and \ref{sec:local}). A
very general approach to parameter estimation is the {\it maximum
  likelihood method}, which shall not be discussed here in detail;
instead we refer to the textbook
literature~\cite{brandt,blobelLohrmann,bevington,eadie,frodesen}.  The
maximum likelihood method can take very general distributions of the
observed variables into account, for example exponential distributions
as they may occur when decay lengths are measured. However, its
application in multi-parameter problems can be very complex, in
particular the error analysis. In cases where the distribution of the
random variables is Gaussian, at least approximately, the {\it least
  squares method} is generally successful. Since many observables in
track reconstruction do at least approximately follow a Gaussian
distribution, this method will be focussed on in the following.

\subsubsection{Least squares estimation}
\label{sec:leastSquares}
If the trajectory of a particle can be described by a closed
expression $f_{\vec \lambda}(\ell)$, where $\vec{\lambda}$ stands for
the set of parameters, $\ell$ is the flight path and $f$ is the
coordinate which could be measured, a set of measurements \{$m_i$\}
with errors \{$\sigma_i$\} will provide an estimate of the parameters
according to the least squares principle
\begin{equation}
X^2 = \sum \frac{ (m_i -  f_{\vec{\lambda}}(\ell_i))^2 } { \sigma_i^2
  } \begin{array}{c} _! \\ = \\ \ \end{array} min
\label{eq:leastSquares}
\end{equation}
One can easily convince oneself that in the case of normally
distributed measurements $m_i$, the above impression is proportional
to the negative logarithm of the corresponding likelihood function,
which shows directly the equivalence of least squares principle and
maximum likelihood principle for this case.

Symbolizing the derivative matrix\footnote{We denote the derivative
  matrix as $\frac{\partial f}{\partial \lambda}$, where
  $\left( \frac{\partial f}{\partial \lambda} \right)_{ij} =
  \frac{\partial f_{\vec{\lambda}}(\ell_i)}{\partial \lambda_j}$.}
 of $f$ with respect to the parameters 
as {\bf F} and the (diagonal) error matrix of the measurements as 
${\bf V}=diag\{\sigma_i^2\}$, the expression to be minimized is
\begin{equation}
(\vec m - F \vec \lambda)^T V^{-1} (\vec m - F \vec \lambda)
\end{equation}
and requiring the derivative to vanish at the minimum leads to the
matrix equation
\begin{equation}
F^T V^{-1} \vec{f} = F^T V^{-1} \vec{m}
\end{equation}
In case of a linear problem, $\vec{f} = F \vec{\lambda}$, the above condition 
can 
be directly inverted
\begin{equation}
\vec \lambda = (F^T V^{-1} F)^{-1} F^T V^{-1} \vec{m}
\label{eq:leastSqParam}
\end{equation}
and the estimated parameters are a linear function of the
measurements. The matrix $(F^T V^{-1} F)^{-1}$ that needs to be
inverted is of the shape $N_\lambda \times N_\lambda$ (where
$N_\lambda$ is the number of parameters describing the particle),
which is inexpensive in terms of computation. Also the covariance
matrix of the parameter estimate can be directly determined as
\begin{equation}
cov(\vec \lambda) = C_\lambda = (F^T V^{-1} F)^{-1}
\label{eq:leastSqCov}
\end{equation}

The popularity of the least squares method can be attributed to its
optimality properties in the linear case:
\begin{itemize}
\item the estimate is unbiased, i.e. the expectation value of the
  estimate is the true value
\item the estimate is {\it efficient}, which means that, of all unbiased 
estimates 
which are linear functions of the observables,
the least squares estimate has the smallest variance. This is called 
the ``Gauss-Markov-Theorem''.
\end{itemize}
Though these properties are strictly guaranteed only for the linear
case, they are still retained in most cases where the function
$f_{\vec{\lambda}}$ can be locally approximated by a linear expansion.

The expression $X^2$ in equation~\ref{eq:leastSquares} will follow a
$\chi^2$ distribution if the function $f_\lambda$ is (sufficiently)
linear and if the measurements $m_i$ follow a normal distribution.
This property can be used for statistical tests. In particular the
second condition should be always kept in mind, as its relevance will
become apparent later.

\subsubsection{The Kalman filter technique}
\label{sec:kalmanfilter}
The least squares parameter estimation as described in the previous
section requires the global availability of all measurements at
fitting time. There are cases when this requirement is not convenient,
for example in real-time tracking of objects, or in pattern
recognition schemes which are based on track following, where it is
not clear a-priori if the hit combination under consideration does
really belong to an actual track.

The Kalman filter technique was developed to determine the trajectory
of the state vector of a dynamical system from a set of measurements
taken at different times~\cite{kalman}. In contrast to a global fit,
the Kalman filter proceeds progressively from one measurement to the
next, improving the knowledge about the trajectory with each new
measurement. Tracking of a ballistic object on a radar screen may
serve as a technical example. With a traditional global fit, this
would require a time consuming complete refit of the trajectory with
each added measurement.

Several properties make the Kalman filter technique an ideal
instrument for track (and vertex)
reconstruction~\cite{billoir84,fruehwirth,billoir}. The {\it
  prediction} step, in which an estimate is made for the next
measurement from the current knowledge of the state vector, is very
useful to discard noise signals and hits from other tracks from the
fit. The {\it filter} step which updates the state vector does not
require inversion of a matrix with dimension of the state vector as in
a global fit, but only with the dimension of the measurement, leading
to a very fast algorithm.  Finally, the problem of random
perturbations on the trajectory, as multiple scattering or energy
loss, can be accounted for in a very efficient way. In its final
result, the Kalman filter process is equivalent to a least squares
fit.

In this article the implementation and nomenclature from
\cite{fruehwirth,grotebock} is used, and these documents are referred
to for a more detailed explanation of the Kalman filter method.  In
this notation, the system state vector {\it at the time} $k$, i.e.
after inclusion of $k$ measurements is denoted by $\tilde x_k$, its
covariance matrix by $C_k$. In our case $\tilde x_k$ contains the
parameters of the fitted track, given at the position of the
$k^{\mbox{th}}$ hit.  The matrix $F_k$ describes the propagation of
the track parameters from the $(k-1)^{\mbox{th}}$ to the
$k^{\mbox{th}}$ hit.\footnote{We assume at this stage a linear system,
  so that $F_k$ and $H_k$ are matrices in the proper sense.  For
  treatment of the non-linear case see below.}  For example, in a
planar geometry with one-dimensional measurements and straight-line
tracks, the propagation takes the form
\begin{equation}
  \left( \begin{tabular}{c} x \\ $t_x$ \end{tabular} \right)_k
= \left( \begin{tabular}{cc} 1 & $z_k - z_{k-1}$\\
                                 0 & 1 \end{tabular} \right)
  \left( \begin{tabular}{c} x \\ $t_x$ \end{tabular} \right)_{k-1}
\end{equation}
where a subset of the track parametrization in
section~\ref{sec:forwardGeometry} has been used.  The coordinate
measured by the $k^{\mbox{th}}$ hit is denoted by $m_k$. In general
$m_k$ is a vector with the dimension of that specific measurement. For
tracking devices measuring only one coordinate, $m_k$ is an ordinary
number.  The measurement error is described by the covariance matrix
$V_k$. The relation between the track parameters $\tilde x_k$ and the
{\it predicted} measurement is described by the projection matrix
$H_k$.  In the example in section~\ref{sec:stereo}, the measured
coordinate in the stereo view $u$ is
\begin{equation}
  H \left( \begin{tabular}{c} x\\ y \end{tabular} \right)
    = \left( \begin{tabular}{cc} $\cos \alpha_{st}$ &
        $-\sin \alpha_{st}$ \end{tabular} \right)
    \left( \begin{tabular}{c} x\\ y \end{tabular} \right)
\end{equation}
with $\alpha_{st}$ as the stereo angle ($45^\circ$ in the example).

In each filter step, the state vector and its covariance matrix are
propagated to the location or {\it time} of the next measurement with
the {\it prediction equations}:

\begin{equation}
 \tilde{x}^{k-1}_k = F_k \tilde{x}_{k-1}\\
 C^{k-1}_k = F_k C_{k-1} F_k^T + Q_k
 \label{eq:transport}
\end{equation}
and the estimated residual becomes
\begin{equation}
 r^{k-1}_k = m_k - H_k \tilde{x}^{k-1}_k\\
 R^{k-1}_k = V_k + H_k C^{k-1}_k H_k^T
 \label{eq:predres}
\end{equation}
Here $Q_k$ denotes the additional error introduced by {\it process
  noise}, i.e. random perturbations of the particle trajectory, for
example multiple scattering. We will see later
(sec.~\ref{sec:multipleScattering}) how this treatment works in
detail.  The updating of the system state vector with the
$k^{\mbox{th}}$ measurement is performed with the {\it filter
  equations}:
\begin{equation}
 K_k = C^{k-1}_k H^T (V_k + H_k C^{k-1}_k H_k^T)^{-1}
\label{eq:filterEquation}
\end{equation}
\begin{displaymath}
 \tilde{x}_k = \tilde{x}^{k-1}_k + K_k (m_k - H_k \tilde{x}^{k-1}_k)\\
\end{displaymath}
\begin{displaymath}
 C_k = (1 - K_k H_k) C^{k-1}_k
\end{displaymath}
with the filtered residuals
\begin{equation}
 r_k = (1 - H_k K_k)\,r^{k-1}_k\\
 R_k = (1 - H_k K_k)\,V_k
\end{equation}
$K_k$ is sometimes called the {\it gain matrix}.
The $\chi^2$ contribution of the filtered point is then given by
\begin{equation}
 \chi^2_{k,F} = r_k^T R_k^{-1}r_k
 \label{eq:filtchi2}
\end{equation}
The system state vector at the last filtered point contains always the
full information from all points. If one needs the full state vector
at every point of the trajectory, the new information has to be passed
upstream with the {\it smoother equations}:
\begin{equation}
 A_k = C_k F_{k+1}^T (C^k_{k+1})^{-1}
\label{eq:kalmansmoother}
\end{equation}
\begin{displaymath}
 \tilde x^n_k = \tilde x_k + A_k (\tilde x^n_{k+1} - \tilde x^k_{k+1})
\end{displaymath}
\begin{displaymath}
 C^n_k = C_k + A_k (C^n_{k+1} - C^k_{k+1}) A_k^T
\end{displaymath}
\begin{displaymath}
 r^n_k = m_k - H_k \tilde x^n_k
\end{displaymath}
\begin{displaymath}
 R^n_k = R_k - H_k A_k (C^n_{k+1} - C^k_{k+1}) A_k^T H_k^T
\end{displaymath}
Thus, smoothing is also a recursive operation which proceeds step by
step in the direction opposite to that of the filter. The quantities
used in each step have been calculated in the preceding filter
process. If process noise is taken into account, e.g. to model
multiple scattering, the smoothed trajectory may in general contain
small kinks and thus reproduce more closely the real path
of the particle.

In the equations above, F and H are just ordinary matrices if both
transport and projection in measurement space are linear operations.
In case of non-linear systems, they have to be replaced by the
corresponding functions and their derivatives:
\begin{equation}
  F_k\tilde x_k \rightarrow f_k(\tilde x_k)\\
  H_k\tilde x_k \rightarrow h_k(\tilde x_k)
\end{equation}
using for covariance matrix transformations
\begin{equation}
  F_k \rightarrow \frac{\partial f_k}{\partial \tilde x_k}\\
  H_k \rightarrow \frac{\partial h_k}{\partial \tilde x_k}
  \label{eq:nonlinear}
\end{equation}
The dependence of $f_k$ and $h_k$ on the state vector estimate will
in general require iteration until the trajectory converges such that
all derivatives are calculated at their proper positions.
We will continue to call $\partial f_k / \partial \tilde x_k$
the transport matrix and $\partial h_k / \partial \tilde x_k$
the projection matrix of our system.

The Kalman filter has also been found to be particularly suited for
implementation in object-oriented programming
language~\cite{browne}.

\subsection{Evaluation of Performance}
When it comes to quantifying the performance of methods in track
pattern recognition, actual numbers will in general strongly depend of
the definition of criteria, which comparisons should take into
account. 

\subsubsection{The reference set}
Assessment of track finding efficiency requires firstly a definition
of a {\it reference set} of tracks that an ideally performing
algorithm should find. Normally tracks will be provided by a Monte
Carlo simulation, and the selection of {\it reference tracks} will
depend on the physics motivation of the experiment. Low momentum
particles arising from secondary interactions in the material are
normally not within the physics scope but merely an obstacle and
should be excluded. Particles travelling outside of the geometrical
acceptance, for example within the beam hole of a collider experiment
cannot be traced by the detector and should be disregarded as well.
Also particles straddling the border of a detector and e.g. traversing
only a small number of tracking layers will often be regarded as
outside of the design tracking volume. A typical convention may be to
regard particles which traverse ${\cal O}$(80\%) of the nominal
tracking layers as constituents of the reference set.

The definition of the reference set can then be regarded as a
definition of effective geometrical acceptance
\begin{equation}
  \epsilon_{geo} = \frac{N_{ref}}{N_{total}}
\end{equation}
with $N$ denoting the number of particles of interest in the reference
set and in total.

\subsubsection{Track finding efficiency}
Definition of the track finding efficiency requires a criterion which
specifies whether a certain particle has been found by the algorithm
or not. There are two rather different concepts:
\begin{description}
\item[Hit matching] This method analyzes the simulated origin of each
  hit in the reconstructed track using the {\it Monte Carlo truth}
  information. If the qualified majority of hits, for example at least
  70\% originates from the same true particle, the track is said to
  {\it reconstruct} this particle. This method is stable in the limit
  of very high track densities, but it requires the Monte Carlo truth
  information to be mapped meticulously through the whole simulation.
\item[Parameter matching] The reconstructed parameters of a track are
  compared with those of all true particles. If the parameter sets
  agree within certain limits (which should be motivated by the
  physics goals of the experiment), the corresponding track is said to
  reconstruct this particle. This method requires less functionality
  from the simulation chain, but it bears the danger of accepting
  random coincidences between true particles and artifacts from the
  pattern recognition algorithm. In extreme cases, this can lead to
  the paradox impression that the track finding efficiency {\it
    improves} with increasing hit density.
\end{description}

The reconstruction efficiency is then defined as
\begin{equation}
  \epsilon_{reco} = \frac{N^{reco}_{ref}}{N_{ref}}
\end{equation}
where $N^{reco}_{ref}$ is the number of reference particles that are
reconstructed by at least one track.  It should be noted that this
definition is such that a value of one cannot be exceeded, and
multiple reconstructions of the same track will not increase the track
finding efficiency. One should also control the abundance of
non-reference tracks which are reconstructed ($N^{reco}_{non-ref}$):
normally the relation
\begin{equation}
\frac{N^{reco}_{non-ref}}{N_{total}-N_{ref}} \ll  \epsilon_{reco}
\label{eq:nonref}
\end{equation}
should hold, otherwise the reference criteria might be too strict.

\subsubsection{Ghosts}
Tracks produced by the pattern recognition algorithm that do not
reconstruct any true particle within or without the reference set are
called {\it ghosts}. A ghost rate can be defined as
\begin{equation}
  \epsilon_{ghost} = \frac{N^{ghost}}{N_{ref}}
\end{equation}
Since the ghost rate may be dominated by a small subset of events with
copious hit multiplicity, it is also informative to specify the mean
number of ghosts per event.

\subsubsection{Clones}
The above definitions for efficiency and ghost rate are intentionally
insensitive to multiple reconstructions of a particle. Such redundant
reconstructions are sometimes called {\it clones}. For a given
particle $m$ with $N_m^{reco}$ tracks reconstructing it, the number of
clones is
\begin{equation}
  N_m^{clone} = \left\{ \begin{array}{l} N_m^{reco} - 1, {\rm if} N_{m}^{reco}>0\\ 
    \ \ \ 0 \ \ \ \ \ \ \ \ ,{\rm otherwise} \end{array} \right.
\end{equation}
and the {\it clone rate} becomes
\begin{equation}
  \epsilon_{clone} = \frac{\sum_m N_m^{clone}}{N_{ref}}
\end{equation}
In practice, clones can usually be eliminated at the end of the
reconstruction chain by means of a {\it compatibility
  analysis}~\cite{compatibility}.

\subsubsection{Parameter resolution}
\label{sec:parameterResolution}
The quality of reconstructed particle parameters and error estimates
from reconstruction in a subdetector is essential for matching and
propagation into another subsystem. For the whole detector, it
determines directly the physics performance. The quality of the
estimate of a track parameter $X_i$ is reflected in the {\it parameter
residual}
\begin{equation}
  \label{eq:parmResid}
  R(X_i) = X_i^{rec} - X_i^{true}
\end{equation}
From the parameter residual distribution, one can then obtain the
parameter estimate bias $\left<R(X_i)\right>$, and the parameter
resolution as a measure of its width. The estimate of the parameter
covariance matrix can be used to define the {\it normalized parameter
  residual}
\begin{equation}
  \label{eq:normParmResid}
  P(X_i) = \frac{X_i^{rec} - X_i^{true}}{\sqrt{C_{ii}}}
\end{equation}
which is often called the {\it pull} of this parameter. Ideally, the
pull should follow a Gaussian distribution with a mean value of zero
and a standard deviation of one.

\subsubsection{Interplay}
Results for the individual performance estimators may very much depend
on the definitions, so it is advisable to always judge several of the
above quantities in combination. For example, the track finding
efficiency should be always seen together with the ghost rate, since a
less strict definition of the criterion if a track reconstructs a
particle will lead to a higher track finding efficiency but also to a
higher ghost rate. Also the parameter resolution will tell if the
reconstruction criterion is correct, because in case of an
inadequately generous assignment, the parameter residuals are likely
to show an increased width, or tails from improperly recognized tracks.
When parameter matching is used, generous definition of the matching
criteria will also increase the track finding efficiency, but reveal
itself in a high clone rate.

Excessive tightening of the reference set criteria can potentially
also ameliorate the visible track finding efficiency, but it will be
at the cost of the effective acceptance, since the total yield of
particles with a certain physical signature is proportional to the
product
\begin{equation}
  \epsilon_{total} = \epsilon_{reco} \cdot \epsilon_{geo}
\end{equation}
always assuming that relation~(\ref{eq:nonref}) holds.

\section{Global Methods of Pattern Recognition}
\label{sec:globalPattern}
\begin{figure}[htbp]
 \begin{center}
  \epsfig{file=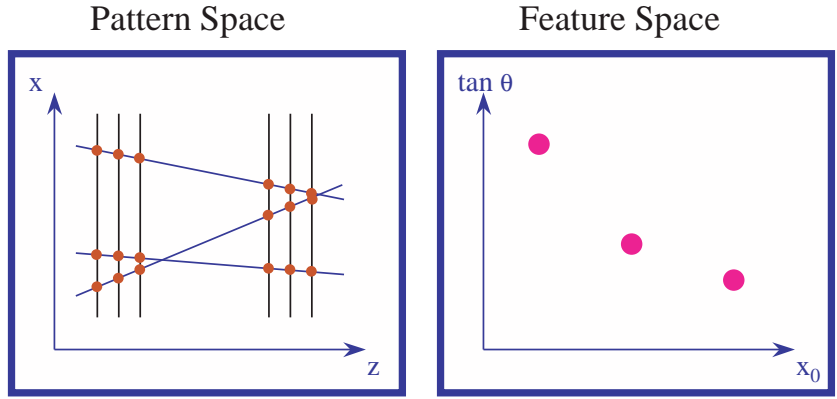,width=13cm,angle=0}
 \end{center}
\caption{Schematic illustration of Pattern Space (left) and Feature
  Space (right)}
\label{fig:patternFeature}
\end{figure}

The task of pattern recognition in general can be described by the
illustration in fig.~\ref{fig:patternFeature}. The physical properties
of the particles that are subject to measurement are described by a
set of parameters, as point of origin, track direction or momentum.
Each particle can therefore be represented by a point in the {\it
  feature space} spanned by these parameters. The signals the particle
leave in the electronic detectors are of a different kind, they are
measured hit coordinates the nature of which is governed by the type
of device.  These coordinates are represented in the {\it pattern
  space}. While the conversion from {\it feature} to {\it pattern} is
done by nature, or by sophisticated simulation algorithms in case of
modelled events, the reverse procedure is the task of the combined
pattern recognition and track fitting process.

\setlength{\unitlength}{1.mm}

Global methods assess the pattern recognition task by treating all
detector hits in a similar way. The result should be independent of
the starting point or the order in which hits are processed.
This is unlike the {\it local methods} that will be discussed in
section~\ref{sec:local}, which depend on suitable {\it seeds} for
track candidates. Global methods aim to avoid any kind of seeding
bias.

\subsection{Template Matching}
The simplest method of pattern recognition can be applied if the
number of possible patterns is finite and the complexity limited
enough to handle them all. In this case, for each possible pattern a
template can be defined, for example a set of drift chamber cells
through which track candidates in a certain area will pass. Such a
technique has been used for the {\it Little Track Finder}, which was
part of the second trigger level of the ARGUS experiment~\cite{ltf},
and which worked by comparing the hits in the drift cells of the axial
layers to masks stored in random access memory. This method allowed
for basic track finding in a 2D parameter space, the track azimuth and
the curvature in the $R/\phi$ projection, within 20~$\mu$s. The
granularity of the ARGUS drift chamber was moderate, which limited the
number of templates that had to be generated.  The concept was later
extended to the ARGUS {\it vertex trigger}~\cite{argusVertexTrigger},
which used the hits of the micro-vertex detector~\cite{argusMicro} and
generalized the algorithm to three dimensions and four parameters
(track curvature being negligible), which allowed to measure the track
origin in $z$ to reject background interactions in the beam pipe. This
algorithm required the definition of more than 245000 masks, where a
five-fold symmetry of the detector had already been exploited.

\setlength{\unitlength}{1.mm}
\begin{figure}[htbp]
 \begin{center}
   \epsfig{file=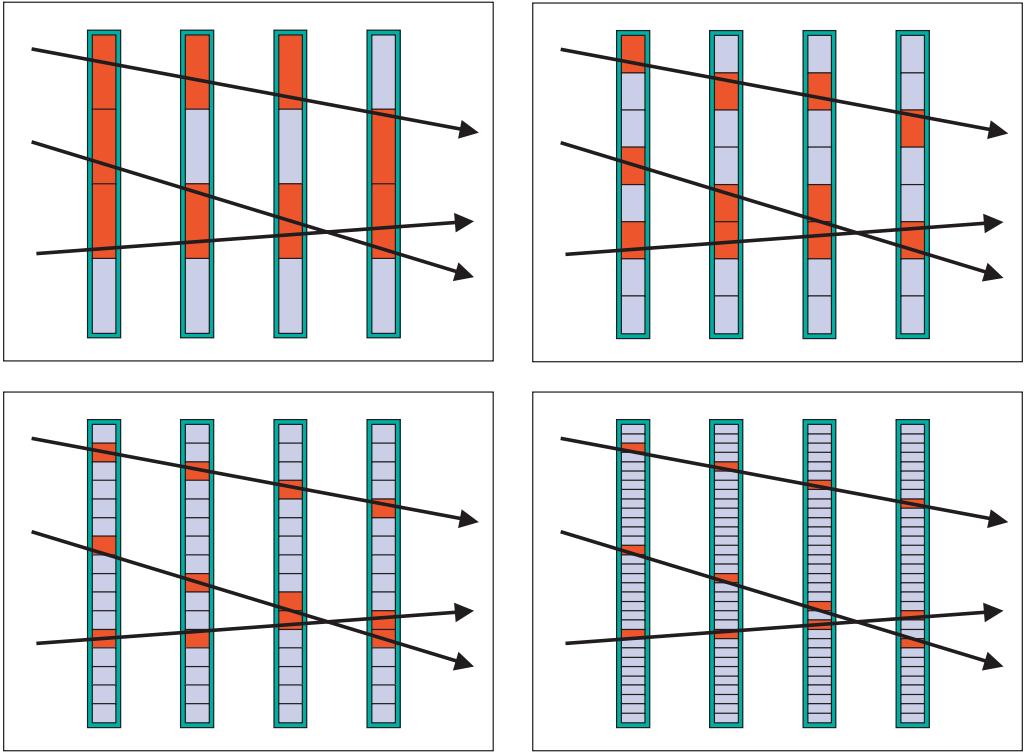,width=10cm,angle=270}
 \end{center}
\caption{Schematic illustration of the tree-search algorithm: in
  several steps (in this case four), the track is matched with
  templates of increasing granularity and resolution. Each step
  descends into the next level of template hierarchy.}
\label{fig:tree}
\end{figure}
Template matching algorithms are mathematically so simple that they
can be hard-wired as {\it track roads}, provided that the hit
efficiency of each element is close to one. Remarkably, the computing
time may be independent of the event complexity, since the number of
templates to be checked is always the same. However, template matching
does not scale very well when the problem requires high dimensionality
or granularity. On one side, with increasing granularity the number of
templates quickly exceeds limits of feasibility already when storing
them. Also the number of computations increases strongly with a finer
resolution of templates. Keeping the granularity low, on the other
hand, means that dense situations cannot be resolved, and other
methods have to be used to disentangle them.

An elegant solution to both problems is the {\it tree-search}
algorithm, which uses templates of increasing structural resolution
that are ordered in a hierarchy~\cite{dellorso,treeprocessor}. In the
first step, the hit structure is viewed at a very coarse resolution
with a small set of templates (fig.~\ref{fig:tree}). For those
templates that have ``fired'', i.e. which match a structure prevalent
in the event, a set of daughter templates with finer granularity is
applied which are all compatible with the first matched template. This
subdivision of templates is iterated until either a matching template
on the finest level of granularity is reached -- indicating that a good
track candidate has been found -- or a pattern matched at a certain
resolution level cannot be resolved at the next level, in which case
it is attributed to a random combination of hits.

The tree-search approach avoids the linear growth of the number of
computations with increasing granularity that would develop in a
purely sequential search; instead, the computing effort, at least for
small occupancy, increases only logarithmically with the number of
detector channels. The algorithm becomes even handier when storage
of all possible templates can be avoided: in many cases symmetries of
the detector can be used to formulate rules how the daughter templates
can be derived from the parent at run-time, and how they are connected
with the event data. The tree-search algorithm is used for example in
the pattern recognition of the HERMES spectrometer, where the final
detector resolution of 250~$\mu$m is reached in 14
steps~\cite{hermes}. Application of tree-search ideally requires
considerable simplicity and symmetry in the detector design, and
therefore cannot be easily used in many complex cases. In particular
inhomogeneous magnetic fields can complicate the application.

\subsection{The Fuzzy Radon Transform}
\label{sec:fuzzyRadon}
In a very general sense, the observed hit density in the event can be
described by a function $\rho(x)$, where $x$ is a very general
description of the measured set of hit quantities. In absence of
stochastic effects, the expected hit density in the pattern space
can be described by an integral
\begin{equation}
        \rho(x) = \int_{P} \rho_{p}(x) D(p) dp
\end{equation}
where $D(p)$ describes the prevalent population of the feature space,
typically a sum of delta functions centred at the parameters of the
particles, and $\rho_{p}(x)$ is the average response function in
pattern space for a particle with parameters p, including all detector
layout and resolution effects~\cite{fuzzyBlom}.

Pattern recognition can then be regarded as an inversion of the above
integral from a stochastically distorted $\rho(x)$. The Fuzzy Radon
transform of the function $\rho_{p}(x)$ is defined as
\begin{equation}
        \tilde D(p) = \int_{X} \rho(x) \rho_{p}(x) dx
\end{equation}
This transformation requires precise knowledge of the response
function, in particular the detector resolution. Track candidates are
then identified by searching local maxima of the function $\tilde
D(p)$.
\setlength{\unitlength}{1cm}
\begin{figure}[htbp]
 \begin{center}
 \begin{picture}(9,19)
   \put(1,9.5){\makebox{
   \epsfig{file=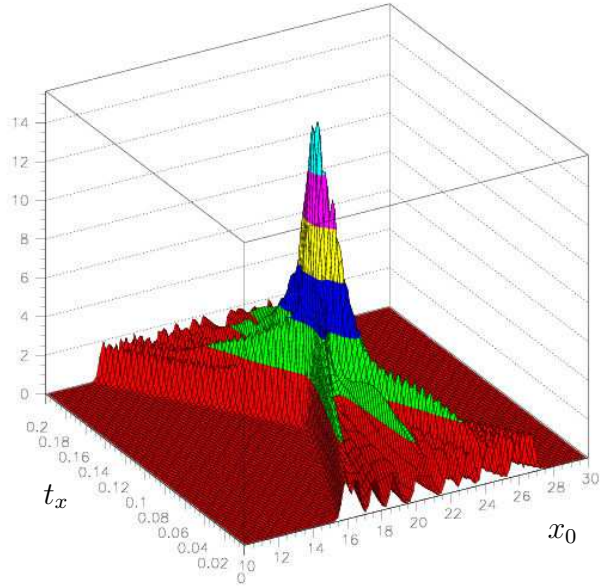,width=9cm,angle=0}
   }}
\put(8.7,10.3){\makebox(1,0.5)[l]{$x_0$}}
\put(2,10.8){\makebox(1,0.5)[l]{$t_x$}}
\put(0,16.5){\makebox(1,0.5)[l]{\Large (a)}}
\put(1,0){\makebox{
    \epsfig{file=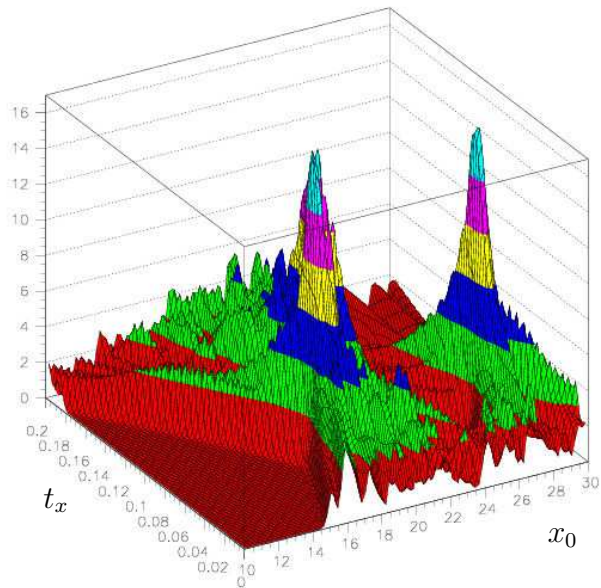,width=9cm,angle=0}
    }}
\put(8.7,0.8){\makebox(1,0.5)[l]{$x_0$}}
\put(2,1.3){\makebox(1,0.5)[l]{$t_x$}}
\put(0,7){\makebox(1,0.5)[l]{\Large (b)}}
 \end{picture}
 \end{center}
\caption{Fuzzy Radon transform $\tilde D(x_0,t_x)$ of the hit signals of a
  single track (a), and in a scenario with three tracks (b),
  where $x_0$ and $t_x$ are the track offset and slope.}
\label{fig:fuzzy}
\end{figure}

This method shall be illustrated in a simple example with a tracking
system consisting of ten equidistant layers in two dimensions without
magnetic field. Tracks are parametrized by an impact parameter $x_0$
and a track slope $t_x = \tan \theta_x$ as defined in
sec.~\ref{sec:forwardGeometry}. As the measurement is one-dimensional,
each hit coordinate gives a linear warp-like constraint in the
parameter plane, where the width of the warp reflects the effect of
the detector resolution (fig.~\ref{fig:fuzzy}a).  For a fictitious
situation with three superimposed tracks, the resulting Fuzzy Radon
transform is shown in fig.~\ref{fig:fuzzy}b. The three peaks are very
pronounced, but development of additional local minima is already
visible even in this clean situation.

\begin{figure}[htbp]
 \begin{center}
   \epsfig{file=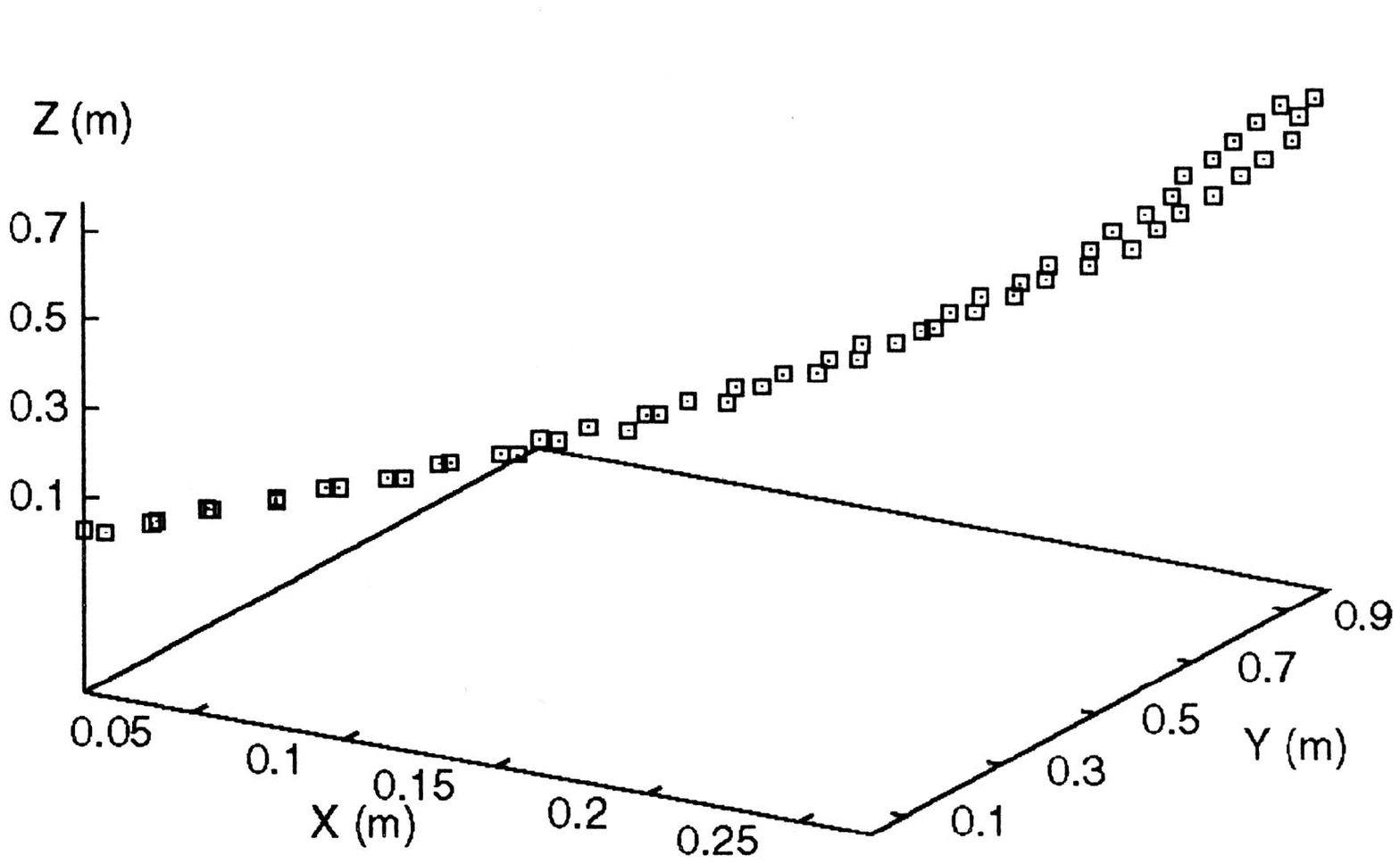,width=6cm,angle=0}
 \end{center}
\caption{Two simulated tracks differing only by curvature
  (taken from~\cite{fuzzyBlom})}
\label{fig:blumTracks}
\end{figure}
\begin{figure}[htbp]
 \begin{center}
   \epsfig{file=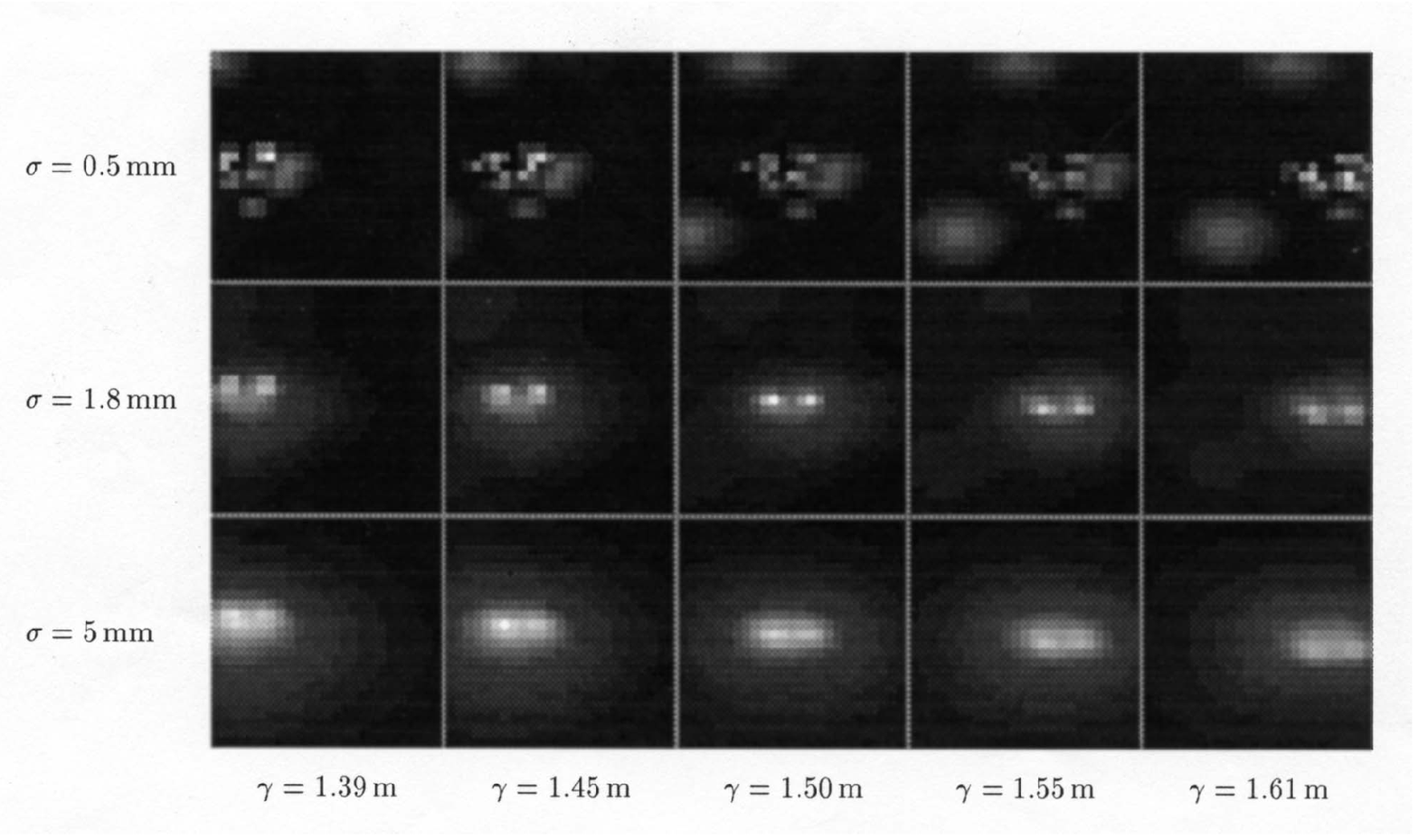,width=13cm,angle=0}
 \end{center}
\caption{Fuzzy Radon transform of the two tracks in
  fig.~\ref{fig:blumTracks} displayed in ($\kappa,\phi$) space, with
  the third track parameter $\gamma$ as described in the text (taken
  from~\cite{fuzzyBlom}). The transform is shown for three values of
  the resolution parameter $\sigma$ in $\rho_p(x)$, where the value in
  the middle row corresponds to the simulated resolution.}
\label{fig:blumTransform}
\end{figure}
In~\cite{fuzzyBlom} this method has been explored for a cylindrical
geometry in the case of two very close tracks which only differ by a
small difference in the curvature value (fig.~\ref{fig:blumTracks}),
with additional noise taken into account.
Figure~\ref{fig:blumTransform} shows the resulting Radon transform
$\tilde D(\kappa,\phi,\gamma)$ as a series of five images around the
central values ($\gamma$ stands for the {\it z speed} of the particle
which is a measure of the dip angle tangent explained in
section~\ref{sec:cylindrical}), where also the resolution parameter
$\sigma$ has been varied. The images show that the individual tracks
can in fact be distinguished (centre image), but it is essential that
the assumed resolution parameter matches the real one.  It should be
noted that automated recognition of the ``track signals'' in such
images would not be a trivial task, and that, for practical purposes,
analysis of fuzzy Radon transforms in multi-dimensional parameter
spaces are in general very demanding in terms of computing power.

Another generalization of the Radon transform has been investigated
in~\cite{harlander}.

\subsection{Histogramming}
\begin{figure}[htbp]
 \begin{center}
  \epsfig{file=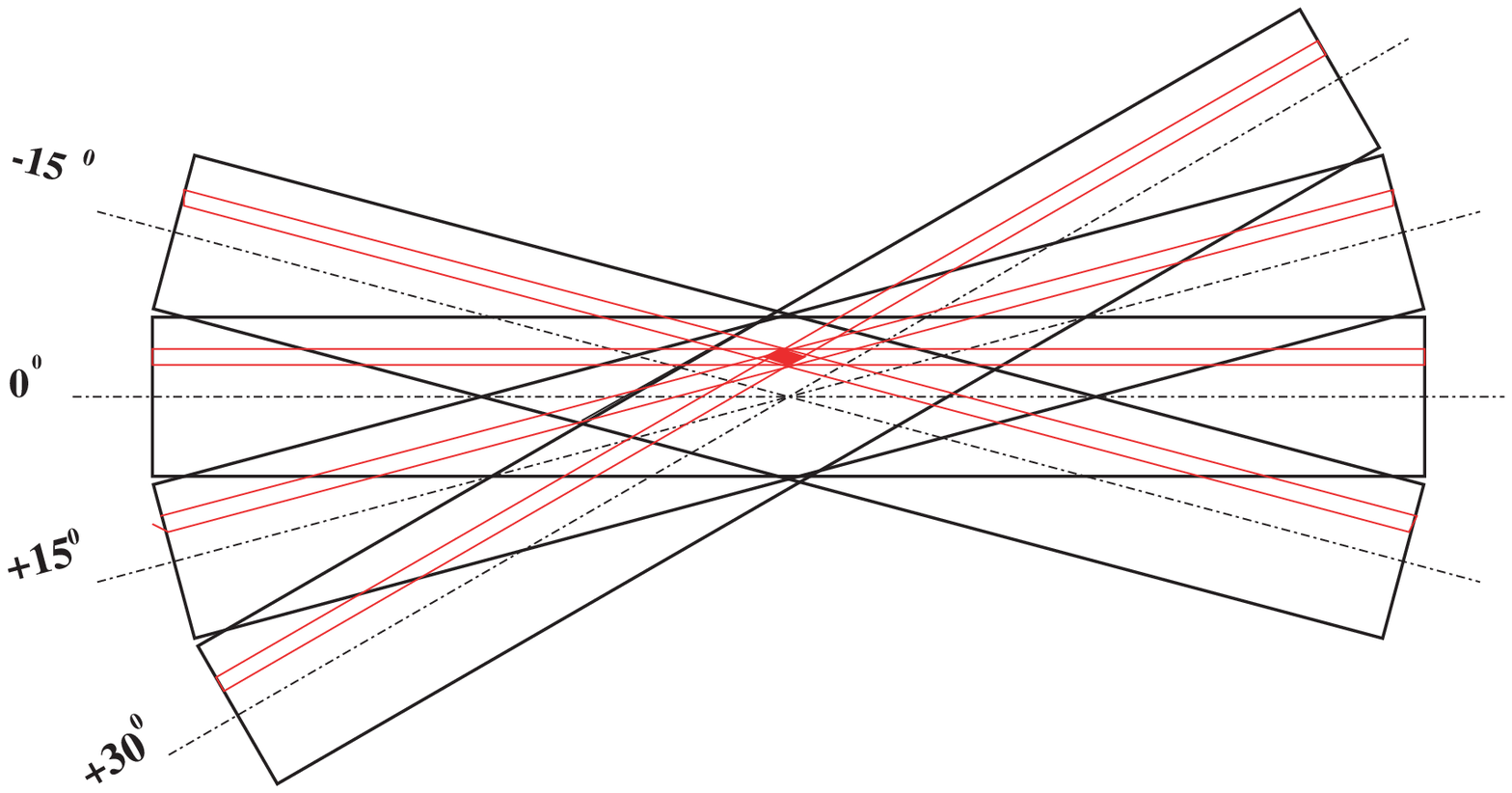,width=13cm,angle=0}
 \end{center}
\caption{Illustration of wire orientations in the ZEUS straw-tube
  tracker. In this representation, the beam is oriented vertical to
  the page, displaced towards the bottom of the page
  (from~\cite{antonov}).}
\label{fig:antonov-views}
\end{figure}
\begin{figure}[htbp]
 \begin{center}
   \epsfig{file=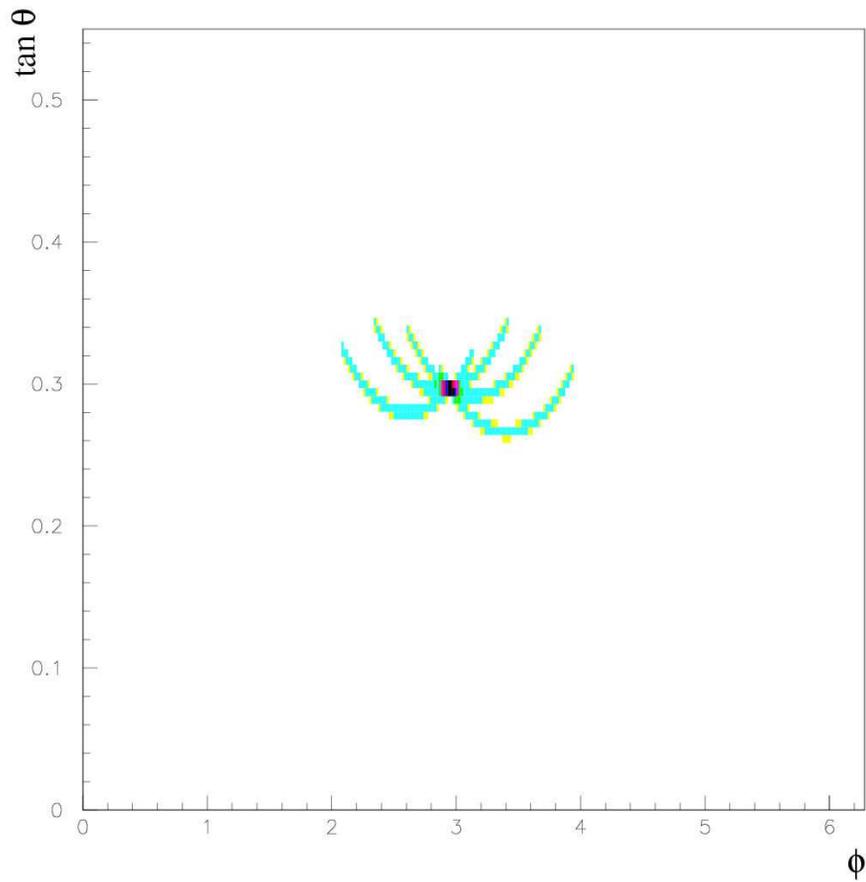,width=13cm,angle=0}
 \end{center}
\caption{Hough transform of a single simulated track in the ZEUS straw-tube
  tracker (from~\cite{antonov}).}
\label{fig:antonov-hough1}
\end{figure}
\begin{figure}[htbp]
 \begin{center}
  \epsfig{file=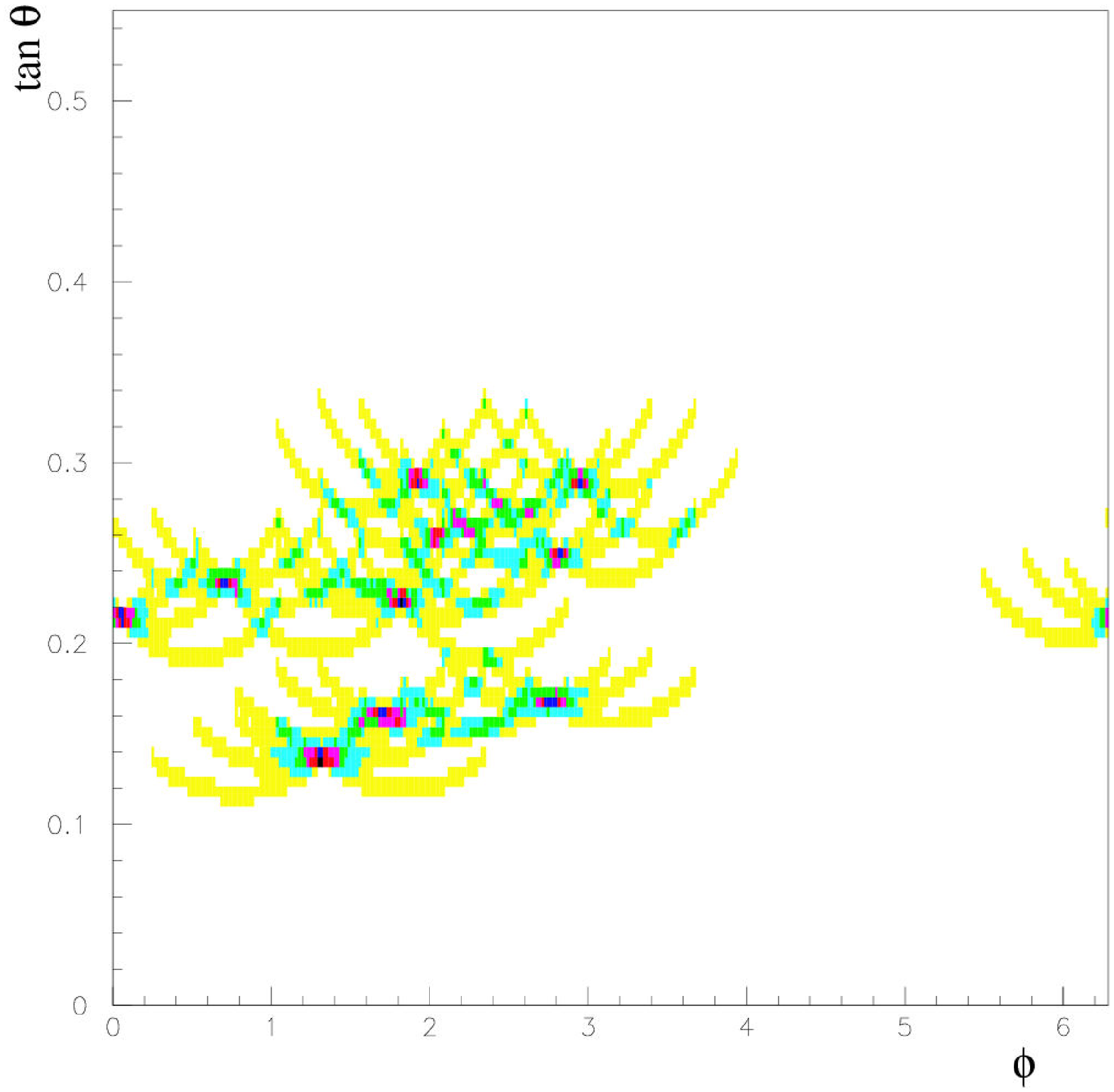,width=13cm,angle=0}
 \end{center}
\caption{Hough transform of a set of simulated tracks in the ZEUS straw-tube 
  tracker (from~\cite{antonov}).}
\label{fig:antonov-houghn}
\end{figure}

As seen in the previous section, the fuzzy Radon transform allows taking
the precise detector resolution into account in an elegant
manner. In cases where effects of the resolution can be neglected, the
response function $\rho_{p}(x)$ only needs to describe the trajectory,
and takes the shape of a delta-function whose argument vanishes for
points on the trajectory. This special form of the Radon transform is
often called {\it Hough} transform~\cite{hough}. The Hough transform
of each point-like hit in two dimensions becomes a line; in more
generality it defines a surface in the feature space. Completion of
the pattern recognition task is thus converted into finding those
points in feature space where many of such lines or surfaces
intersect, or at least approach each other closely in shape of
knots~\cite{hough}.

Histogramming can be regarded as a discrete implementation of the
Hough transform. Hit information is converted to a constraint in a
binned feature space, and the frequency of entries in a bin above a
certain limit is indicative for a track candidate. However, in most
tracking devices a single measurement is not sufficient to constrain
all track parameters. One solution is then to convert each measurement
into a discretized curve or surface in parameter space, and to sample
the contribution of all hits in corresponding accumulator cells. An
example for such an implementation is shown for the straw-tube tracker
(STT) of the ZEUS experiment~\cite{antonov}. This detector system is
used as a forward tracker and consists of two superlayers with eight
layers of straw tubes each. The straws are arranged in the four
different stereo views $0^\circ$, $\pm 15^\circ$ and $30^\circ$, as
illustrated in fig.~\ref{fig:antonov-views}. The $0^\circ$ straws are
oriented such that the point of closest approach to the beam line is
in the middle of the straw.  Taking the beam spot into account and
neglecting the curvature of the segment within the confines of the
straw tube tracker, each hit provides an arc-like constraint in the
parameter space spanned by polar angle $\theta$ and the azimuth angle
$\phi$.  This structure is displayed in the histogram from four views
for a single track in fig.~\ref{fig:antonov-hough1}. The hits from the
$0^\circ$ straws give a transform which is symmetric in azimuth, while
the yields from the other views are slightly skewed in correspondence
to the stereo angle.  The parameters of the track are clearly
indicated by the intersection of the four constraints. The resulting
histogram is already much more complex in a sample with 10~simulated
tracks, where combinatorial overlaps occur
(fig.~\ref{fig:antonov-houghn}).

Another popular way of avoiding the underconstrained case is to
combine several hits to track segments before applying the Hough
transform. For example, in a 2D {\it pattern space} without magnetic
field, two measured coordinates in the same projection from nearby
hits in different detector layers give a straight track segment which
represents a point in the {\it feature space}. Histogramming all
segment entries in the feature space should then reveal track
candidates as local maxima. This procedure is often referred to as
{\it local Hough transform}~\cite{ohlsson}.

\setlength{\unitlength}{1cm}
\begin{figure}[htbp]
  \begin{center}
    \begin{picture}(9,9)
      \put(0,0){\makebox{
          \epsfig{file=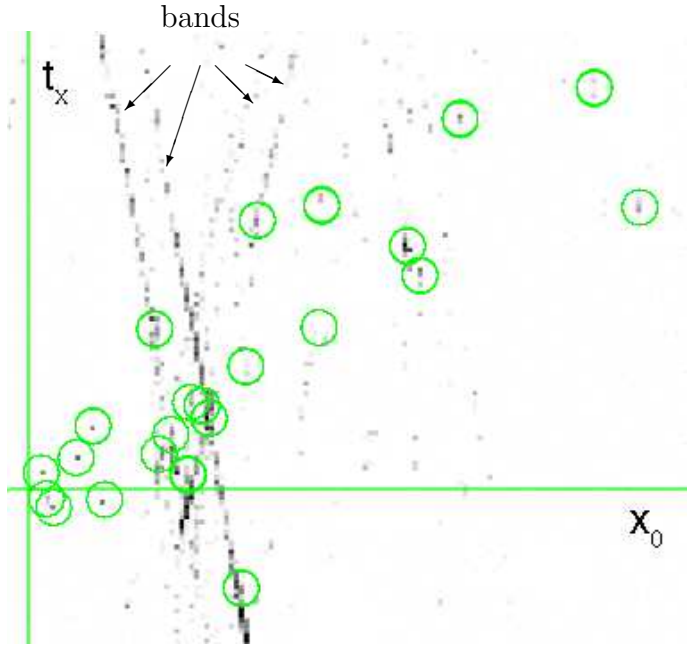,width=9cm,angle=0}
          }}
      \put(2.3,7.7){\vector(-1,-1){0.6}}
      \put(2.7,7.7){\vector(-1,-3){0.45}}
      \put(2.9,7.7){\vector(1,-1){0.5}}
      \put(3.3,7.7){\vector(2,-1){0.5}}
      \put(1.7,8.1){\makebox(2,0.5)[c]{bands}}
    \end{picture}
  \end{center}
  \caption{Local Hough transform in a simulated event with five
  interactions, in the feature space spanned by impact parameter $x_0$
  and track slope $t_x = \tan \theta_x$ (from~\cite{borg}). The
  parameters of true particles are illustrated by circles. The colour
  intensity in each pixel corresponds to the count of segments falling
  into this square. While the histogram shows the expected
  enhancements at the {\it true} parameters of most simulated
  particles, it also displays artificial structures, indicated as
  {\it bands} in the plot that complicate the analysis.}
\label{fig:houghHerab}
\end{figure}
In general, a price has to be paid for this artificial construction of
a higher dimension of measurement, since random combinations of hits
of different origin lead to ghost segments. The abundance of such
contaminations depends strongly on the hit and particle density. A
practical example illustrating this problem is shown in
fig.~\ref{fig:houghHerab} (taken from~\cite{borg}). The geometry
corresponds to the ``PC'' part of the HERA-B spectrometer (see
fig.~\ref{fig:herabDetector}), which consists of four tracking
superlayers, as indicated in fig.~\ref{fig:localBands}a, though in the
latter the drawing has been simplified from six to three individual
layers per superlayer.  A simulated high-multiplicity event with five
simultaneous $pN$ interactions has been passed through a local Hough
transform, from which a closeup is shown in fig.~\ref{fig:houghHerab}.
The genuine tracks as generated by the Monte Carlo are indicated as
circles in the feature space. While enhancements on the histogram are
clearly seen at the track parameters of the true particles (indicated
by circles), the histogram shows a significant number of bands which
are caused by the interference of track patterns. Such interference
occurs when several tracks cross the same superlayer of the tracking
system within a close distance, as illustrated in
fig.~\ref{fig:localBands}b for four intersecting tracks: the proximity
gives rise to a multitude of combinatorial segments, which have
roughly the correct spatial information ($x_{SL3}$), but a wide range
of deviating slopes shadowing the entries with the proper value.
These segments enter the histogram with their spatial coordinate
transformed to the reference plane relative to which all impact
parameters are defined (in this case given by $z=z_{ref}$) in the
manner
\begin{equation}
  x_0 = x_{SL3} + (z_{ref}-z_{SL3}) \cdot \tan \theta_x
\end{equation}
The wide spread in the slope $\tan \theta_x$ results in a band in the
parameter space, where the tilt of the band
\begin{equation}
  \frac{d\tan \theta_x}{dx_0} = \frac{1}{z_{ref}-z_{SL3}}
\end{equation}
reflects the distance of the superlayer (at $z_{SLi}$) from the
reference plane (at $z_{ref}$). It is therefore not surprising that in
the given detector example with four superlayers, bands of four
different slopes can occur.

\setlength{\unitlength}{1cm}
\begin{figure}[htbp]
  \begin{center}
    \begin{picture}(13,6)
      \put(-0.5,5.2){\makebox(1,0.5)[l]{\Large (a)}}
      \put(   7,5.2){\makebox(1,0.5)[l]{\Large (b)}}
      \put(0,5)
      {\makebox{\epsfig{file=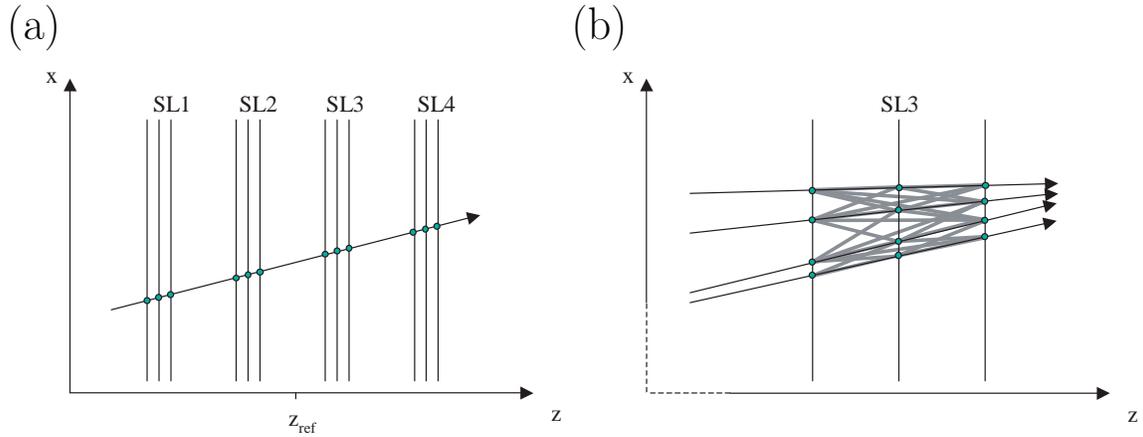,width=5cm,angle=270}}}
    \end{picture}
  \end{center}
  \caption{(a) Illustration of the model detector with four tracking
    superlayers discussed in the text, with the response of a single
    passing track. (b) Schematic illustration of track segments for a
    local Hough transform generated from four tracks intersecting in
    superlayer SL3, showing the abundance of ghost segments compared
    to the proper ones.}
\label{fig:localBands}
\end{figure}
Even in absence of ghost segments from track overlap, the pattern of
track signals in the discretized feature space will in general reflect
the underlying layer structure of the tracking system. The local Hough
transform is usually based on {\it short segments}, i.e.  those
composed of hits in subsequent or at least nearby layers, which has
the advantage that the line topology of the track is exploited and the
background from random hit combinations is still relatively small.
However, due to the small leverage, the angular error can be sizeable,
which may impose additional difficulty in identifying the track
candidates in the Hough transform. {\it Long segments} spanning across
many layers of the tracking system have the principal advantage of
better angular resolution. However, a wide variety of hits have to be
combined, so that the number of random combinations increases
accordingly. The performance of different approaches has been
analyzed in detail in~\cite{schober}. For the individual application,
the optimal choice will depend on the relative importance of
resolution and multiple scattering effects.

\subsection{Neural Network Techniques}
The human brain is particularly skilled in recognizing patterns. It is
capable of analyzing patterns in a global manner; it is
self-organizing, adaptive and fault-tolerant. It is therefore not
surprising that methods have been sought for which aim at solving
pattern recognition problems by means of artificial neural networks.
Another intriguing aspect of the human brain is the massively parallel
processing of information, which raises hopes that algorithms can be
derived which can take full advantage of inherently parallel computing
architectures. Because of the wide scope of this subject, this article
cannot give a full introduction into this field. A collection of
classic papers reprinted is available in~\cite{reprint}.

An artificial neuron manifests a simple processing unit, which
evaluates a number of input signals and produces an output signal. A
neural network consists of many neurons interacting with each other -
the output signal of a neuron is fed into the input of many other
neurons. While many classification problems can be attacked with
simplified layouts, the {\it feed-forward} networks, track pattern
recognition in general uses fully coupled topologies.

\subsubsection{The Hopfield neuron}
\begin{figure}[htbp]
 \begin{center}
  \epsfig{file=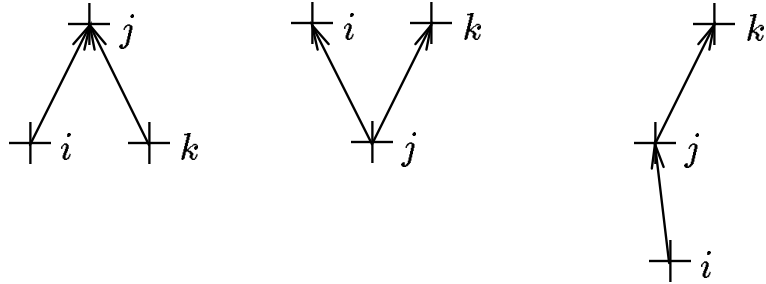,width=10cm,angle=0}
 \end{center}
\caption{Three typical cases for adjacent track segments in the
  Denby-Peterson algorithm. The first two combinations correspond to
  incompatible segments, in the third case, both segments are likely
  to come from the same track. (from~\cite{borg})}
\label{fig:dp-verzweig}
\end{figure}
In the Hopfield model~\cite{hopfield}, each neuron is in general
interacting with every other neuron. All interactions are symmetric,
and the state of each neuron, expressed by its activation $S_i$, can
only be either {\it active} (1) or {\it inactive} (0). The interaction is
simulated by updating the state of a neuron according to the
activations of all other neurons. The update rule in the Hopfield
model sets the new state of a neuron to
\begin{equation}
S_i = \Theta \left( \sum_j (w_{ij} S_j - s_i) \right)
\end{equation}
where the weights $w_{ij}$ determine the strength of each interaction,
$s_i$ are threshold values. The theta function $\Theta(\ldots)$, whose
value is zero for negative arguments and one otherwise, is only the
simplest example of an {\it activation function}, which relates the
updated activation to the weighted sum of the other activations. It
can be shown~\cite{hopfield} that such interactions characterize a
system with an energy function
\begin{equation}
E = -\frac{1}{2} \left( \sum_{ij} w_{ij} S_i S_j - 2 \sum_i s_i S_i
\right)
\end{equation}
and that the interaction leads to a final state which corresponds to
the minimum of the energy function~\cite{hopfield,shrivastava}.

\subsubsection{The Denby-Peterson method}
An adaptation of Hopfield networks to track finding has been developed
by Denby~\cite{denby} and Peterson~\cite{peterson}. The basic idea is
to associate each possible connection between two hits with a neuron.
Activation of such a neuron means that both hits are part of the same
track. It is then essential to define an interaction such that in the
global energy minimum only neurons corresponding to valid connections
will be active. Interaction is only meaningful with neurons that have
one hit in common. An approach to such an energy function is
illustrated in fig.~\ref{fig:dp-verzweig}~\cite{borg}: while in the
first two cases the neurons (ij) and (jk) represent segments
incompatible with the same track and therefore must have a repulsive
interaction, the third case is much more track-like and should have an
attractive interaction. This desired behaviour can be obtained
by an energy function
\begin{eqnarray}
E & = & - \frac{1}{2} \sum \delta_{jk} \frac{-\cos^m
  \theta_{ijl}}{d_{ij}+d_{jl}} S_{ij} S_{kl} \nonumber \\
  & & + \frac{1}{2} \alpha \left( \sum_{l \neq j} S_{ij} S_{il}
       + \sum_{k \neq i} S_{ij} S_{kj} \right)
       + \frac{1}{2} \delta \left( \sum S_{kl} - N \right) ^2
\label{eq:energyDenbyPeterson}
\end{eqnarray}

where $S_{ij}$ is the activation of the neuron associated with the
segment (ij), i.e. the connection between hits $i$ and $j$, and
$\theta_{ijl}$ is the angle between the segments ($ij$) and ($jl$).
The variables $\alpha$ and $\delta$ are Lagrange multipliers preceding
terms that suppress unwanted combinations as the first two cases in
fig.~\ref{fig:dp-verzweig}, and fix the number of active segments to
the number of hits, $N$. Track finding is then reduced to finding the
global minimum of this multivariate energy function. The interaction
is simulated by recalculating the activity of each neuron with the
{\it update rule}, which takes the activations of all other neurons
into account.

It is remarkable that the Denby-Peterson method works without actual
knowledge of a track model -- it favours series of hits that can be
connected by a line as straight as possible, but also allows small
bending angles from one segment to the next. Thus also curved tracks
can be found, provided that a sufficient number of intermediate
measurements exists which split the track into a large number of
almost collinear segments. The Denby-Peterson algorithm is in
particular indifferent about the global shape of the track - a circle
and a wavy track with the same local bending angles but alternating
directions are of equal value.

One of the first explorations of the Denby-Peterson method has been
performed on track coordinates measured by the ALEPH
TPC~\cite{stimpflAbele}. The algorithm found tracks in hadronic $Z^0$
decays rather accurately, which may be at least partially attributed
to three favourable circumstances: pattern recognition benefits
considerably from the the 3D nature of the hits measured in the TPC,
and equally from the clean event structure and the low occupancy.
Moreover, the algorithm is applied such that the initialization
activates only neurons that already correspond to plausible
connections of hits. The authors of~\cite{stimpflAbele} have also
investigated the behaviour of the method on events with much higher
track numbers, simulated by piling up Monte-Carlo events, and found
that the total CPU time of the neural network algorithm is dominated
by the initialization of the neurons, which indicates the degree of
selection already involved at this stage.

\setlength{\unitlength}{1.mm}
\begin{figure}[htbp]
 \begin{center}
  \epsfig{file=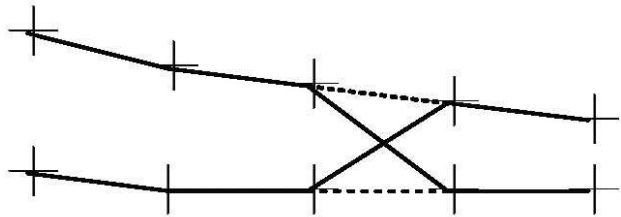,width=9cm,angle=0}
 \end{center}
\caption{Wrong activations in the case of nearby tracks (from~\cite{borg}).}
\label{fig:BorgCrossing}
\end{figure}
The behaviour of the Denby-Peterson method under high track densities
has been further investigated in~\cite{borg} by applying it to a four
superlayer geometry resembling the ``PC'' part of the HERA-B tracker
(see fig.~\ref{fig:herabDetector}). These studies found that the
classical Denby-Peterson method cannot be relied on to converge safely
in cases of nearby parallel tracks. This behaviour is explained in
fig.~\ref{fig:BorgCrossing}: there is no possibility of resolving a
cross-wise misassignment, since the system has reached a local energy
minimum, and no additional segment can be attached because it would
temporarily lead to an illicit branching of the track according to the
rules illustrated in fig.~\ref{fig:dp-verzweig} and formulated in
eq.~\ref{eq:energyDenbyPeterson}.

\setlength{\unitlength}{1.mm}
\begin{figure}[htbp]
 \begin{center}
  \begin{tabular}{cc}
   \epsfig{file=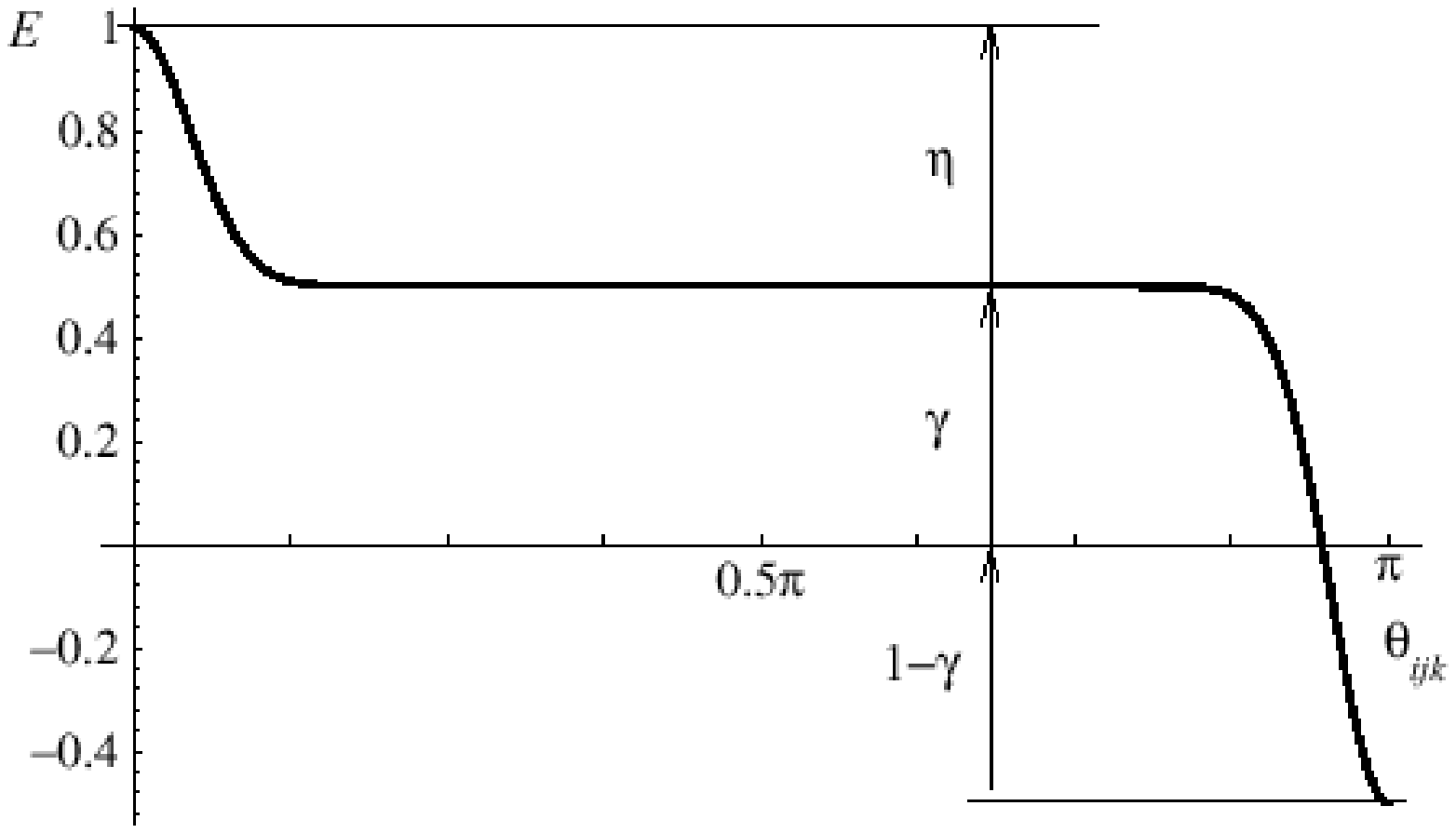,width=7.1cm,angle=0} &
   \epsfig{file=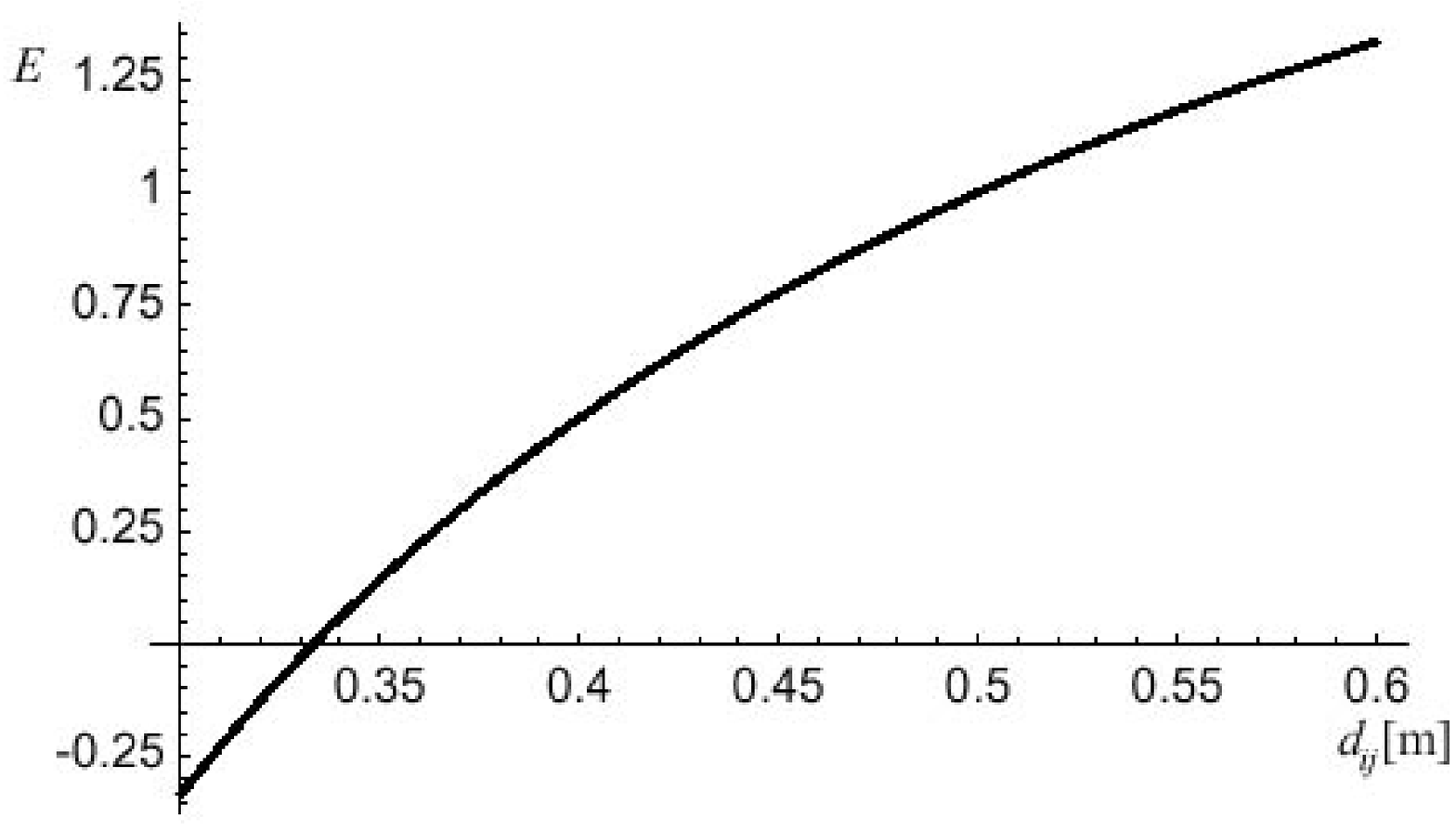,width=7.1cm,angle=0}
  \end{tabular}
 \end{center}
\caption{Modified energy function versus angle $\theta_{ijk}$ (left)
  and generalized segment length $d_{ij}$ (right) as used
  in~\cite{borg}.}
\label{fig:BorgEnergy}
\end{figure}
The situation can be improved, as shown in~\cite{borg} by dropping
the branching restriction and instead accounting for undesired angles
in the cost function, by the replacement
\begin{equation}
 \frac{-\cos^m \theta_{ijl}}{d_{ij}+d_{jl}}
 \rightarrow f(\cos^m \theta_{ij,kl}) 
\end{equation}
where the angle-dependent part is chosen such that only segments with
angles close to $180^{\circ} $ give a strong negative contribution,
and by adding a term proportional to
$ \left( \delta - 1/d_{ij} \right)$
for each neuron, which introduces a typical inverse segment length
$\delta$ into the energy function, where the length of an individual
segment $d_{ij}$ is generalized such that the superlayer
structure of the tracker is taken into account. (The full definitions
are given in~\cite{borg}.)
The energy as function of segment angle and length
is displayed in fig.~\ref{fig:BorgEnergy}.

\begin{figure}[htbp]
 \begin{center}
  \epsfig{file=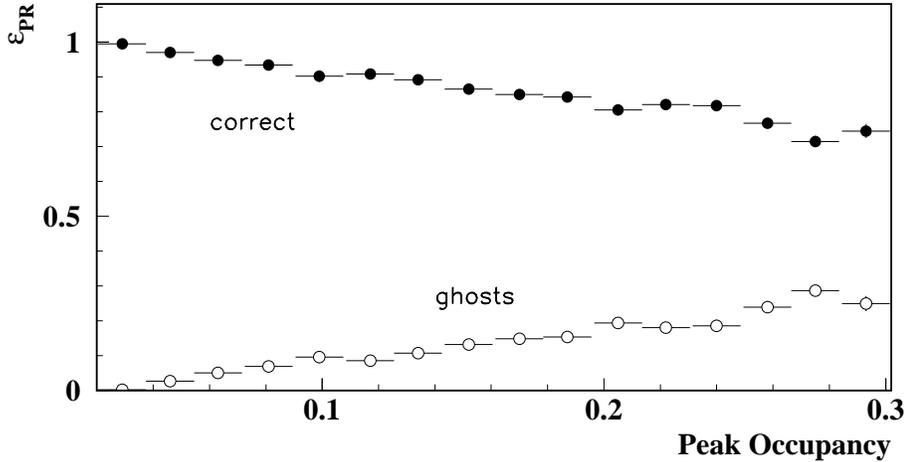,width=13cm,angle=0}
 \end{center}
\caption{Efficiency and ghost rate for pattern recognition using the
  modified 2D Denby-Peterson algorithm on simulated events in a fixed
  target geometry (from~\cite{mankelChep}).}
\label{fig:dp-effvsocc}
\end{figure}

\setlength{\unitlength}{1.mm}
\begin{figure}[htbp]
 \begin{center}
  \epsfig{file=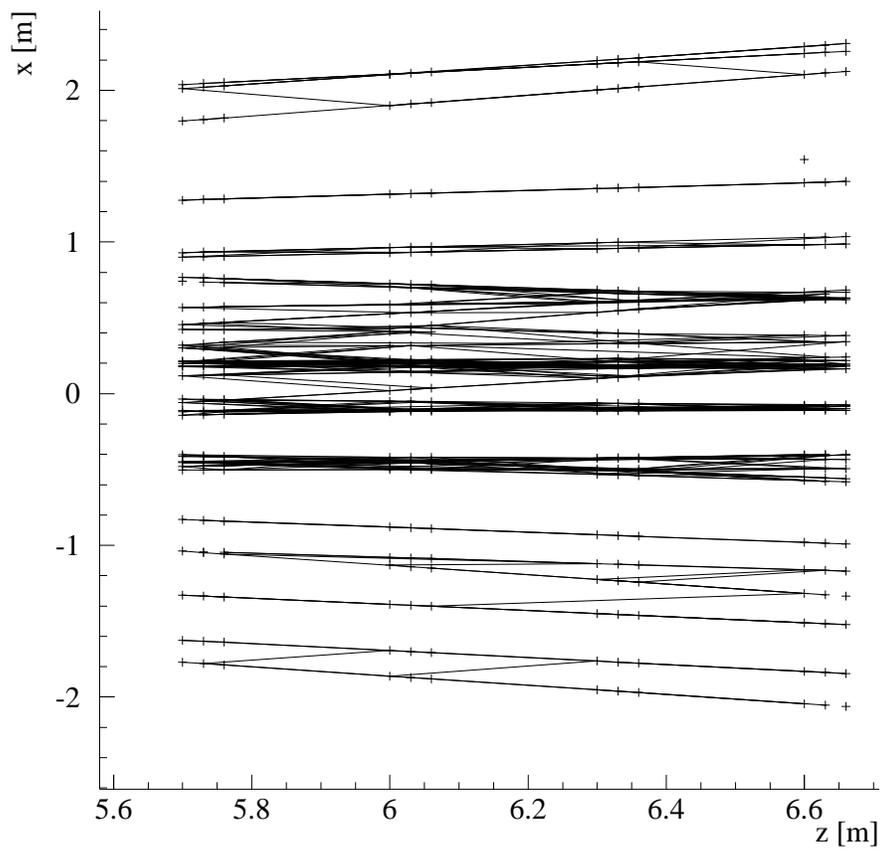,width=13cm,angle=0}
 \end{center}
\caption{State of the network after one iteration~\cite{borg}. Crosses
  denote the locations of the simulated hits.}
\label{fig:BorgOneIt}
\end{figure}
The effect of this variation of the method is visible in
fig.~\ref{fig:BorgOneIt}, which shows the system after one iteration
applied to an event with low track multiplicity. At this point, there
are still branchings that would not be allowed in the classical
Denby-Peterson approach, and which disappear under further iteration.
With these modifications the algorithm obtains reasonable efficiency
and ghost rate values~\cite{mankelChep,borg}, as displayed in
fig.~\ref{fig:dp-effvsocc}.

\setlength{\unitlength}{1.mm}
Several properties of the Denby-Peterson algorithm limit its
application at production scale in the general case. The fact that it
does not take any explicit track model into account lets it ignore
valuable information, which could otherwise help to resolve ambiguous
situations. A straight track with random perturbations e.g. is
equivalent to a slightly curved track. Neither is there a way to take
explicitly the resolution of the detector into account. The computing
time per event increases with the third power of the track density,
since the number of neurons that have to be generated is proportional
to the number of hits squared and the number of non-zero elements in
the weight matrix increases with the number of neurons in the vicinity
of the track. Perhaps the dominant shortcoming of the Denby-Peterson
method is the fact that it does not have a direct extension for
finding 3D tracks on the basis of single-coordinate measurements
(see~\ref{sec:singleCoord}), though it is in principle possible to
circumvent this problem by first forming space points or segments out
of the hits, provided that the ghost combinations are properly
eliminated later. Such an approach has been successfully followed
in~\cite{abtPC}, where a method resembling a discrete form of a
Denby-Peterson net, referred to as {\it cellular
  automaton}~\cite{abtSI}, was used to select optimal combinations of
space points, complemented by a subsequent track following step.

\subsubsection{Elastic arms and deformable templates}
\label{sec:elasticArms}
The above-mentioned limitations of the Denby-Peterson algorithm are
overcome with the {\it elastic arms}
algorithm~\cite{ohlsson,ohlsson2}, which was introduced by Ohlsson,
Peterson and Yuille in 1992. The basic idea can be described as
follows: a set of M {\it deformable templates} is created, which
correspond to valid parametrizations of tracks with parameters
\{$t_1$, ...  $t_M$\}. The number $M$ must be adjusted to the
approximate number of tracks in the event. The algorithm should then
move and deform these templates such that they fit the pattern given
by the positions of N detector hits, which are represented by
\{$\xi_1$ ... $\xi_N$\}.

As in the Denby-Peterson case, the approach proceeds by formulation of
an energy function, whose absolute minimum is at the set of parameters
which solve the pattern recognition problem. This requires two
elements: an activation-like quantity $S_{ia}$ whose value is one if hit
$i$ is assigned to track $a$, and zero otherwise, and a function
$M_{ia}(\xi_i,t_a)$ describing a metric between track template and
hit, typically the square of the spatial distance. The energy function
can then be defined as
\begin{equation}
\tilde E(S,\xi,t) = \sum_{i=1}^N \sum_{a=1}^M S_{ia} M_{ia}(\xi_i,t_a)
\label{eq:etilde}
\end{equation}
To avoid trivial solutions, it is necessary to introduce the condition
that each hit must be assigned to some template in the form
\begin{equation}
\sum_{a=1}^M S_{ia} = 1
\end{equation}
for each hit $i$. This requirement is called {\it Potts
  condition}~\cite{potts}. One immediate consequence of this condition
is the necessity to introduce a special template to which noise hits
can be assigned.

The main challenge is then to find the global minimum of the energy
function. Since this function tends to be very spiky, as will be
illustrated in more detail below, this problem is usually tackled by
extending the energy function according to a {\it stochastic model},
which simulates a thermal motion in the system and smoothens out the
spike structure.  Search of the minimum starts then at high
temperature, and the temperature is successively lowered.  At zero
temperature, the extended energy function becomes identical to the
original one. This technique is called {\it simulated annealing}.

Instead of the temperature $T$, normally its inverse $\beta = 1/T$ is
used.  At finite temperature, the $S_{ia}$ are replaced by their
thermal mean values $V_{ia}$, which take continuous values and lead to
a fuzzy hit-to-track assignment. They can be derived from the metric
function as
\begin{equation}
  V_{ia} = \frac{e^{-\beta M_{ia}}}
    {e^{-\beta \lambda} + \sum_{b=1}^M e^{-\beta M_{ib}}}
\label{eq:potts}
\end{equation}
where the index $b$ in the sum in the denominator runs over all
templates except for the noise template. $V_{ia}$ is called the {\it
  Potts factor}. The temperature determines the range of influence for
a hit: at zero temperature ($\beta \rightarrow \infty$), the hit is
assigned only to the nearest template, with the corresponding $V_{ia}$
equal to one.  At higher temperature, the degree of the assignment
decreases smoothly with increasing distance.  The noise parameter
$\lambda$ represents the symbolic {\it noise template} which, in the
limit of zero temperature, takes over hits that are further than
$\sqrt{\lambda}$ away from the nearest genuine template. It is
therefore logical to set $\lambda$ in correspondence to the detector
resolution, typically as three or five standard deviations. The term
$e^{-\beta \lambda}$ accounts for assignments to the noise template.
The Potts factor of the noise template is calculated as
\begin{equation}
  V_{i0} = 1 - \sum_{a \neq 0} V_{ia}
\end{equation}
instead of eq.~\ref{eq:potts}, since the concept of a distance does
not make sense here.

The only remaining steps necessary to solve the pattern recognition
problem are
\begin{enumerate}
\item to find a suitable initialization for the templates, and
\item to find the absolute minimum of the energy function.
\end{enumerate}
It turns out that both are non-trivial in practical applications.
Before turning to realistic scenarios, it is very instructive to look
at the shape of the energy function in a very trivial example (taken
from~\cite{borg}), which consists of a detector measuring only one
spatial coordinate, named $x$, and a track model consisting only of
one parameter for each template. Two hits are considered with
coordinates $x_1$ and $x_2$, and two templates with parameters $x_a$
and $x_b$.

\begin{figure}[htbp]
 \begin{center}
  \epsfig{file=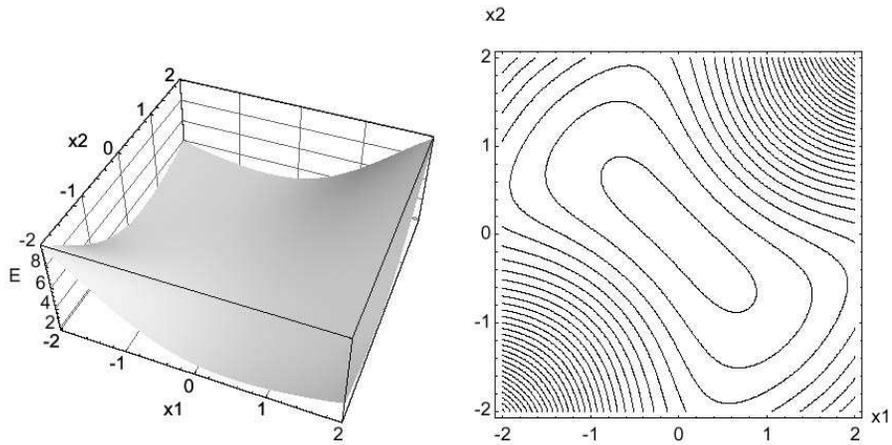,width=12cm,angle=0}
 \end{center}
\caption{Representations of the energy function of a one-dimensional
  detector with two hits, as a function of the parameters of two
  templates $x_a$ and $x_b$ at high temperature~\cite{borg}.}
\label{fig:borgEnergy1}
\end{figure}
\begin{figure}[htbp]
 \begin{center}
  \epsfig{file=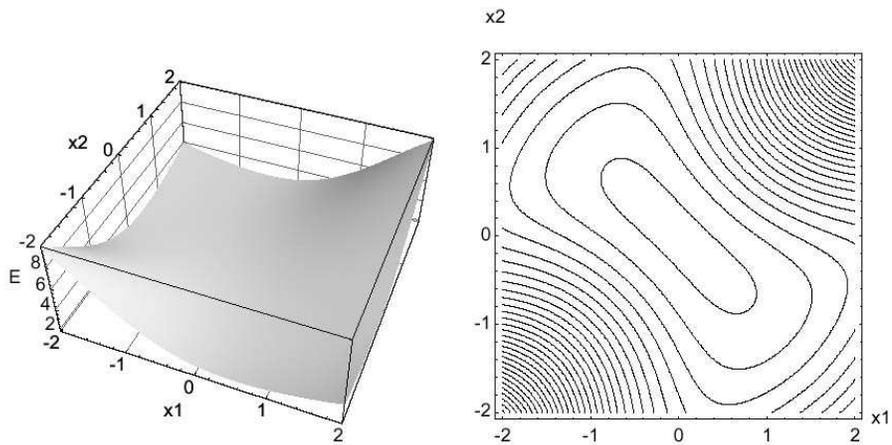,width=12cm,angle=0}
 \end{center}
\caption{Energy function at critical temperature~\cite{borg}.}
\label{fig:borgEnergy2}
\end{figure}
\begin{figure}[htbp]
 \begin{center}
  \epsfig{file=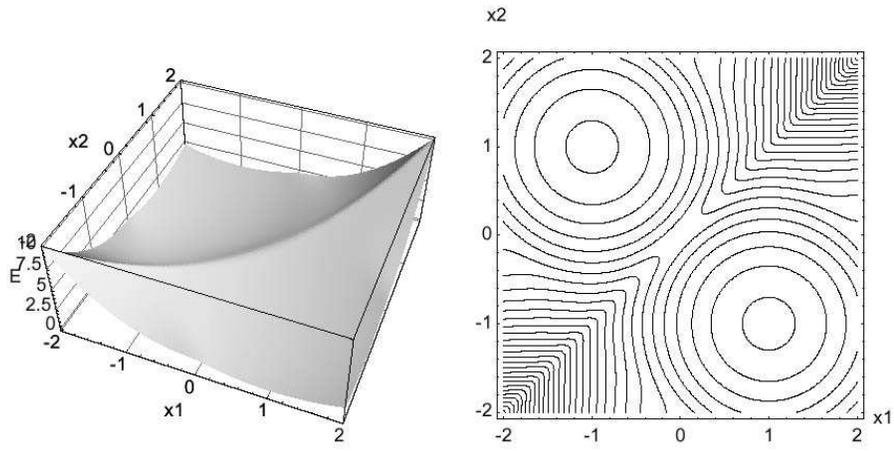,width=12cm,angle=0}
 \end{center}
\caption{Energy function at low temperature~\cite{borg}.}
\label{fig:borgEnergy3}
\end{figure}
\begin{figure}[htbp]
 \begin{center}
  \epsfig{file=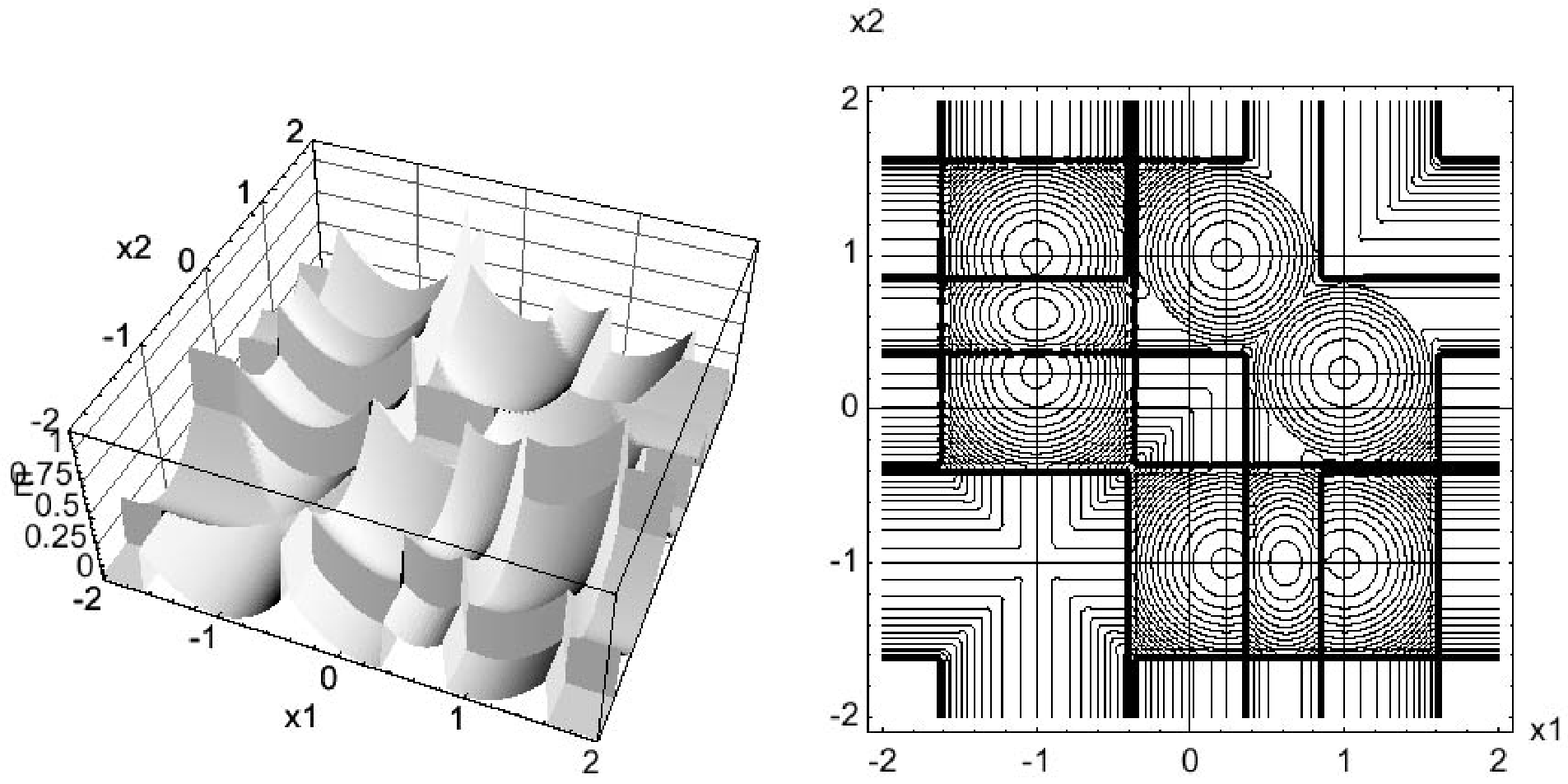,width=12cm,angle=0}
 \end{center}
\caption{Energy function with three hits at low temperature, with
  $\lambda = 0.4$~\cite{borg}}
\label{fig:borgEnergy4}
\end{figure}
The energy as a function of the template parameters is shown in
fig.~\ref{fig:borgEnergy1} at a high temperature (the hits being at
coordinates $x_a=-1$ and $x_b=+1$). At this temperature, the templates
perceive only a blurred image of the hit pattern. The global minimum
is at the coordinates in the centre between the hits. When the
temperature is lowered to a critical temperature $T_c$, a saddle point
develops (fig.~\ref{fig:borgEnergy2}), and the previous single minimum
splits into two. The critical temperature is related to the
coordinates as
\begin{equation}
T_c = \frac{1}{\beta_c} = \left( \frac{x_a - x_b}{2} \right)
\end{equation}
At very low temperature (fig.~\ref{fig:borgEnergy3}), two minima have
developed at positions corresponding to the two equally valid
solutions, $x_a=x_1\ \wedge\ x_b=x_2$ and $x_a=x_2\ \wedge\ x_b=x_1$. The
potential ridge at the line $x_a=x_b$ can be interpreted as a
repulsive force between the templates~\cite{ohlsson}.

The presence of the noise template parameter $\lambda$ introduces
further local minima into the energy function. An example is shown in
fig.~\ref{fig:borgEnergy4} with three hits (with $x_c=0.24$) and
$\lambda =0.4$. While the previous solutions are still valid,
additional minima appear that correspond to either one or two of
the genuine hits being attributed to noise.

The complexity of the energy function for this very simple example is
already staggering, and illustrates why initialization and convergence
are serious issues.

In their initial study, Ohlsson, Peterson and Yuille~\cite{ohlsson}
applied the method to hits from the DELPHI TPC. Reconstruction was
restricted to tracks coming from a vertex spot common to all events,
so that track candidates were described by only three parameters,
which simplified the situation considerably.  The initialization was
obtained with a local Hough transform. The moderate hit density
allowed performing first the Hough transform in the projection
transverse to the magnetic field, searching for track candidates in
the space of curvature and azimuth. For each candidate found as a
narrow peak in this projection, all hits within a certain
neighbourhood were used to calculate the longitudinal tilt angle,
which was again histogrammed.

The elastic arms phase then used gradient descent to minimize the
energy function at a given temperature. The temperature was lowered by
5\% in each step. The Hough transform produced an abundance of
templates. The excessive templates were either attracted to noise, or
converged to tracks that had already templates associated with them;
these had to be weeded out at the end. The result was found to be
rather independent of algorithm parameters. The CPU time per event was
dominated by the elastic arms step (1 min on a contemporary computer),
in contrast to the Hough transform initialization (1~s).

Once more one has to note that pattern recognition in the TPC (here
DELPHI's) benefits strongly from the clean event structure with a
moderate track density, and the remarkable 3D measurement capabilities
of the chamber. An interesting study targeted at much more dense
events with 2D measurements has been performed in
1995~\cite{lindstroem}. The algorithm was applied to the barrel part
of the Transition Radiation Tracker (TRT) of the ATLAS detector, with
40 layers of straw drift-tubes with a diameter of 4~mm and a hit
resolution of 150~$\mu$m. Since the required hit resolution could only
be obtained using the drift-time measurement, the left-right ambiguity
had to be resolved.  This problem was approached with the elegant
method from~\cite{blancenbecler}, which introduces energy terms for
both left-right assignments (in the nomenclature of
eq.~\ref{eq:etilde})
\begin{equation}
  \label{eq:blanc}
  \tilde E(S,\xi,t) = \sum_{i=1}^N 
  \sum_{a=1}^M S_{ia} \left( s_{ia}^+M_{ia}^+(\xi_i,t_a)
  + s_{ia}^-M_{ia}^-(\xi_i,t_a) \right)
\end{equation}
where the left-right assignment parameters $s_{ia}^\pm$, which satisfy
the condition $s_{ia}^+ + s_{ia}^- = 1$, introduce a repulsive
interaction between the alternative left-right assignments, so that a
track can only be assigned to one of the two ambiguities of a hit.

The initialization again used a local Hough transform. The
minimization phase of the elastic arms step at a given temperature,
however, did not rely on simple gradient descent, but used the {\it
  Hessian} matrix, i.e. the second derivative of the energy with
respect to the parameters, in a multidimensional generalization of the
Newton method. The efficiency was found to be 85\% for fast tracks
completely contained in the barrel TRT. The efficiency was practically
identical to the one of the Hough transform itself, indicating that
the elastic arms part did not find any new tracks that had not been
properly covered by the initialization. The main application of the
elastic arms part was therefore to verify track candidates found by
the Hough transform and resolve the hit associations.

\begin{figure}[htbp]
 \begin{center}
  \epsfig{file=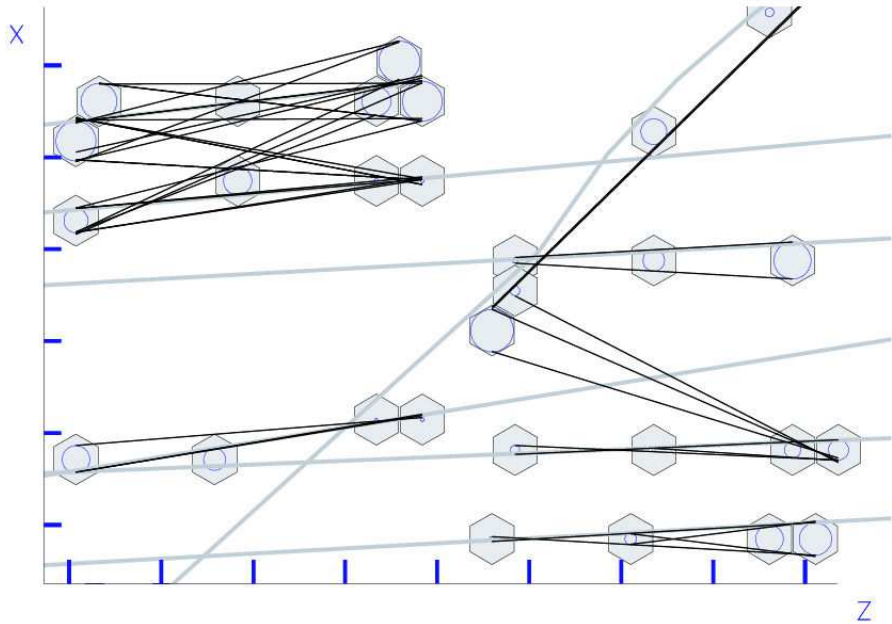,width=12cm,angle=0,
    bbllx=20,bblly=99,bburx=574,bbury=501}
\end{center}
\caption{Illustration of segment initialization in the $zx$
  projection. The circles are drift distance isochrones of each hit
  with the drift cell indicated by a surrounding hexagon. The light
  grey lines are the simulated particles, the black straight lines
  connecting the hits are the segments produced to initialize the
  elastic arms algorithm~\cite{borg}.}
\label{fig:borgSegments}
\end{figure}
The track finding capabilities of elastic arms have been further
investigated in~\cite{borg} and~\cite{paus} with events passed through
a full Geant simulation of the ``PC'' area of the HERA-B spectrometer
(see fig.~\ref{fig:herabDetector}). Since the interpretation of the
Hough transform turned out problematic in the fixed target geometry
under study, a different approach was followed. Track candidates were
initialized by searching hit triplets in the $0^{\circ}$ projection in
each of four superlayers (fig.~\ref{fig:borgSegments}). All triplets
with a straight-line-fit yielding $\chi^2 < 3.8$ were accepted, and then
matched according to their track parameters. Combinations with triplet
segments from all four superlayers were used to initialize the templates
in the horizontal plane. The elastic arms algorithm was
then used
\begin{figure}[htbp]
 \begin{center}
  \epsfig{file=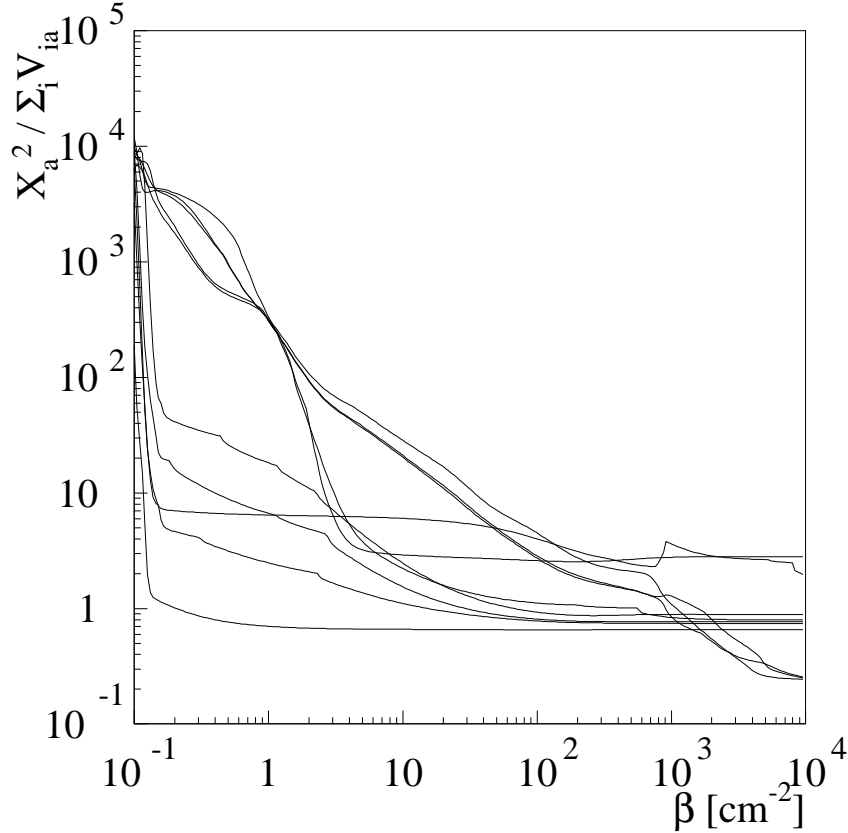,width=12cm,angle=0}
 \end{center}
\caption{Development of $\chi^2$ with increasing $\beta$
  (corresponding to decreasing temperature) with 10 muon
  tracks~\cite{borg}.}
\label{fig:borgChi2Mu}
\end{figure}
to perform the pattern recognition together with the stereo layers,
arranging the tracks vertically. Proper operation of this method was
shown with test events with ten muon tracks, where the convergence of
the tracks in the annealing from a temperature parameter of $0.1 \ 
{\rm cm}^{-2}$ to $10^4\ {\rm cm}^{-2}$ is illustrated in
fig.~\ref{fig:borgChi2Mu} by the decrease of the $\chi^2$ per track.
While the algorithm was actually performing the task of {\it vertical
  pattern recognition} after horizontal initialization, the computing
time for the annealing with 10 tracks turned out to be already about
4000s, and it increased at least with the second power of the number
of templates. For this reason, dense events with 100 and more track
candidates could not be seriously addressed with this method.

\begin{figure}[htbp]
 \begin{center}
   \epsfig{file=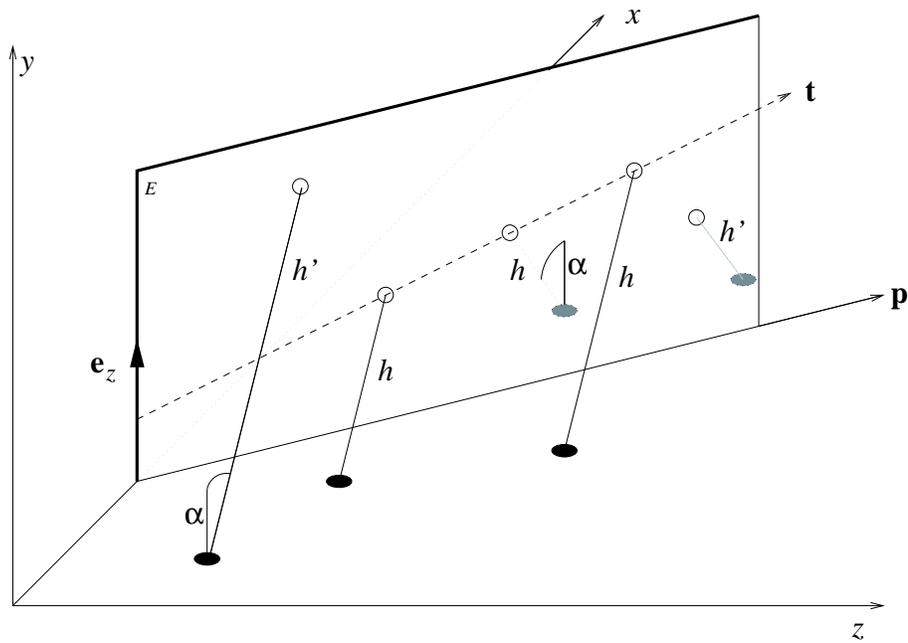,width=12cm,angle=0}
 \end{center}
\caption{Scheme of converting information from stereo layers to vertical
  ($y$) coordinates, by using the horizontal projection track
  candidate that is indicated by {\bf p}. The true track is indicated
  by a dotted arrow labelled {\bf t}. The ovals filled either in black
  or grey are coordinates measured under the stereo angle $\pm \alpha$
  projected into the $xz$ plane. The lines labelled $h$ are stereo
  hits stemming from the true track, these fall on the trajectory {\bf
    t} when the $y$ coordinate is inferred with
  eq.~\ref{eq:convertVertical}. Other hits of different origin
  (labelled $h'$) lead to background hits in the vertical plane (from
  ~\cite{paus}).}
\label{fig:pausVertical}
\end{figure}
For this reason, a subsequent study~\cite{paus} focused on the
reduction of the processing effort. The first major step was the
extension of the segment initialisation to 3D. This was achieved by
using the segments found from triplets in the $xz$ projection to
convert the information from the stereo layers to 3D coordinates: the
segment in the projection defined a vertical plane in which the track
candidate had to be contained (fig.~\ref{fig:pausVertical}).
Intersections of stereo wires with this plane lead to indirect
measurements in the vertical coordinate $y$; the measurement equation
\begin{equation}
  u [v] = x \cos \alpha - (\pm) y \sin \alpha
\end{equation}
was inverted to
\begin{equation}
  y = \pm \frac{x \cos \alpha - u [v]}{\sin \alpha}
  \label{eq:convertVertical}
\end{equation}
and the triplet and segment finding proceeded with the stereo layers
in a similar fashion. The stereo coordinates $u$ and $v$ took drift
distance measurements into account, which improved the resolution but
lead to left-right ambiguities also in the vertical segment finding.

The second crucial improvement concerned the minimization algorithm
within each annealing step. The simplicity of the gradient descent
method has made it highly popular for neural network applications, but
as already observed in~\cite{lindstroem}, it is by far not the most
efficient method. One of its main drawbacks is the fact that its
convergence slows down as it approaches the minimum where the surface
of the energy function flattens out. On the other hand, large
gradients as they can easily occur at lower temperatures (see
fig.~\ref{fig:borgEnergy4}) tend to increase the step size drastically
and throw the algorithm completely off the mark.  These effects
contribute largely to the high computing demands.

It is therefore promising to explore more efficient minimization
techniques for high-dimensioned functions~\cite{paus}. The {\it
  Quickprop} algorithm~\cite{quickprop} parametrizes the dependence of
the energy function on a template parameter $t_a^{(k)}$ (where $a$ is
the identifier of the template and $k$ the index of the template
parameter) in second order
\begin{equation}
E\left( t^{(k)}_a \right)_{\bf \{t_a\}}=
c_0 + c_1 t^{(k)}_a + c_2 \left( t^{(k)}_a \right)^2
\end{equation}
and replaces the parameter in iteration step $(i+1)$ with the value at
the minimum of the parabola, which is calculated using the gradients
of the two previous steps $i$ and $(i-1)$:
\begin{equation}
  \Delta t^{(k)}_{a,i+1} =
  \frac{-\frac{\partial E}{\partial t^{(k)}_a}{\Big |}_i}
  {\frac{\partial E}{\partial t^{(k)}_a}{\Big |}_i
    - \frac{\partial E}{\partial t^{(k)}_a}{\Big |}_{i-1}} \Delta
  t^{(k)}_{a,i}
\end{equation}
Another more sophisticated minimization method, the {\it RPROP
  algorithm}~\cite{rprop}, eliminates entirely the dependence of the
step width of the gradient by using only its sign.  Each component of
the template parameter set has its own step width, which is reduced in
each step if the sign of the partial derivative has not changed, and
somewhat increased if the sign has changed, indicating a step across
the minimum.

In application to fully simulated events, the RPROP algorithm turned
out to be ten times faster than simple gradient descent.  The
Quickprop algorithm reduced the computing time by yet another factor
of two, but failed to converge properly on about 10\% of the tracks,
so that the RPROP algorithm was finally chosen for further
study~\cite{paus}.

\begin{table}[htbp]
\centering
\begin{tabular}{|c|rrr|rrr|}
\hline
$N_{int}$ & \multicolumn{3}{c|}{Segment initialization}
   & \multicolumn{3}{c|}{Elastic arms (incl. initialization)}\\
 & Efficiency & Ghostrate & CPU time & Efficiency & Ghostrate & CPU time\\
\hline
1 & 91\% &   38\% &   4s & 90\% & 3.7\% &  15s\\
2 & 91\% &  100\% &  14s & 89\% & 5.9\% &  40s\\
3 & 89\% &  240\% &  47s & 87\% & 7.5\% &  105s\\
4 & 87\% &  440\% & 107s & 86\% & 10\%  &  198s\\
5 & 85\% & 1100\% & 234s & 83\% & 13\%  &  371s\\
\hline
\end{tabular}
\caption{Efficiency of segment initialization and elastic arms
  algorithm as compiled from~\cite{paus}, as a function of the number
  of superimposed interactions, $N_{int}$. In the elastic arms section
  of the table, efficiency, ghost rate and CPU time include the effects of
  the segment initialization.}
\label{tab:pausEfficiency}
\end{table}
The segment initialization achieved a track efficiency of 91\% for
single interactions, which dropped to 85\% for five superimposed
interactions in an event (tab.~\ref{tab:pausEfficiency}). The relative
efficiency of the subsequent elastic arms phase was always better than
98\%, indicating that hardly any of the good tracks the initialization
had found were lost. On the other hand, the elastic arms algorithm
strongly reduced the rate of ghost tracks prevalent in the
initialization. The CPU time consumption, determined on a HP9000/735
processor with 125~MHz clock rate, was still relatively high, but with
slightly more than 2 minutes for five simultaneous interactions
already in a feasible range. With increasing track density the CPU
fraction of the initialization increased steadily and exceeded that of
the elastic arms part beyond three superimposed interactions.

The investigations underline that elastic arms can in principle be
employed in an efficient manner, but require a very good
initialization of the track candidates. This has lead to the general
perception that elastic arms should not be used for track finding from
scratch, but should rather be seen as a tool to optimize assignment of
hits to tracks, to resolve left-right or other ambiguities, or to
detect and eliminate outlier hits. A similar philosophy is followed
in~\cite{abtPC}. A very interesting development in this context is the
{\it deterministic annealing filter} (DAF)~\cite{daf1,daf2}, which
extends the track fit with the Kalman filter with a fuzzy hit
assignment and obtains a mathematical equivalent of the elastic arms
procedure.

\section{Local Methods of Pattern Recognition}
\label{sec:local}
While global methods of pattern recognition have the common property
to treat all hit information in an equal and unbiased way,
simultaneous consideration of all hits can be very inefficient in
terms of speed. In fact many detector layouts provide sufficiently
continuous measurements so that the sheer proximity of hits makes it
already likely that they belong to the same track. This is one of the
reasons why {\it local} methods of track pattern recognition, often
called {\it track following}, are the workhorses of many
reconstruction programs in high energy physics.

Track following methods are essentially based on three elements:
\begin{itemize}
\item A parametric track model, which connects a particle trajectory
  with a set of track parameters and provides a method of {\it
    transport}, i.e. extrapolation along the trajectory
\item A method to generate {\it track seeds}, i.e. rudimentary initial
  track candidates formed by just a minimal set of hits which serve as
  starting point for the track following procedure
\item A quality criterion, which allows distinguishing good track
  candidates from ghosts so that the latter can be discarded
\end{itemize}
A related variant of track following is the {\it propagation} of a
track candidate found in one part of the tracking system into another,
collecting suitable hits on the way. In this case the initial track
candidate takes the r\^ole of the seed.

\subsection{Seeds}
\begin{figure}[htbp]
  \begin{center}
    \begin{tabular}{cc}
      \epsfig{file=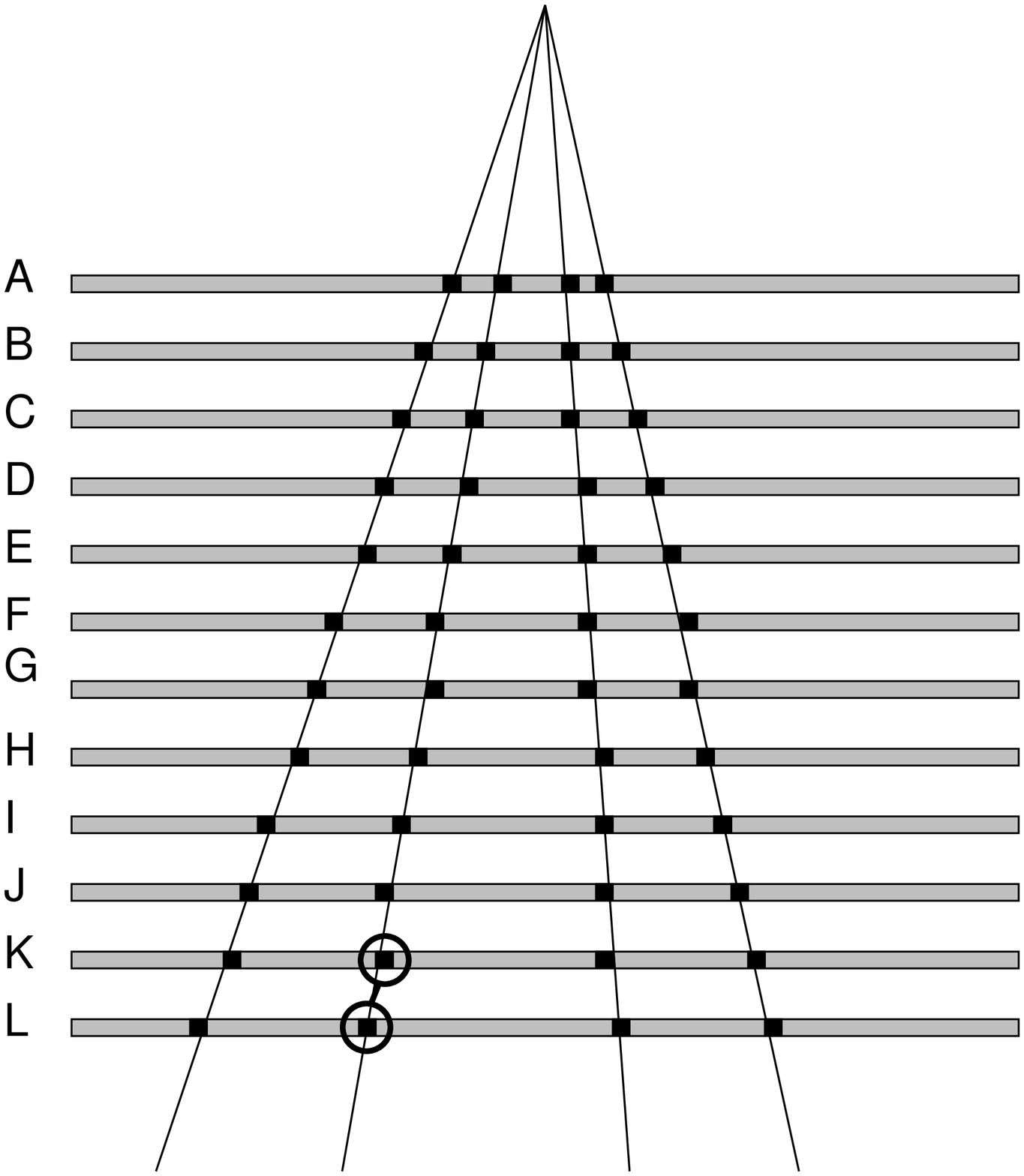,width=6cm,angle=0} &
      \epsfig{file=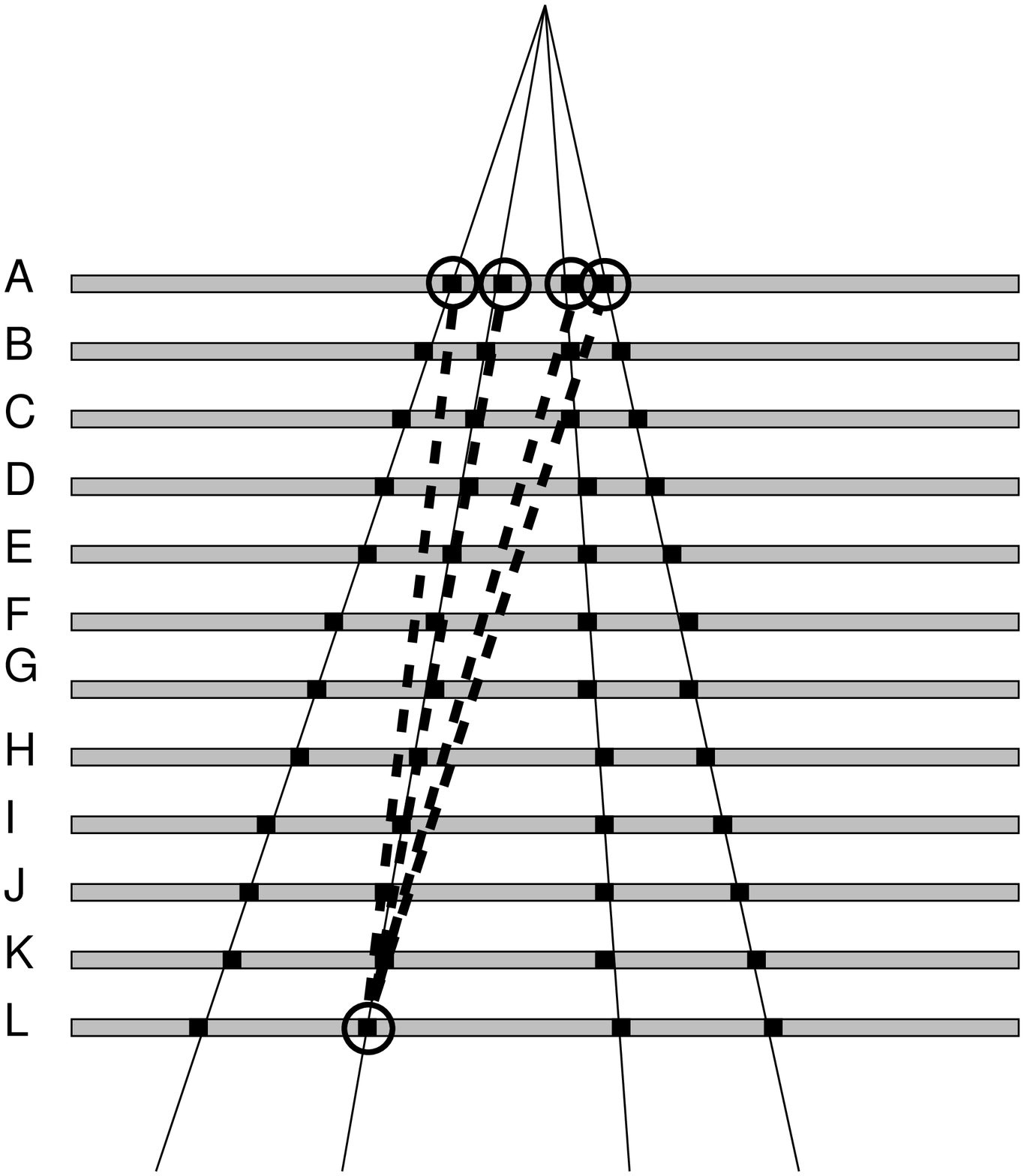,width=6cm,angle=0}
    \end{tabular}
  \end{center}
  \caption{Seeding schemes with nearby (left) and distant layers
    (right).}
  \label{fig:seedingScheme}
\end{figure}

There are different possible philosophies how seeds can be
constructed.  This is illustrated in fig.~\ref{fig:seedingScheme},
which shows schematically a tracking system with equidistant layers.
Starting from the last layer L, where the hit density is lowest, seeds
can be obtained by combining the hit with suitable others in the
neighbouring layer K (left side). This is the natural choice which
exploits the local proximity of hits as a selection criterion.  The
angular precision of such a short segment is in general limited
because of the small leverage, but the rate of fake seeds is
relatively small, since most wrong combinations tend to obtain a steep
slope that is incompatible with the relevant physical tracks and can
be discarded immediately. A completely different alternative is to
combine hits for example from the distant layers K and A to construct
seeds. These seeds have potentially a much better precision in angle,
but the number of choices to be considered is also much higher.  The
gain of precision can in fact be very limited if the material
within the tracker introduces sizable multiple scattering dilution.
For the latter reasons, seed combinations from nearby layers are often
preferred in practical applications.

\begin{figure} 
\begin{center}
\epsfig{file=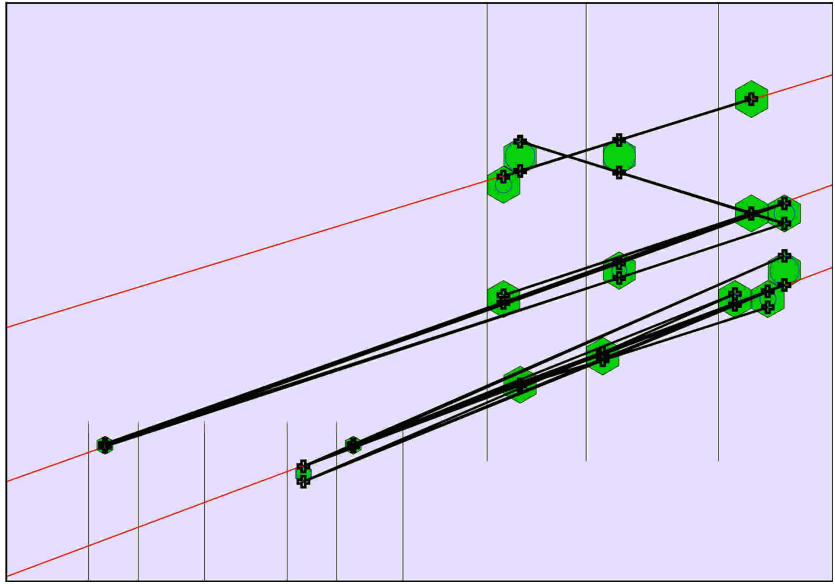,width=11cm}
\end{center}
\caption{Creating seeds from drift chamber hit triplets. The style of
  displayed items is similar to fig.~\ref{fig:borgSegments}.  Crosses
  indicate the hit coordinates used to construct the
  triplets(from~\cite{conc}).}
\label{fig:seed}
\end{figure}

Though the number of hits required for a seed is in general dictated
by the dimensionality of the parameter space, additional hits can very
efficiently decrease the ghost rate of the seeds.
Figure~\ref{fig:seed} shows the construction of seeds consisting of
three drift chamber hits each~\cite{conc}. In this example without
magnetic field, only two hits would be minimally needed to define a
seed, but the example shows that using hit triplets reduces the
combinatorics considerably.

\subsection{2D Versus 3D propagation}
\begin{figure} 
  \begin{center}
    \epsfig{file=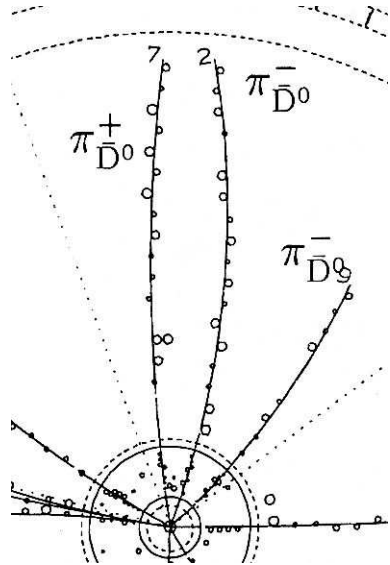,width=5cm}
  \end{center}
  \caption{Close-up of the drift chamber area from the ARGUS event
    display in fig.~\ref{fig:argusAtlas}~\cite{argusRareB}. The tracks
    are obtained by a track-following algorithm that proceeds from the
    outer towards the inner layers.}
  \label{fig:argusClose}
\end{figure}
Many detector layouts allow track following in a projection. For
example, drift chambers with many wires parallel and of same length
may allow separation of a pattern space that is measured in a plane
orthogonal to the wires. This means that parameter propagation during
the track following process is far less costly in terms of
computations, and that the seeds can be constructed from only two
measured hits in the case of a field-free area, or from three hits
within a magnetic field. It should be noted that in presence of a
magnetic field, a 2D treatment is only possible if the field is
oriented parallel to the wire, and homogeneous in wire direction. An
example for such an application is the pattern recognition in the
ARGUS drift chamber (fig.~\ref{fig:argusClose}), where the seeds are
constructed from three hits in the outer layers, and the track
following proceeds towards the beam line~\cite{albrecht}.

However, pattern recognition in projections cannot avoid that at some
point, 3D information must be inferred. This can be achieved by
performing track finding independently in all available projections,
and then merging compatible projected track candidates into a 3D
object. For an unbiased tracking, at least three independent views
must exist (see sect.~\ref{sec:stereo}), and each view must possess
enough hit information to find the track by itself with good
efficiency. A typical symmetric arrangement consists of three views
with $0^{\circ}$, $120^{\circ}$ and $240^{\circ}$ stereo angle, among
which all layers are evenly distributed. This approach leads to
virtually azimuth-independent track parameter resolutions.

A more economic alternative is a design with an asymmetric layer
distribution which is less costly in terms of the number of channels
but requires suitable pattern recognition algorithms.  It is possible
to perform first the pattern recognition in the $0^{\circ}$
projection, and then use the resulting track candidate to convert the
measurements in the $+\alpha$ and $-\alpha$ layers into the vertical
coordinate~\cite{albrecht}, as already illustrated in a different
context in fig.~\ref{fig:pausVertical}. The next step then proceeds
with track finding in the vertical projection. In this case, only the
$0^{\circ}$ projection needs to be equipped with enough layers for a
standalone track finding, while the two stereo views are combined and
thus the number of layers per stereo view can be smaller. A reasonable
scenario for this design comprises 50\% of the layers oriented at
$0^{\circ}$, 25\% in the $+\alpha$ and 25\% in the $-\alpha$
projection.

In the case of genuine 3D measurements, 3D seeds can be easily
constructed from two hits in the field-free case, and from three hits
in the case with magnetic field, which normally will hardly lead to
combinatorial problems. This is the situation in the barrel part of
the CMS inner detector~\cite{cmsPixel,cmsStrip}, where three layers of
silicon pixel detectors with 150~$\mu m$ pixel size will be used to
initiate track seeds, or in TPCs. In case of intrinsically 2D
measurements, 3D seeding has the general disadvantage that the seeds
will become rather complex, consisting of 4--5 measurements and under
high particle density also many false seeds will be generated. Also
left-right ambiguities have a strong impact here: a seed constructed
from five drift chamber hits yields 32 ambiguous track parameter sets
upon expanson of all possible left-right assignments. Once the seed is
constructed, the {\it track following} step involves many
extrapolations of the track parameters which are more costly with the
full set of parameters, in particular if the covariance matrix is to
be transported as well.

On the other hand, 3D propagation is easier to apply in the sense that
the full coordinate information is always available, so that e.g. the
decision if the track candidate intersects a particular detector
volume or not can be made unambiguously and multiple scattering
effects can be accounted for with good precision. The issue of merging
the different projections is also avoided.

\subsection{Na\"ive Track Following}
The na\"ive variant shall be discussed here essentially to allow for
comparison with the more sophisticated approaches. Starting from a
seed, the trajectory is extrapolated to the detector part where the
next hit is expected. If a suitable hit is found, it is appended to
the track candidate. Where several hits are at disposal, na\"ive track
following selects the one closest to the extrapolated trajectory.
This procedure is continued until the end of the tracking area is
reached, or no further suitable hit can be found.

Na\"ive track following is relatively easy to apply to tracking
scenarios with moderate track density and often leads to a reasonable
computational effort since the number of hits to be considered is
roughly proportional to both the number of layers and the number of
tracks. The application to situations with large hit density soon
reaches its limitations, since in dense environments, track following
runs the risk of losing its trail whenever several possible
continuations exist. The main complications can be summarized as
follows:
\begin{enumerate}
\item Some expected hits may be missing because of limited device
  efficiency, which will be called a {\it track fault} in the
  following. This also includes the case where the hit is existing,
  but out of expected coordinate bounds, for example because of delta
  electrons created by the impact of the particle. In drift chambers
  with single hit readout, the drift time measurement of the followed
  track can be superseded by another particle passing the same cell
  closer to the signal wire.
\item Wrong hits may be closer to the presumed trajectory than the
  proper hits and be picked up in their stead. This can happen easily
  just after the seeding phase when the precision of the track
  parameters is still limited, or when some false hits have already
  been accumulated. A wrong hit may stem from another reconstructable
  track, from a non-reconstructable low-energy particle, or from
  detector noise.
\item Left-right ambiguities in wire drift chambers double the number
  of choices. Especially in small drift cells, e.g. in straw tube
  trackers, wrong left-right assignments are to some degree
  unavoidable and need to be coped with.
\end{enumerate}
These aspects can pose a particular problem if the track density is
subject to strong variations, e.g. due to a fluctuating number of
simultaneous interactions under LHC-like conditions.

\subsection{Combinatorial Track Following}
This variant is aware of possible ambiguities, and in each track
following step, each continuation hit which is possible within a wide
tolerance gives rise to a new branch of the procedure, so that in
general a whole tree of track candidates emerges. The final selection
of the best candidate must be done in a subsequent step, which may
involve a full track fit on each candidate. This kind of method is
potentially unbeatable in terms of track efficiency, but in general
highly resource consuming and therefore only used in special cases
with limited combinatorics.

\subsection{Use of The Kalman Filter}
\label{sec:trackFollowKalman}
All track following approaches have to evaluate if a certain hit is
compatible with the presumed trajectory and thus suitable to be added
to the track candidate. The suitability of a hit should be based on
criteria which exploit all the knowledge based on those hits that have
been accumulated so far. Not only the track parameters themselves,
also their precision needs to be known. The ideal tool in this
situation is the progressive fit implemented by the Kalman filter,
which has been discussed in section~\ref{sec:kalmanfilter}.

The Kalman filter prediction already provides an excellent criterion
for hit selection. When a hit is considered to be appended to the
track, first the {\it predicted residual} $r^{k-1}_k$ from
equation~\ref{eq:predres} can be used as a rough criterion. After
passing a hit through the filter process (see
eq.~\ref{eq:filterEquation}), the {\it filtered $\chi^2$} defined in
equation~\ref{eq:filtchi2} is an even more precise measure. In
general, the decision power will increase when more and more hits are
accumulated in the track candidate. Once the full track is available,
the result of the Kalman smoother (eq.~\ref{eq:kalmansmoother}) can be
used to detect and remove further outlier hits.

\subsection{Arbitration}
In practical applications of track following, means are required to
reduce its dependency on the starting point, and to decrease its
vulnerability against stochastic influences. This process is called
{\it arbitration}. For example, it is mandatory not to depend on a
single option of seeding tracks, which would lead to loss of a track
if one of the seeding layer happens to be inefficient, but one will
normally use several combinations of layers for seeding.  Such
redundancy increases the probability to obtain a seed for a track even
in presence of device inefficiency. When an expected hit appears to be
missing in a layer during propagation, it may be advisable not to
discard the candidate immediately, but to proceed further until a
fault limit is exceeded. In a case where more than one hit could
present a suitable continuation for a track, one might want not to
decide immediately for the closest hit but create branches into
different candidates which are pursued independently. When a hit
appears to be fine for a continuation, the algorithm should account
for the possibility that this hit is wrong and the right hit has
disappeared for some reason. However, na\"ively applied, all these
extensions lead to either vast combinatorics, which will explode with
increasing hit density, or suffer from ad-hoc limitations. A method to
overcome these problems will be detailed in the following.

\newpage
\subsection{An Example for Arbitrated Track Following}
This section discusses the {\it concurrent track evolution} algorithm
as an example for an approach to track following with arbitration,
which is in detail described in ~\cite{conc,concmag}.

\subsubsection{Algorithm}
The basic idea is to allow for concurrency of a certain number of
track candidates at any time during the propagation of a certain seed,
or even a set of seeds. These tracks are propagated in a synchronized
manner from one sensitive tracking volume to the next. At each
propagation step for each track candidate, branching into several
paths is possible and will in general occur. Multiple branches appear
when several continuation hits are consistent with the present
knowledge of track parameters, or when more than one tracking volume
is within reach.  Also the possibility that the expected hit is simply
missing, e.g.  because of device inefficiency, gives rise to a new
branch. Thus the procedure explores the {\it available paths} for all
track candidates {\it concurrently} which leads to a rapid creation of
new track candidates. On the other hand, the number of track
candidates should not grow beyond control. This is achieved by
applying a quality selection on the whole set of concurrent track
candidates after each round of propagation, using suitable estimators
for the {\it quality} of a track. This leads to a favourable timing
behaviour even for high multiplicity events.  Concurrent track
evolution can thus be regarded as a variant of {\it deferred
  arbitration}~\cite{deferred}. The actual propagation is based on the
Kalman filter.
\begin{figure} 
\begin{center}
\epsfig{file=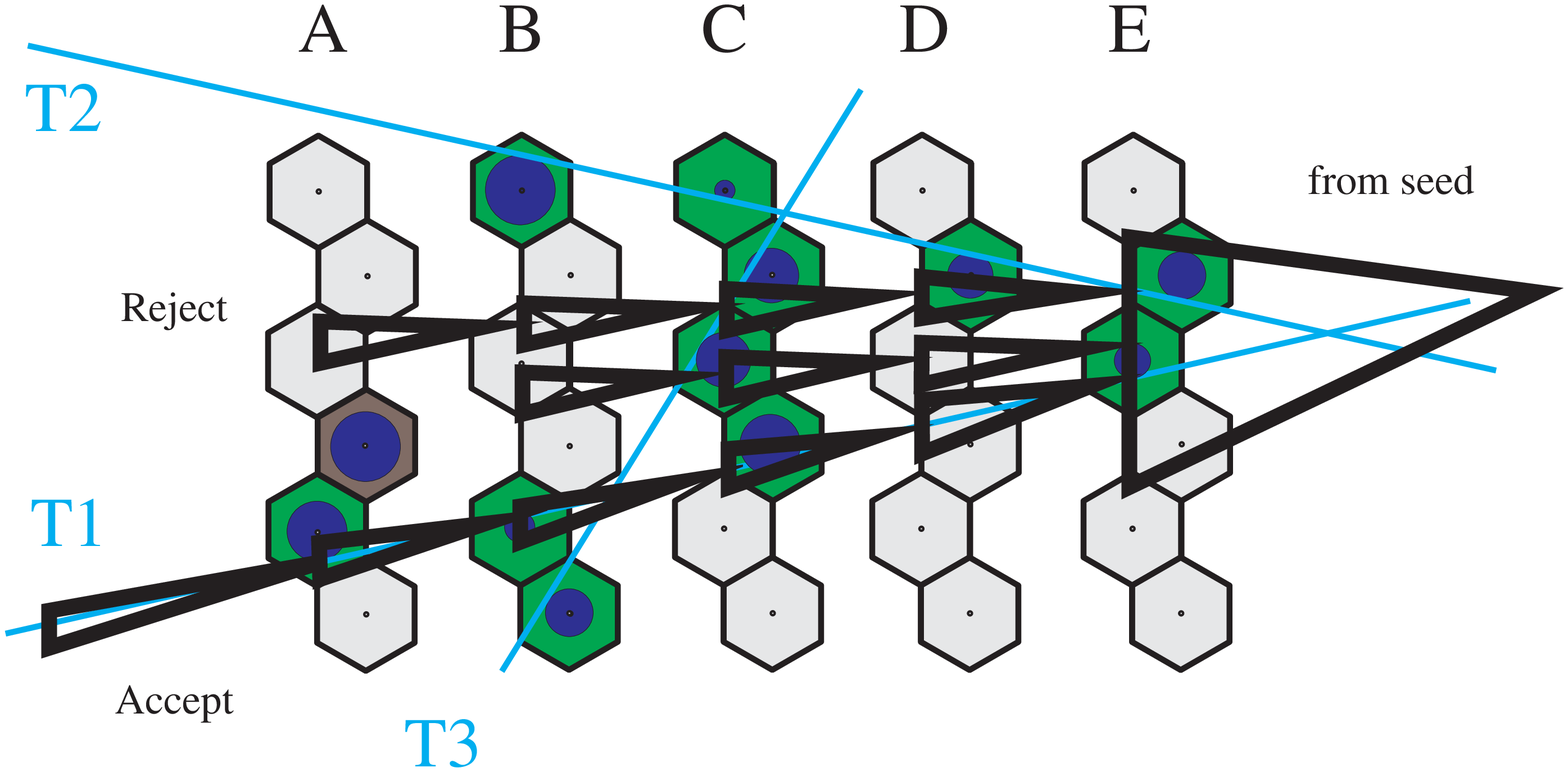,width=15cm,clip=,angle=0}
\end{center}
\caption{Schematic view of concurrent track evolution in a five-layered
  part of a tracking system with hexagonal drift cells, which is
  traversed by three particles, labelled T1, T2 and T3. The simulated
  drift time isochrones are indicated by circles. The propagation
  proceeds upstream from the right to the left and starts with a seed
  of hits from track T1 outside of the picture (from~\cite{conc}).}
\label{fig:concurrent}
\end{figure} 

An illustration of this strategy is shown in fig.~\ref{fig:concurrent}
taken from~\cite{conc}, which shows a potentially ambiguous situation
caused by two nearby tracks T1 and T2 plus a large angle track T3 in
five layers of honeycomb drift chambers.  For simplicity, it is
assumed here that the algorithm discards track candidates with more
than one missing hit ({\it fault}) in a row, and that the maximum
number of concurrent candidates is three -- in reality, higher limits
may be used.  It is also assumed that a seed of hits from track T1 has
been formed on the right side outside of the figure. The propagation
proceeds upstream from right to left. The illustration shows how three
parallel candidates arise from different left-right assignments to the
two drift chamber hits in layer E, which are propagated through layers
D and C -- including the tolerance of a fault on track T1 in layer D.
In layers B and A, the false paths are discarded because of
accumulating too many faults, and the proper reconstruction of track
T1 is retained. Track T2 should then be found later with a different
seed, while track T3 is likely to be non-reconstructable.

Track following in the na\"ive sense would always accept the hit with
the smallest $\chi^2$ contribution, possibly a good solution when the
hit density is small. In the presence of multiple scattering and high
hit densities, a wrong hit will frequently have a smaller $\chi^2$
contribution than the proper one, or replace a proper hit which
is missing due to detector inefficiency, or shadowed by another
track passing the same cell. On the other hand, full evaluation of all
possible hit combinations would exceed all bounds of computing
resources when applied to dense events. Thus, the concurrent track
evolution strategy combines the virtues of track following and
combinatorial approaches. As will be shown below, the optimization in
each evolution step using a quality estimator provides an elegant
means to deal with the main problems in high occupancy track
propagation.

\subsubsection{Parameters}
The algorithm is controlled by parameters which determine the
selection of hits for propagation of candidates, and for optimization
of concurrent candidates on each level. The parameter ${\rm
  \delta_u^{max}}$ is the range around the predicted coordinate in the
next considered tracking layer, in which continuation hits are
searched. The parameter ${\rm \delta \chi^2_{max}}$ stands for the
maximum tolerable {\it filtered $\chi^2$} increment according to
eq.~\ref{eq:filtchi2}.  Missing hits (faults) are in general tolerated
but only a certain number of subsequent faults (${\rm
  N_{Faults}^{max}}$) are accepted.  The pruning of track candidates
after each evolution step is then regulated with absolute and relative
cuts. The {\it quality} of a track candidate can be estimated with a
function of the form
\begin{equation}
        Q = f({\rm N_{Steps}}, {\rm N_{Faults}}, \chi^2_i,...)
\end{equation}
where ${\rm N_{Steps}}$ is the number of evolution steps passed so
far, and $\chi^2_i$ stands for the contribution of the accumulated hit
$i$ to the total $\chi^2$. If needed, also a bias from the track
parameters could be introduced here, which suppresses e.g. tracks that
are very steep or have very low momentum. A convenient simple quality
estimator is
\begin{equation}
   Q = {\rm N_{Steps}} - {\rm N_{Faults}} -  w_{\chi^2} \cdot \sum_i \chi^2_i 
\end{equation}
which applies a certain malus (in this case 1) for each missing hit,
which is equivalent to an ill-matching hit with a $\chi^2$
contribution of $1/w_{\chi^2}$ (in the configuration of
tab.~\ref{tab:paramtable} equal to 10). Furthermore, cuts are applied
relative to the {\it best} candidate currently in the set:
candidates whose quality differs from the best candidate by more than
$\delta {\rm q_{min}}$ are discarded. Finally, all concurrent track
candidates are ranked in decreasing order of quality, and only the
first ${\cal R}_{max}$ candidates in rank are retained. If propagation
cannot be continued though the end of the tracking system is not
reached, this may have a natural reason, e.g. the particle may have
been stopped or decayed in flight. In such cases, the best remaining
track candidate on the last level is kept if it comprises at least a
certain minimum number of hits, $N_{Hits}^{min}$.

\subsection{Track Following And Impact of Detector Design Parameters}
The practical behaviour of such an algorithm, as it has been developed
for the HERA-B spectrometer has been studied in~\cite{conc}, including
an investigation of the impact of detector design and performance on
the pattern recognition capability. As the experiment has never
routinely taken physics data at the high design interaction rate of
40~MHz, the results have been obtained from a full Geant simulation
with on average five superimposed $pN$ interactions, one of them
containing beauty hadrons.  As seen in fig.~\ref{fig:herabDetector},
the inner part of the HERA-B main spectrometer acceptance within about
$25~\rm{cm}$ radius from the beam line is covered by micro-strip
gaseous chambers (MSGC), while the outer part is instrumented with
Honeycomb drift
chambers~\cite{herabBeauty94,herabDesign,herabVancouver}.  The {\it
  pattern tracker} consists of four superlayers outside of the
magnetic field, which consist of 6 individual layers each (the area
marked ``PC'' in fig.~\ref{fig:herabDetector}), except for the inner
part of the two middle superlayers that have only four layers each.
Half of the layers measure a horizontal coordinate ($0^\circ$
orientation), the other half are arranged at $\pm 100$~mrad stereo
angle. The seeds were produced from hit triplets in the hindmost two
superlayers for upstream, and in the foremost two superlayers for
downstream propagation (fig.~\ref{fig:seed}). Track finding was
performed first in the $0^\circ$ projection, then continued in the
combined stereo layers, where the vertical coordinates were determined
using the horizontal projection of the track candidate with the method
explained in sec.~\ref{sec:elasticArms} (see
eq.~\ref{eq:convertVertical} and fig.~\ref{fig:pausVertical}).

\begin{table}
\centering
\begin{tabular}{|l|l||l|l|}
\hline
Parameter & Value & Parameter & Value\\
\hline
\hline
${\rm N_{Hits}^{min}(x)}$  &  9  &
             ${\rm N_{Faults}^{max}(x)}$  & 2 \\
${\rm N_{Hits}^{min}(y)}$  &  9  &
             ${\rm N_{Faults}^{max}(y)}$  & 2 \\
${\rm \delta \chi^2_{max}(x)}$  & 8 &
             ${\cal R}_{\rm max}          $      & 5 \\
${\rm \delta \chi^2_{max}(y)}$  & 16 &
             ${\rm w_{\chi^2}}$          & 0.1 \\
${\rm \delta q_{min}}$     & $-1$ & 
                                         &            \\
\hline
\end{tabular}
\caption{Table of parameters used in the implementation in~\cite{conc}}
\label{tab:paramtable}
\end{table}
The algorithm parameters used are summarized in
table~\ref{tab:paramtable}. The parameters allow for a
delicate adjustment of balance between the extremes of na\"ive track
following (${\cal R}_{\rm max}=1$), where always the apparently best
path is followed, and combinatorial track following (${\cal
  R}_{\rm max}=\infty$), which retains all paths. The detailed
simulation allowed to study some principal effects of tracking system
properties on pattern recognition parameters which will be shown in
the following.

\begin{figure}
\begin{center}
\epsfig{file=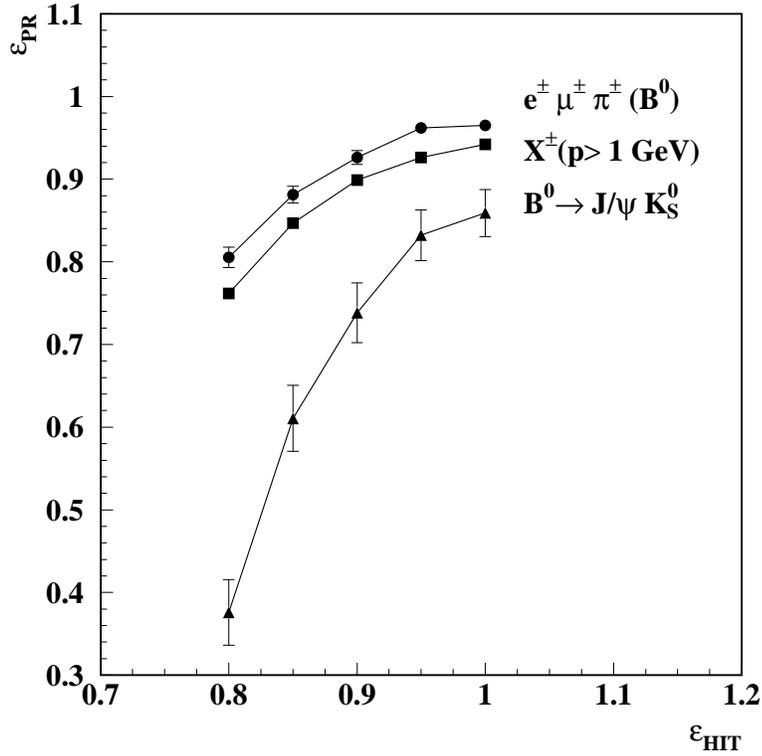,width=11cm,clip=,angle=0}
\end{center}

\caption{Pattern
  recognition efficiency for different values of the hit efficiency on
  simulated events consisting of one $pN$ interaction with a $B^0$
  meson with the decay chain $B^0 \rightarrow J/\psi K^0_S \rightarrow
  \ell^+\ell^- \pi^+ \pi^-$, where $\ell^+\ell^-$ can be a pair of
  muons or electrons, superimposed with on average four unbiased
  inelastic interactions. The filled squares show the track finding
  efficiency for charged particles with momentum above 1~GeV, the
  filled circles are for particles from the $B$ decay. The triangles
  indicate the combined efficiency of all four $B$ decay
  particles~\cite{conc}.}
\label{eff-vs-ehit}
\end{figure}
\subsubsection{Influence of detector efficiency}
Figure~\ref{eff-vs-ehit} shows how the hit efficiency of the detector
devices affects the pattern recognition performance on tracks emerging
from B decays. Above $\epsilon_{\rm HIT}=95\%$, the hit inefficiency
is well compensated by the algorithm (operating with ${\rm
  N_{Faults}}=2$), resulting in an excellent track finding
performance. Smaller hit efficiency leads to sizeable loss in the
fraction of detected particles.

\begin{figure}
\begin{center}
\unitlength1cm
\begin{picture}(14,8)
\put(0.,0.){\epsfig{file=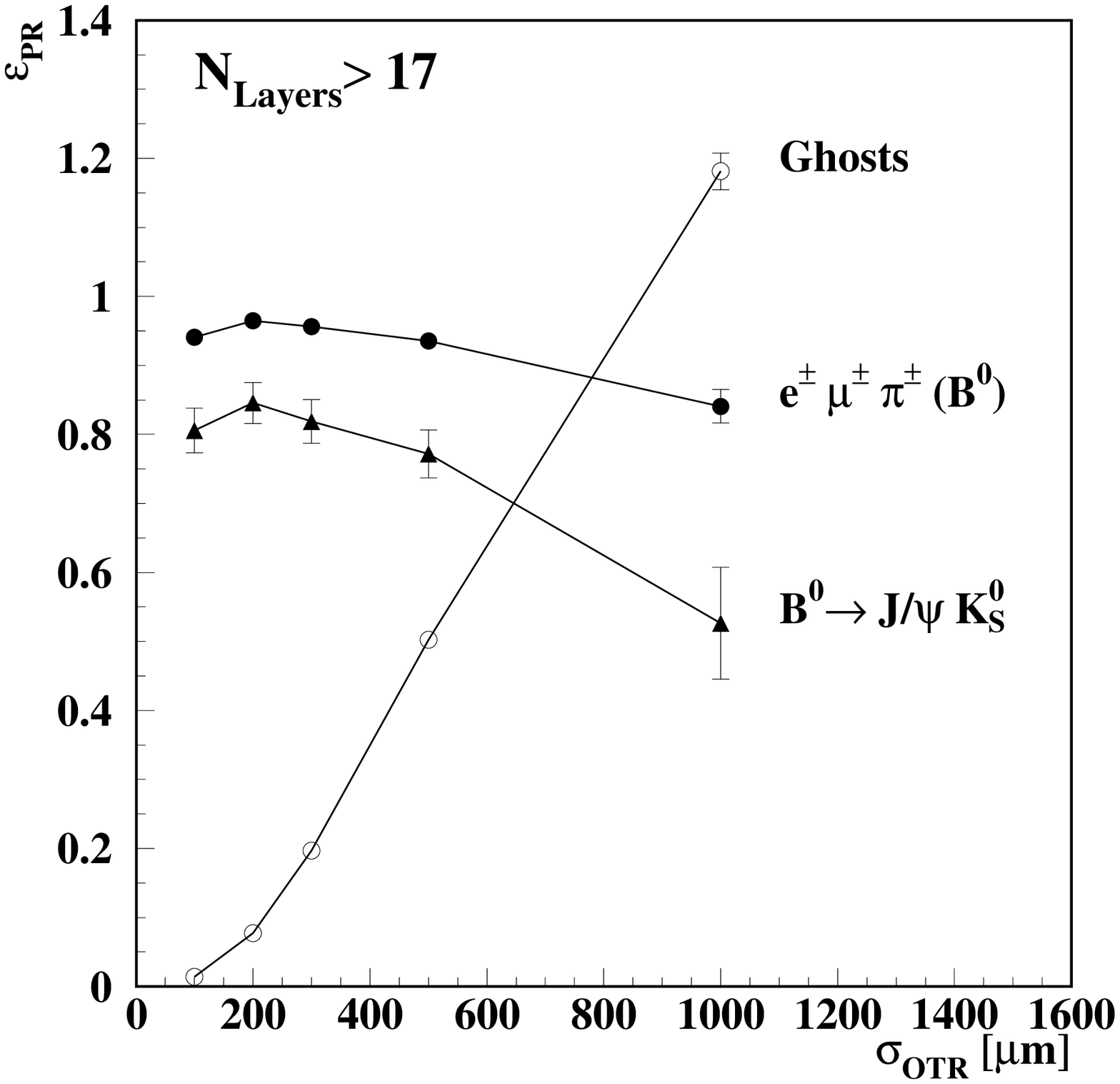,width=7.0cm}}
\put(5.8,6.1){\makebox(0,0)[t]{\large (a)}}
\put(7.0,0.){\epsfig{file=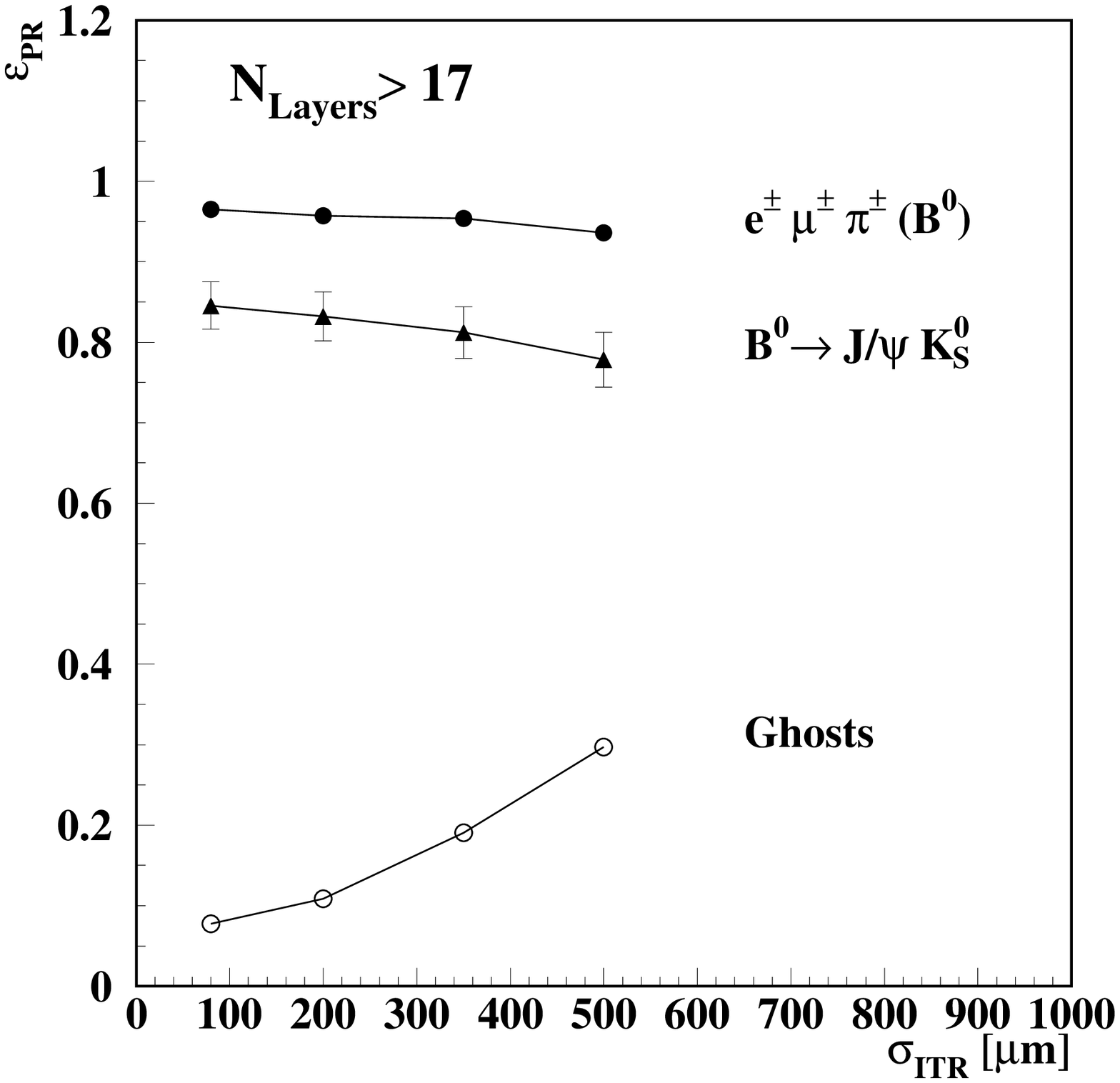,width=7.0cm}}
\put(12.8,6.1){\makebox(0,0)[t]{\large (b)}}
\end{picture}
\end{center}
\caption{Pattern recognition efficiency for different outer (a) and
  inner tracker resolutions (b), for particles from the $B^0
  \rightarrow J/\psi K^0_S$ decay mode as detailed in the caption of
  fig.~\ref{eff-vs-acc}. Only tracks passing at least 17 out of 24
  possible tracking layers were considered. Also the ghost rate is
  displayed (open circles)~\cite{conc}.}
\label{eff-vs-acc}
\end{figure}
\subsubsection{Effect of detector resolution}
The influence of the spatial resolution is shown in
fig.~\ref{eff-vs-acc}.  The
simulated resolutions of outer and inner tracking system were
varied independently. It is interesting to see that the efficiency
degrades only slowly with the resolution being increased up to 1~mm. The
slight drop in efficiency at 100~$\mu$m in fig.~\ref{eff-vs-acc}a is
an artifact due to numerical approximations. Both figures indicate
that the effect of resolution on track finding efficiency should not
be overrated. Much stronger is the effect on the ghost rate, the plots
underline that a good resolution helps considerably to suppress fake
reconstructions.

\begin{figure}
\begin{center}
\epsfig{file=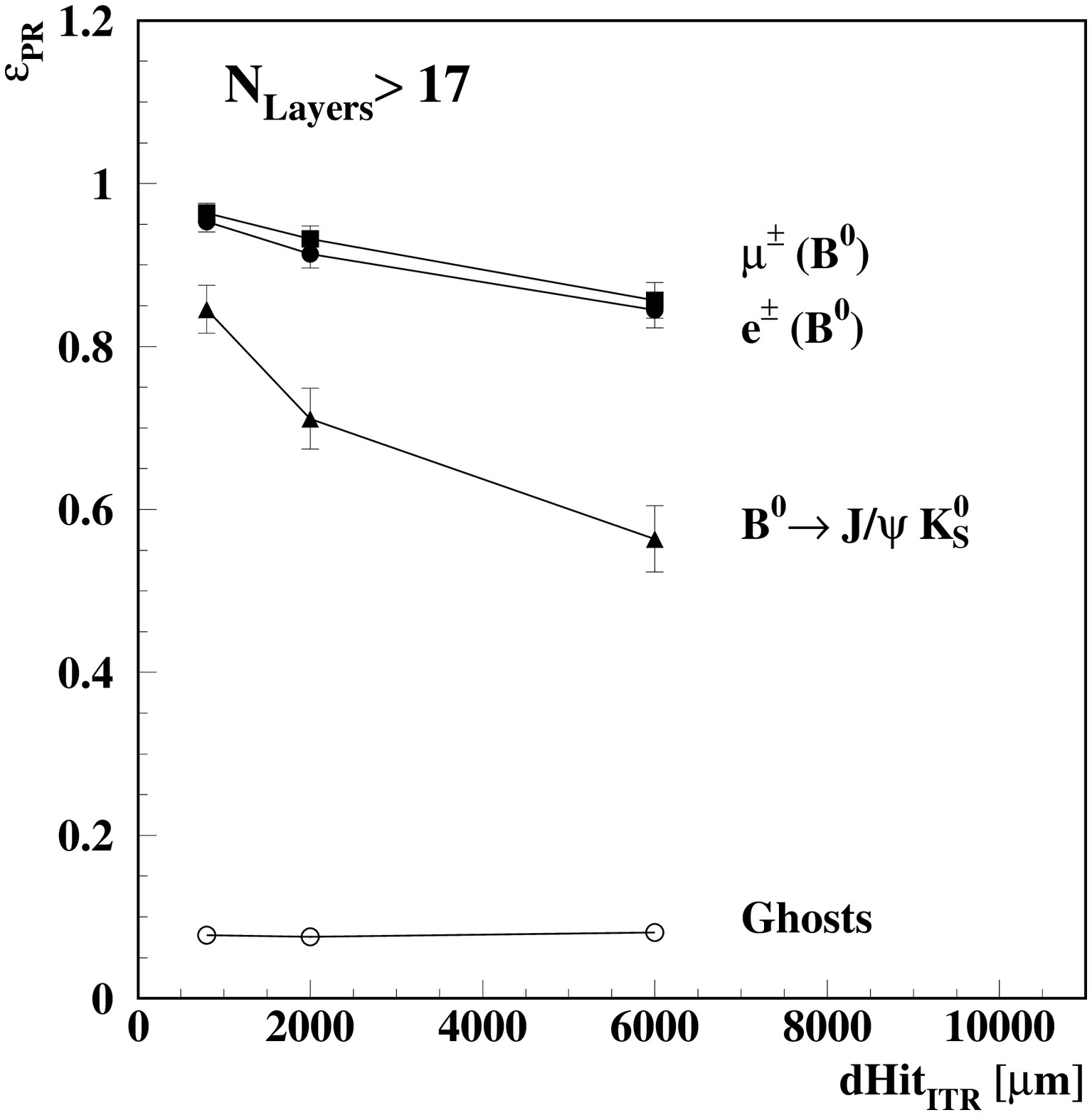,width=11cm,clip=,angle=0}
\end{center}
\caption{Pattern
  recognition efficiency for different double hit resolutions of the
  inner tracker for particles from the $B^0 \rightarrow J/\psi K^0_S$
  decay mode as detailed in the caption of fig.~\ref{eff-vs-acc}. Also
  the ghost rate is shown (open circles). Only tracks passing at least
  17 out of 24 possible tracking layers were considered~\cite{conc}.}
\label{eff-vs-dod}
\end{figure}
\subsubsection{Influence of double track separation}
The simulation of the inner tracker devices allowed varying of the
double track resolution, i.e. the distance down to which nearby tracks
can be resolved as individual hits in a device. In micro-strip gaseous
chambers (MSGC) as they are used by HERA-B, the double track
separation distance is in general larger than the resolution, since it
depends on the cluster sizes. As visible in fig.~\ref{eff-vs-dod}, the
efficiency drops significantly with double track resolutions worse
than 800~$\mu$m.

\begin{figure}
\begin{center}
\epsfig{file=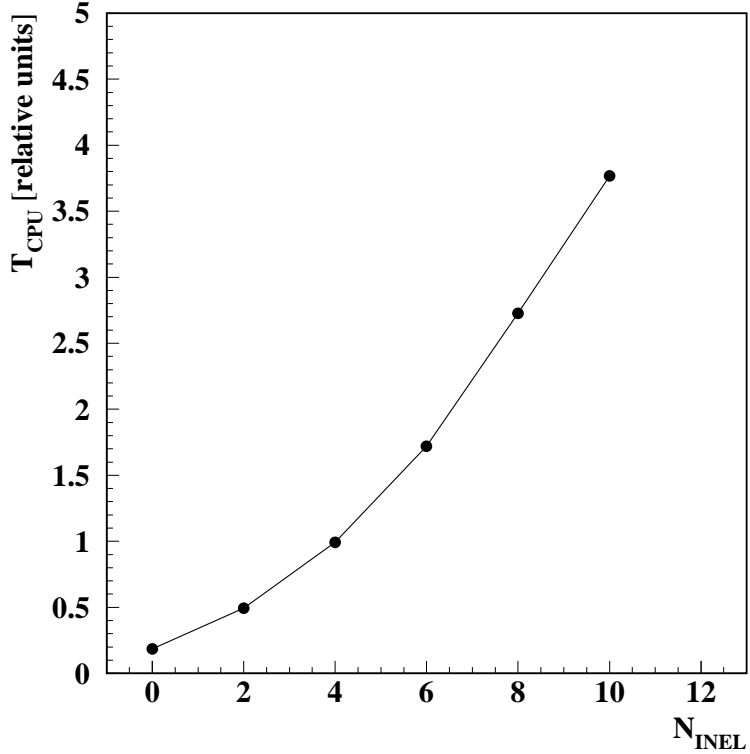,width=11cm,clip=,angle=0}
\end{center}
\caption{Mean
  computing time per event as a function of the number of inelastic
  interactions superimposed to one $b\bar b$ interaction~\cite{conc},
  normalized to the value at $N_{INEL}=4$.}
\label{timescal-vs-nint} 
\end{figure}
\subsubsection{Execution speed}
As already seen in sec.~\ref{sec:elasticArms}, the CPU time
consumption is an essential aspect for the selection of a pattern
recognition algorithm. The concurrent track evolution algorithm was
tested on the same geometry and event type as the elastic arms
algorithm implementation (see tab.~\ref{tab:pausEfficiency}), and
required on average 4s per event with four superimposed inelastic
interactions, compared to several minutes for the elastic arms method
on the same type of processor. Also the behaviour with increasing
track density is important, since steep increases with a sizable power
of the track multiplicity, as they may arise from combinatorics, can
have a very negative impact on use of a reconstruction program at
production scale.  Figure~\ref{timescal-vs-nint} shows the average
computing time per event normalized to that for the nominal four
superimposed inelastic interactions. At high interaction multiplicity,
the computing time per event settled rather gracefully on a roughly
linear dependence, indicating a constant amount of time per track, at
an acceptable loss of efficiency, which can be considered a
good-natured behaviour. With the speed shown above, the algorithm is
fast enough to be used in quasi-online
reconstruction~\cite{mankelOnline}.

\subsection{Track Propagation in a Magnetic Field}
In general the above track following strategy can be applied also
within a magnetic field. The main difference is that the transport
function in eq.~\ref{eq:transport} becomes non-linear, and the
transport matrix becomes a local derivative as displayed in
eq.~\ref{eq:nonlinear}.  If the field is homogeneous, or if
inhomogeneity can at least be neglected within typical transport
distances, the transport function and matrix can usually be expressed
analytically.

In many cases, however, the field is neither homogeneous nor
describable in an analytic expression, instead, it is parametrized in
terms of a field map, which has been measured with Hall probes, or
computed by means of a field simulation program.  In this case,
numerical methods have to be used to derive the transport function.  A
very suitable method is the Runge-Kutta procedure~\cite{recipes},
which integrates the equations of motion by expanding the trajectory
up to a certain order and sampling the field at a series of
intermediate points, which are chosen and weighted such that all
powers of the errors below a certain order cancel.  Even this
procedure meets considerable challenges when the field varies strongly
and a very high precision, matching the detector resolution, must be
warranted.  In this situation, an embedded Runge-Kutta method with
adaptive step size can help: the next highest order of Runge-Kutta is
compared with the preceding one and the difference serves as an error
estimate, which is then used to adjust the step size.

Application of the Kalman filter does not only require a transport
function for the track parameters, but also the derivative matrix of
the new parameters with respect to the old is needed (see
eq.~\ref{eq:nonlinear}). Calculation of this derivative matrix can be
efficiently performed within the same Runge-Kutta framework that is
used for the parameter transport itself~\cite{oest}.

\begin{figure}
\begin{center}
\unitlength0.9cm
\begin{picture}(13,19)
\put(0.0,9.5){\epsfig{file=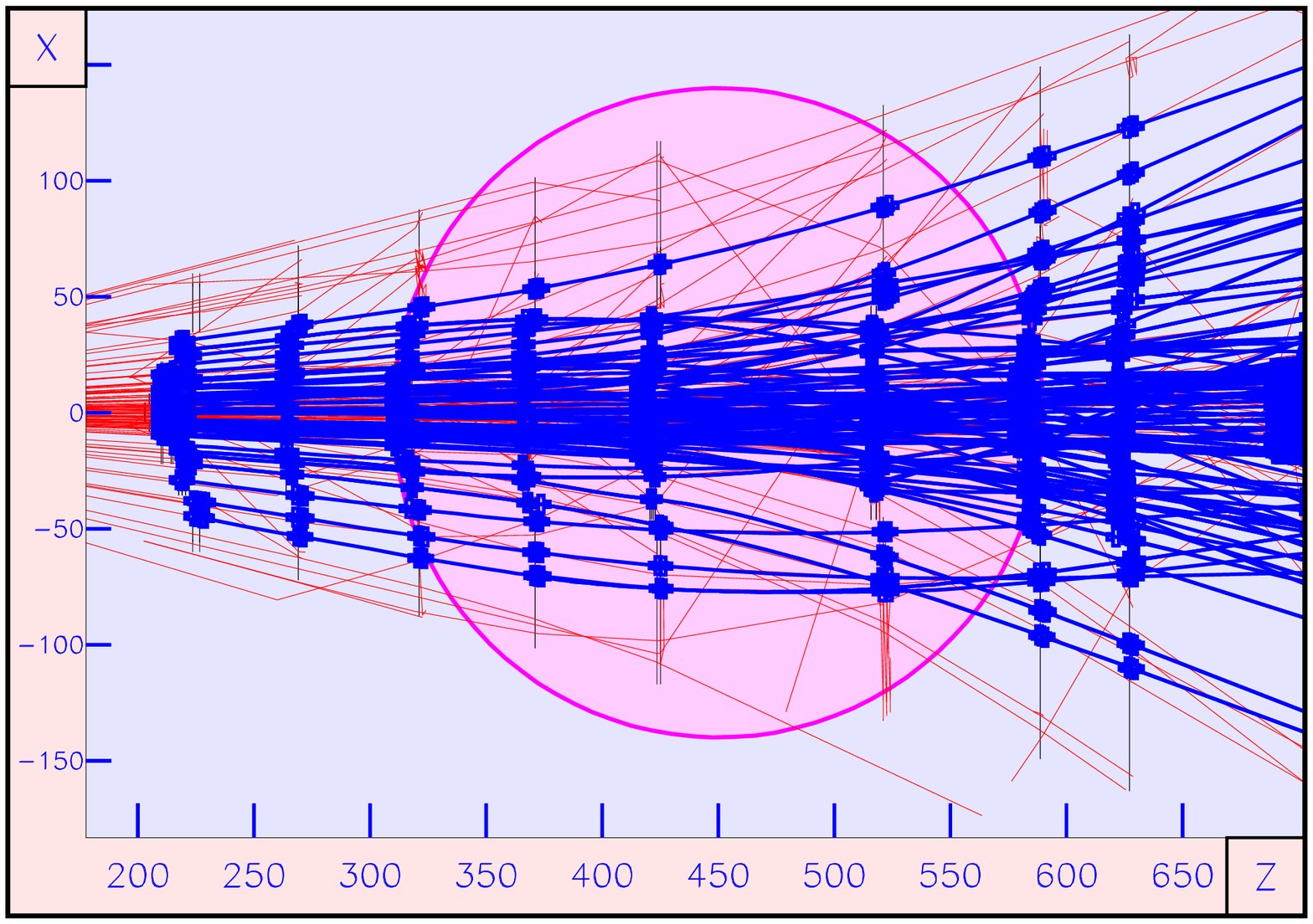,width=11.7cm}}
\put(1.9,18.1){\makebox(0,0)[t]{\huge (a)}}
\put(0.0,0.0){\epsfig{file=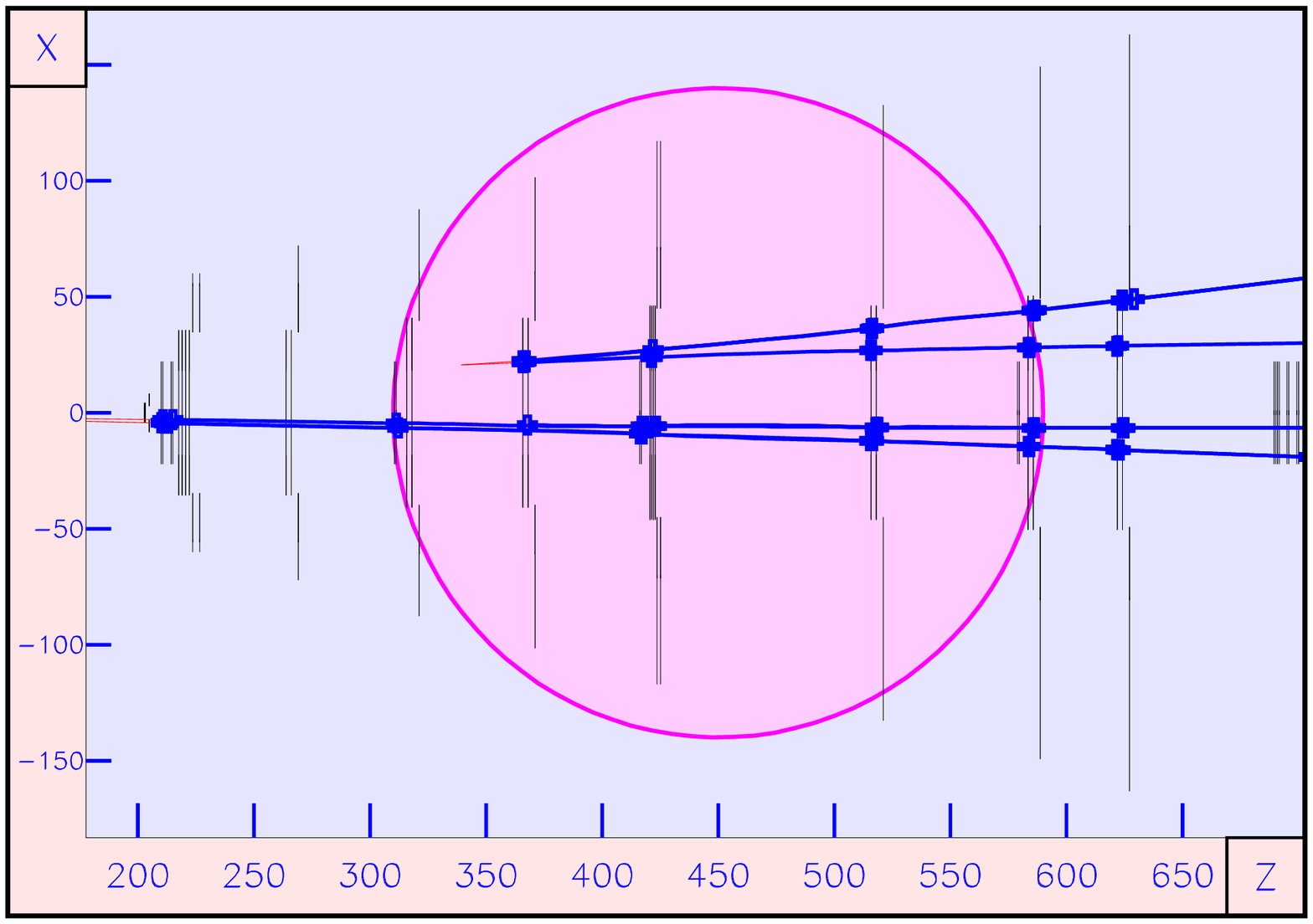,width=11.7cm}}
\put(1.9, 8.5){\makebox(0,0)[t]{\huge (b)}}

\put(12.3, 6.9){\makebox(0,0)[t]{\large $\pi^+$}}
\put(12.3, 6.2){\makebox(0,0)[t]{\large $\pi^-$}}
\put(12.3, 5.5){\makebox(0,0)[t]{\large $\mu^+$}}
\put(12.3, 4.6){\makebox(0,0)[t]{\large $\mu^-$}}
\end{picture}
\end{center}
\caption{(a) Display of a simulated event with one interaction containing the
  golden B decay and six superimposed inelastic interactions, focussed
  on the magnet area, where the pole shoe of the magnet is indicated
  by the large circle~\cite{concmag}. Both the Monte Carlo tracks
  (light grey) and the reconstructed tracks (thick dark lines) are
  show (reconstructed hit points denoted by crosses).  (b) Same event,
  with the display restricted to particles from the golden $B$ decay.}
\label{fig:concmag}
\end{figure}

An extension of the concurrent track evolution algorithm for track
following in the magnetic field has been developed and tested on the
HERA-B geometry in~\cite{concmag}. Track segments found in the
field-free part of the spectrometer were followed upstream through the
inhomogeneous field of the magnet tracker. Figure~\ref{fig:concmag}
shows an event display with simulated tracks including a $B$ decay
reconstructed with this method. The algorithm achieved a high track
propagation efficiency in spite of the large track density.

\section{Fitting of Particle Trajectories}
\label{sec:fitting}
After pattern recognition has done its work, the detector hits are
separated into sets each of which, ideally, contains manifestations of
one specific particle. It is then the task of the track fit to
evaluate the track parameters and thus the kinematical properties of
the particle with optimal precision. Even if the pattern recognition
itself is already providing track parameters and covariance matrices
to some degree, obtained for example by means of the Kalman filter, it
will in general be left to a final track fit to take all necessary
effects into account which are often neglected at the track finding
stage because they are costly to apply under the full combinatorics of
pattern recognition.

\subsection{Random Perturbations}
\begin{figure} 
\begin{center}
\epsfig{file=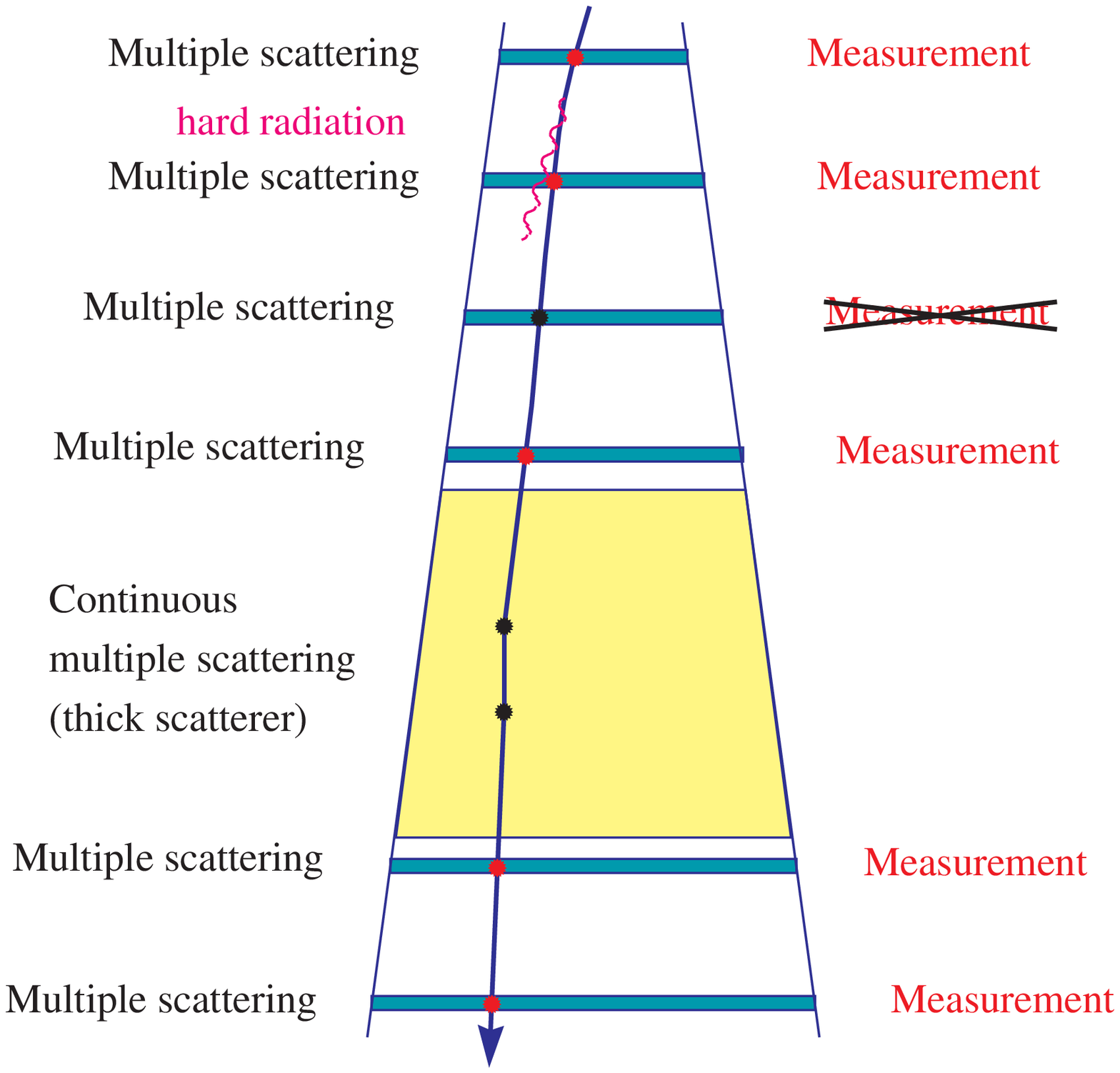,width=11cm,clip=,angle=0,bbllx=2092,
  bblly=2121,bburx=2595,bbury=2595}
\end{center}
\caption{Schematic view of the information flow in the track
  fit. Elements shown are measurements (hits) in the tracking layers,
  in one case a missing hit which is e.g. not found by the pattern
  recognition procedure, and random perturbations like multiple
  scattering and photon radiation.}
\label{fig:fitscheme}
\end{figure}

In the easiest case, track parameters could be derived from the
measurements by applying the least squares fit formulas from
eq.~\ref{eq:leastSqParam} and~\ref{eq:leastSqCov} in
sec.~\ref{sec:leastSquares}. In realistic applications, the problem is
usually more involved because of the way the trajectory of the
particle is influenced by random perturbations that dilute the
information content of the measurements, most commonly multiple
scattering and ionization or radiative energy loss. Their influence is
schematically displayed in fig.~\ref{fig:fitscheme}. One can interpret
the diagram in such a way that, from step to step, the measurements,
labelled on the right side, improve the degree of amount of
information about the kinematical properties of the particle, while
the perturbations labelled on the left side reduce it.

\subsection{Treatment of Multiple Scattering}
\label{sec:multipleScattering}
Multiple scattering occurs through the elastic scattering of charged
particles in the Coulomb field of the nuclei in the detector material.
Since the nuclei are usually much heavier than the traversing
particles, the absolute momentum of the latter remains unaffected,
while the direction is changed. If the longitudinal extension of the
traversed material block can be neglected (this is normally referred
to as {\it thin scatterer approximation}), only track parameters
related to particle direction are affected directly, for example the
track slopes $t_x=\tan \theta_x$ and $t_y=\tan \theta_y$ introduced in
section~\ref{sec:forwardGeometry}. The stochastic nature of multiple
scattering is that of a Markov process.

The distribution of the deflection angle follows a bell-like shape,
though it cannot be accurately described by a Gaussian because of its
pronounced tails.  The variance of the projected multiple scattering
angle is calculated within Moli\`ere
theory~\cite{moliere,moliere2,moliere4} as
\begin{equation}
  C_{MS} = \left( \frac{13.6 \, \mbox{MeV}}{\beta p c} \right)^2 t 
  \left[ 1 +  0.038\, \ln\, t \right]^2
\end{equation}
where $t$ is the traversed path length in terms of radiation lengths
$x_R$, usually called {\it radiation thickness}. (While the radiation
length is frequently abbreviated as $x_0$ in the literature, the
symbol $x_R$ is used here instead to avoid confusion with other uses
of $x_0$ throughout this article.) For a planar object arranged in a
plane vertical to the $z$ axis, the radiation thickness along $z$ is
given by
\begin{equation}
  \tilde t = \int \frac{dz}{x_R(z)}
\label{eq:radThickness}
\end{equation}

Taking the track inclination against the $z$ axis into account,
one obtains the effective radiation thickness
\begin{equation}
  t = \tilde t \sqrt{1 + t_x^2 + t_y^2}
\end{equation}
so that the final formula becomes (assuming $\beta \approx 1$)
\begin{equation}
  C_{MS} = \left( \frac{13.6 \, \mbox{MeV}}{pc} \right)^2
  \sqrt{1 + t_x^2 + t_y^2}\, \tilde t \left[ 1 + 0.038\,
  \ln\,\sqrt{1 + t_x^2 + t_y^2}\, \tilde t \,\right]^2 \nonumber
\end{equation}
In general, multiple scattering could be treated in the track fit by
expressing the angular uncertainty of each thin scatterer as an
additional contribution to the error of each affected measurement.
Since a multiple scattering deflection will influence all downstream
measurement errors in a correlated way, this introduces artificial
correlations into the hitherto uncorrelated measurements, so that the
matrix ${\bf V}$ in section~\ref{sec:leastSquares} is no longer
diagonal. Evaluation of eq.~\ref{eq:leastSqParam} requires then
inversion of non-trivial matrices whose dimension is not only the
number of parameters but the number of measurements. Straight-forward
solutions of this problem have been devised~\cite{lutz}, which
intrinsically treat all multiple scattering angles as free parameters.
In many practical situations however, where the number of parameters
may be five and the number of measurements perhaps as large as 70,
this can lead to serious problems.

The generally accepted solution for the above problem is provided by
the Kalman filter technique. The multiple scattering dilution is added
as {\it process noise} (represented by the matrix $Q_k$ in the
transport equation, eq.~\ref{eq:transport}) at the very position in
the trajectory where it originates.  The Kalman filter normally
proceeds in the inverse flight direction along the path of the
particle and takes the influences illustrated in
fig.~\ref{fig:fitscheme} into account. Mathematically, the result
will be identical to a straight-forward least squares fit as described
in the previous paragraph, but the detailed procedure avoids handling
of huge matrices.

In Kalman filter language, the resulting covariance matrix
contribution for thin scatterers is
\begin{eqnarray}
\label{eq:thin}
  {\rm cov}(t_x,t_x) &=& (1+t_x^2) (1+t_x^2+t_y^2) C_{MS}\\ {\rm
    cov}(t_y,t_y) &=& (1+t_y^2) (1+t_x^2+t_y^2) C_{MS} \nonumber \\ 
  {\rm cov}(t_x,t_y) &=& \ \ \ t_x t_y \ \ 
  (1 + t_x^2 + t_y^2) C_{MS} \nonumber
\end{eqnarray}
(These and related formulas and their derivation can be found
in~\cite{wolin}).

It may be interesting to see how such a fit works in practice. In the
following, results of a study are shown which has been performed on
basis of simulated events in the HERA-B geometry
(fig.~\ref{fig:herabDetector}), applying a Kalman filter based track
fit to the simulated hits~\cite{rangerFit}.  This kind of geometry is
typical for modern forward spectrometers, and generally similar to
COMPASS~\cite{compass} or the planned LHCb~\cite{lhcb} and
bTEV~\cite{btev}. The study was based on detector design resolutions
and not intended to make quantitative statements on the actual
spectrometer performance, but to provide insight into the effects of
combining various different detector types, the sizable number of hits
per track, and the considerable amounts of material in the tracking
area that make an accurate treatment of multiple scattering essential.

\subsubsection{Impact parameter and angular resolutions}
\begin{figure} 
\begin{center}
\unitlength0.8cm
\begin{picture}(13,13)
\put(0,0){\makebox{\epsfig{file=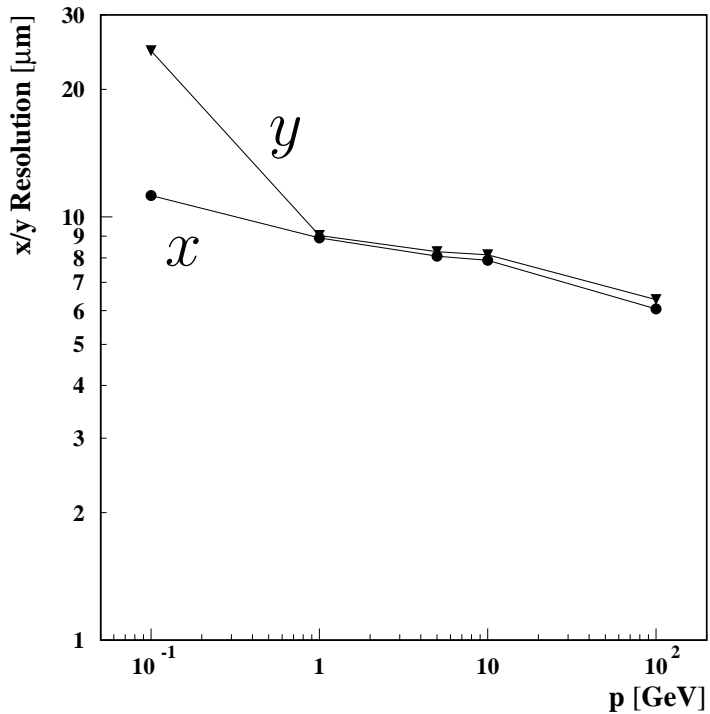,width=10.4cm,clip=,angle=0}}}
\put(3,   8){\makebox(0,0)[t]{\huge \bf $x$}}
\put(4.7,10){\makebox(0,0)[t]{\huge \bf $y$}}
\end{picture}
\end{center}
\caption{Impact parameter resolution at the first track point,
separately for the coordinate in the bending plane ($x$, circles) and
the non-bending plane ($y$, triangles).}
\label{impact}
\end{figure} 
\begin{figure} 
\begin{center}
\unitlength0.8cm
\begin{picture}(13,13)
\put(0,0){\makebox{\epsfig{file=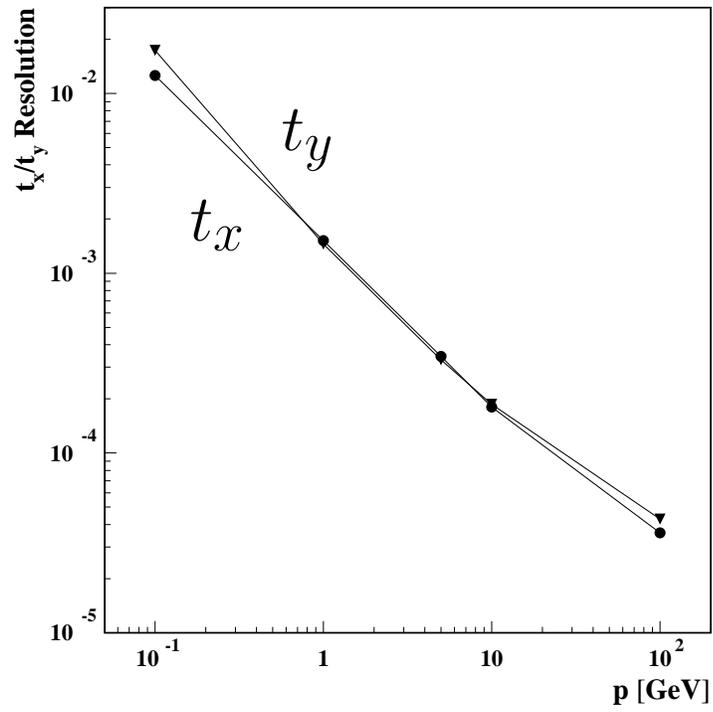,width=10.4cm,clip=,angle=0}}}
\put(3.5,8.5){\makebox(0,0)[t]{\huge \bf $t_x$}}
\put(5,   10){\makebox(0,0)[t]{\huge \bf $t_y$}}
\end{picture}
\end{center}
\caption{Resolution of the slope parameters $t_x = \tan \theta_x$
(circles) and $t_y = \tan \theta_y$ (triangles).}
\label{slope}
\end{figure}
The visible track parameter resolution was obtained by calculating the
{\it track parameter residual} for each track using the Monte Carlo
truth, and applying a Gaussian fit to the distribution. (The term {\it
  visible} is used to distinguish this resolution from the one
estimated by the fit.) The impact parameter resolution for tracks
passing the Silicon micro-vertex detector and the outer tracker as a
function of momentum is shown in fig.~\ref{impact}. Since this impact
parameter is defined with respect to the position of the first hit of
the track counting from the interaction point, the resolution is
governed by the error of the first coordinate and only weakly
dependent on momentum. Multiple scattering acts like a filter which
dilutes the information from the following layers, only at higher
momentum their contribution to the resolution at the first point
becomes visible.

Since the vertex detector measurement accuracy is approximately
isotropic, horizontal and vertical resolution are almost identical,
the deviation at $p=100~{\rm MeV}$ is explained by the fact that the
strips in the first vertex detector layer are oriented almost parallel
to the $y$ axis. The resolution of track slopes is shown in
fig.~\ref{slope} and turns out to be dominated by the pronounced
$\propto 1/p$ behaviour expected in a multiple scattering-dominated
regime. At high momentum, the onset of coordinate resolution effects
appears to be just visible, where the slightly better resolution of
the horizontal slope ($t_x$) may be due to the dominantly vertical
orientation (parallel to $y$) of the wires in the main tracking
system.

The impact parameter resolution given above should not be confused
with the quantities relevant for physics performance where assignment
to vertices is important. In the latter case, the track parameters
must be extrapolated from the first track point to the interaction
area. With extrapolation distances of typically $\cal O$(10~cm), the
resolution of the {\it extrapolated} impact parameters will generally
be fully dominated by the {\it angular} resolution rather than the
impact parameter resolution at the first point.

\subsubsection{Momentum resolution}
\label{sec:momres}
\begin{figure} 
\begin{center}
\unitlength1cm
\begin{picture}(13,13)
\put(0,0){\makebox{\epsfig{file=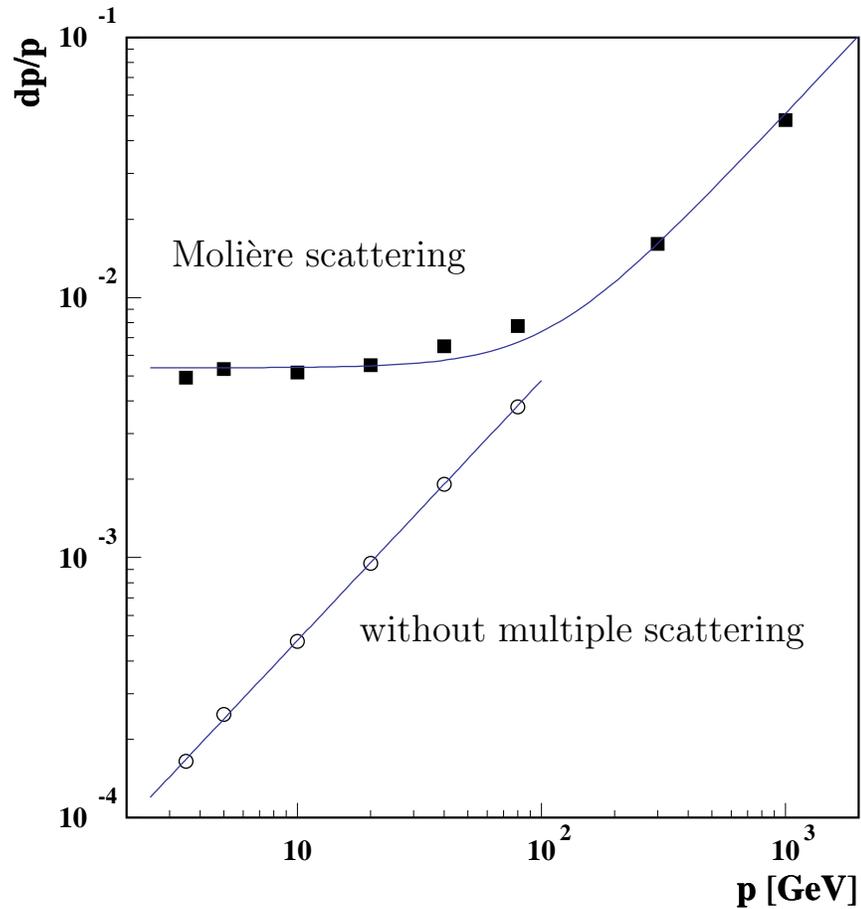,width=13cm,clip=,angle=0}}}
\put(4.5, 9.0){\makebox(0,0)[t]{\large Moli\`ere scattering}}
\put(8.0, 4.0){\makebox(0,0)[t]{\large without multiple scattering}}
\end{picture}
\end{center}
\caption{Visible momentum resolution (filled squares) for simulated
  muons (filled squares) together with the fit of the parametrization
  described in the text (upper solid line). In the lower part, the
  open squares show the visible resolution with multiple scattering
  switched off in the simulation, with a similar function fitted. The
  open circles show the pure coordinate resolution as estimated by the
  track fit, with a linear function fitted to it.}
\label{mom}
\end{figure} 
A very central design issue in spectrometers is resolution of
momentum, since it determines the rejection power against background
in particle spectrometry. The relative momentum resolution, labelled
$dp/p$, as a function of momentum is shown in fig.~\ref{mom} for
particles traversing the areas {\it SI}, {\it MC} and {\it PC} of the
spectrometer (see fig.~\ref{fig:herabDetector} for definition) in the
polar angle area $0.1<\theta<0.15$. The circle symbols show the
relative momentum resolution that results with multiple scattering
switched off in the simulation, leading to a strictly linear
dependence on $p$. This behaviour is expected since the resolution is
then only determined by the coordinate resolution and the geometrical
layout of the spectrometer - size and number of layers - that
provides the leverage for momentum measurement together with the
magnetic field. The result reflects the fact that the curvature
$\kappa$, which is the inverse of the radius of curvature, can be
measured with a precision that is independent of its actual value,
hence $\delta \kappa$=const. On the other hand the curvature is
inversely proportional to the momentum, so that $dp/p \propto
p$.  In presence of multiple scattering, the resolution shows a
multiple scattering-dominated regime below momenta of $\approx 50\ 
\rm{GeV}$, and a transition into a linear rise at high momentum.
Superimposed is a fit with a constant and a linear resolution term
added in quadrature. This parametrization, which corresponds to a
commonly used function introduced by Gluckstern~\cite{gluckstern} for
an even spacing of tracking stations does not fit the visible
resolution very well in the momentum mid-range, which can be
attributed to the uneven distribution of measurements, resolutions,
material and magnetic field strength in the spectrometer.

\subsubsection{Effects of fit non-linearity}
The presence of the inhomogeneous magnetic field introduces particular
effects of non-linearity into the fitting problem. The least squares
fit technique, which the Kalman filter is built on, can still be
applied, with the transport matrices now obtained as derivatives of
the transport function. As already noted in
sec.~\ref{sec:leastSquares}, the optimal properties of the least
squares technique are still retained on the condition that the
derivatives are taken at the position of the final trajectory. Since
this is initially not necessarily the case, the fit must be repeated
iteratively until the procedure converges.

\begin{figure} 
\begin{center}
\epsfig{file=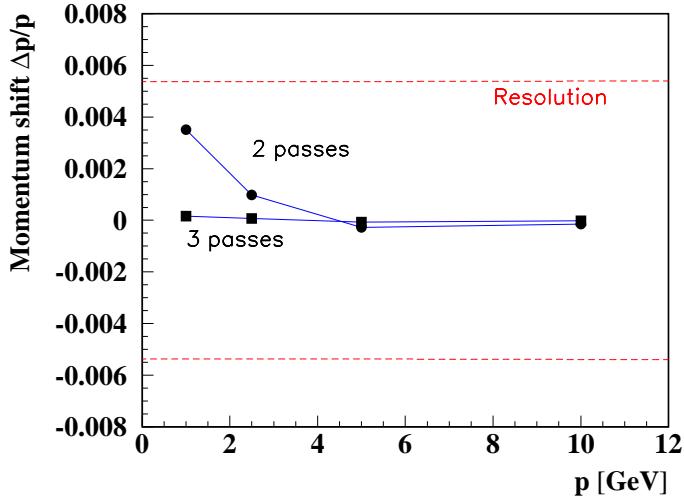,width=10cm,clip=,angle=0}
\end{center}
\caption{Residual of the momentum parameter (see
  eq.~\ref{eq:normParmResid}) normalized to the momentum itself for
  two and three passes of the fit. The relative momentum resolution is
  indicated by the dashed lines for comparison.}
\label{niter}
\end{figure} 
The practical implications of non-linearity are visible in
fig.~\ref{niter}, which shows the mean relative deviation of the
reconstructed from the true momentum value. Small systematic shifts of
reconstructed momentum are observed for momentum below 5~GeV with two
fit passes applied. These shifts reflect the convergence behaviour of
the fit due to non-linearity. They are found to be virtually removed
when a third pass is applied.

\subsubsection{Contributions of different parts of the spectrometer}
\begin{figure} 
\begin{center}
\epsfig{file=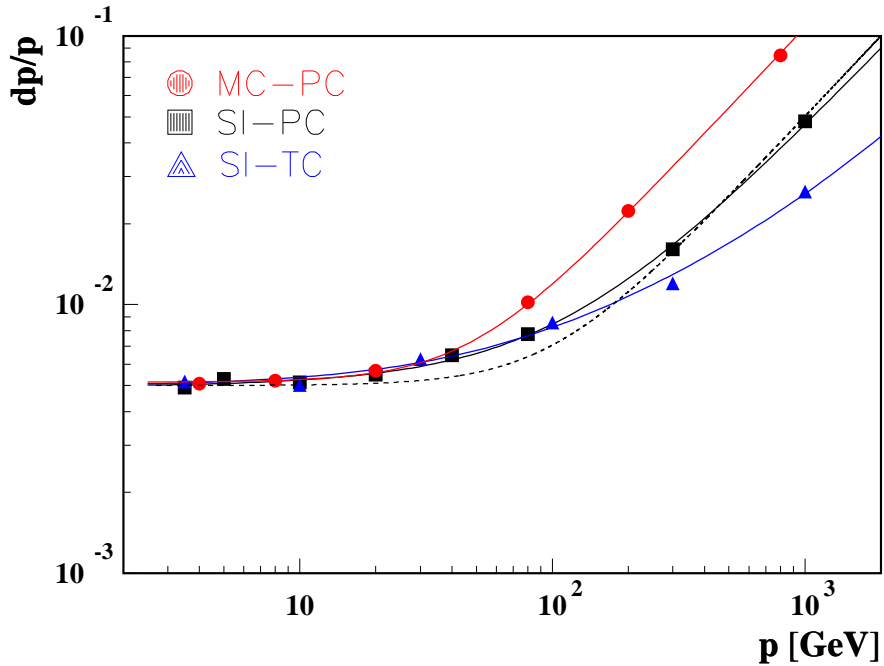,width=13cm,clip=,angle=0}
\end{center}
\caption{Relative momentum resolution in the MC--PC (circles), SI--PC
  (squares) and SI--TC (triangles) spectrometer ranges, together
  with the fits described in the text. The dashed line is the upper
  fit in fig.~\ref{mom}.}
\label{momtc}
\end{figure} 
For understanding detector design, it is also important to investigate
how much different parts of the spectrometer contribute to the
momentum measurement. In the HERA-B geometry
(fig.\ref{fig:herabDetector}), the tracking system is grouped into the
vertex detector (SI), the chambers within the magnet (MC), the
chambers just behind the magnet (PC) and the so-called {\it trigger
  chambers} (TC), which are separated from the PC part by the
ring-imaging \v{C}erenkov detector (RICH). In order to separate the
contributions of the different spectrometer parts, the range of the
fit was modified by omitting the vertex detector hits (labelled
MC--PC range) and by adding the hits from the tracking chambers at the
end of the main tracking system (SI--TC range).  The resulting
momentum resolutions are displayed in fig.~\ref{momtc}.  It turns out
that without including the vertex detector (MC--PC), the momentum
resolution is well described by a constant and a linear term added in
quadrature. In the regime of linear rise, the poorer coordinate
resolution is reflected in comparison to the system including the
vertex detector.  When the fit on the other hand is extended into the
``TC'' region which is mainly designed to support the trigger
(SI--TC), these additional measurements with their huge lever arm are
expected to improve the coordinate contribution of the resolution.
Such an improvement is visible in fig.~\ref{momtc} for $p \geq
100~GeV$, where it is hardly relevant for the physics scope of the
experiment. A third term proportional to the square-root of the
momentum had to be added in quadrature to fit the resolution for the
latter two ranges.

\subsubsection{Parameter covariance matrix estimation}
\begin{figure} 
\begin{center}
\begin{tabular}{cc}
   \unitlength1cm
   \begin{picture}(8,6)
        \put(0,0){\makebox{\epsfig{file=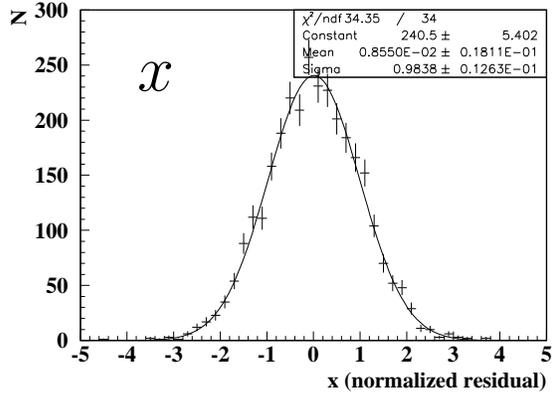,width=8cm}}}
        \put(2,4.5){\makebox(0,0)[t]{\huge \bf $x$}}
   \end{picture}
                        &
   \unitlength1cm
   \begin{picture}(8,6)
        \put(0,0){\makebox{\epsfig{file=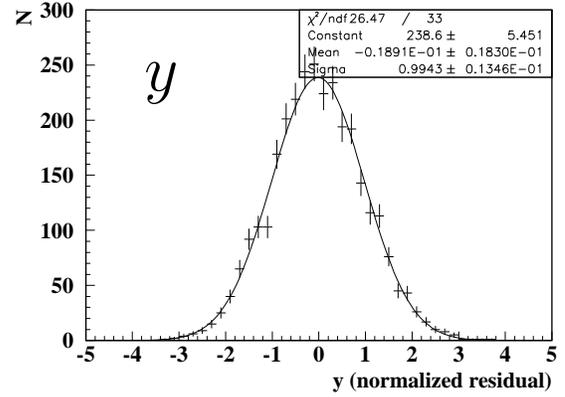,width=8cm}}}
        \put(2,4.5){\makebox(0,0)[t]{\huge \bf $y$}}
   \end{picture}
                         \\
   \unitlength1cm
   \begin{picture}(8,6)
        \put(0,0){\makebox{\epsfig{file=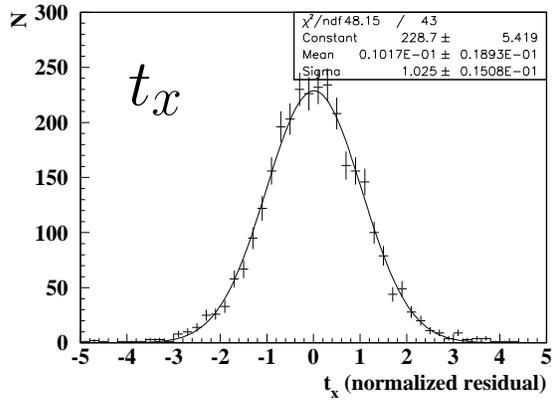,width=8cm}}}
        \put(2,4.5){\makebox(0,0)[t]{\huge \bf $t_x$}}
   \end{picture}
                        &  
   \unitlength1cm
   \begin{picture}(8,6)
        \put(0,0){\makebox{\epsfig{file=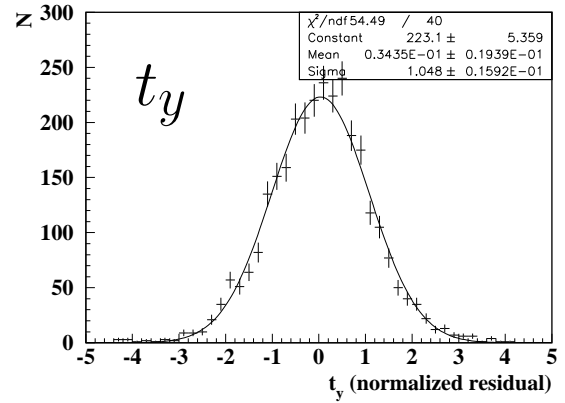,width=8cm}}}
        \put(2,4.5){\makebox(0,0)[t]{\huge \bf $t_y$}}
   \end{picture}
                          \\
   \unitlength1cm
   \begin{picture}(8,6)
        \put(0,0){\makebox{\epsfig{file=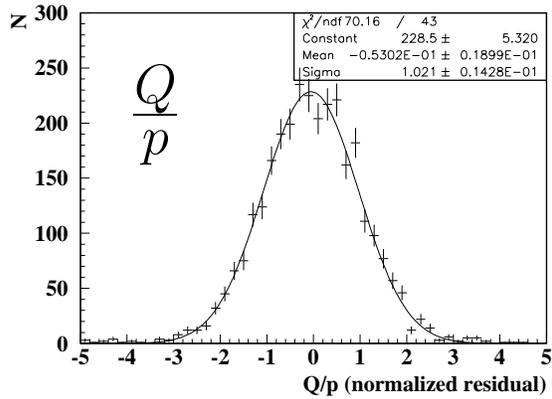,width=8cm}}}
        \put(2,4.5){\makebox(0,0)[t]{\huge \bf $\frac{Q}{p}$}}
   \end{picture}
\end{tabular}
\caption{Normalized parameter residual distributions for muons of
  10~GeV, based on 3000 simulated tracks.}
\label{pulls10}
\end{center}
\end{figure}
A very important task of the track fit is the quantification of the
covariance matrix of the estimated track parameters. The reliability
of parameter error estimation can be studied by investigating
distributions of {\it normalized parameter residuals} (see
eq.~\ref{eq:normParmResid} in sec.~\ref{sec:parameterResolution}),
which use the estimated error for normalization. In the example at
hand, the resulting pull distributions are shown in
fig.~\ref{pulls10}, where unbiased fits with a Gaussian function are
superimposed. Distortions of the parameter estimates would show up as
deviations of the mean values from zero, which are however not present
in this case. The Gaussian cores of the pulls agree in all cases with
unity width, indicating a reliable estimate of the covariance matrix.
One should note that only mean value and variance of the pull
distribution are indicators of the quality of the estimate. The actual
{\it shape} of the distribution, e.g.  whether it is Gaussian or not,
reflects the underlying structure of the problem, as will be more
clearly visible in the next section.

\subsubsection{Goodness of fit}
Since the Kalman filter is mathematically equivalent to a least-squares 
estimator, the sum of the {\it filtered $\chi^2$} contributions will 
follow a $\chi^2$ distribution, provided that the random variables 
entering into the fit have Gaussian distributions. In
this case the {\it $\chi^2$ probability} 
$$ 
P_{\chi^2} = \int\limits_{-\infty}^{\chi^2} 
       f(\tilde{\chi}^2)\,d\tilde{\chi}^2 
$$
where $f(\tilde{\chi}^2)$ is the standard $\chi^2$ distribution for
the appropriate number of degrees of freedom, should be evenly
distributed between 0 and 1. ($P_{\chi^2}$ is often called {\it
  confidence level}.)  This prerequisite is not strictly
fulfilled in case of Moli\`ere scattering, so that deviations are to
be expected. These effects have potentially large influence in modern
radiation hard drift chambers, where the drift cells are enclosed in
a multitude of small gas volumes and a considerable amount of material is
introduced into the tracking area.

\begin{figure} 
\begin{center}
\unitlength1cm
\begin{picture}(12,12)
  \put(0,6){\makebox{\epsfig{file=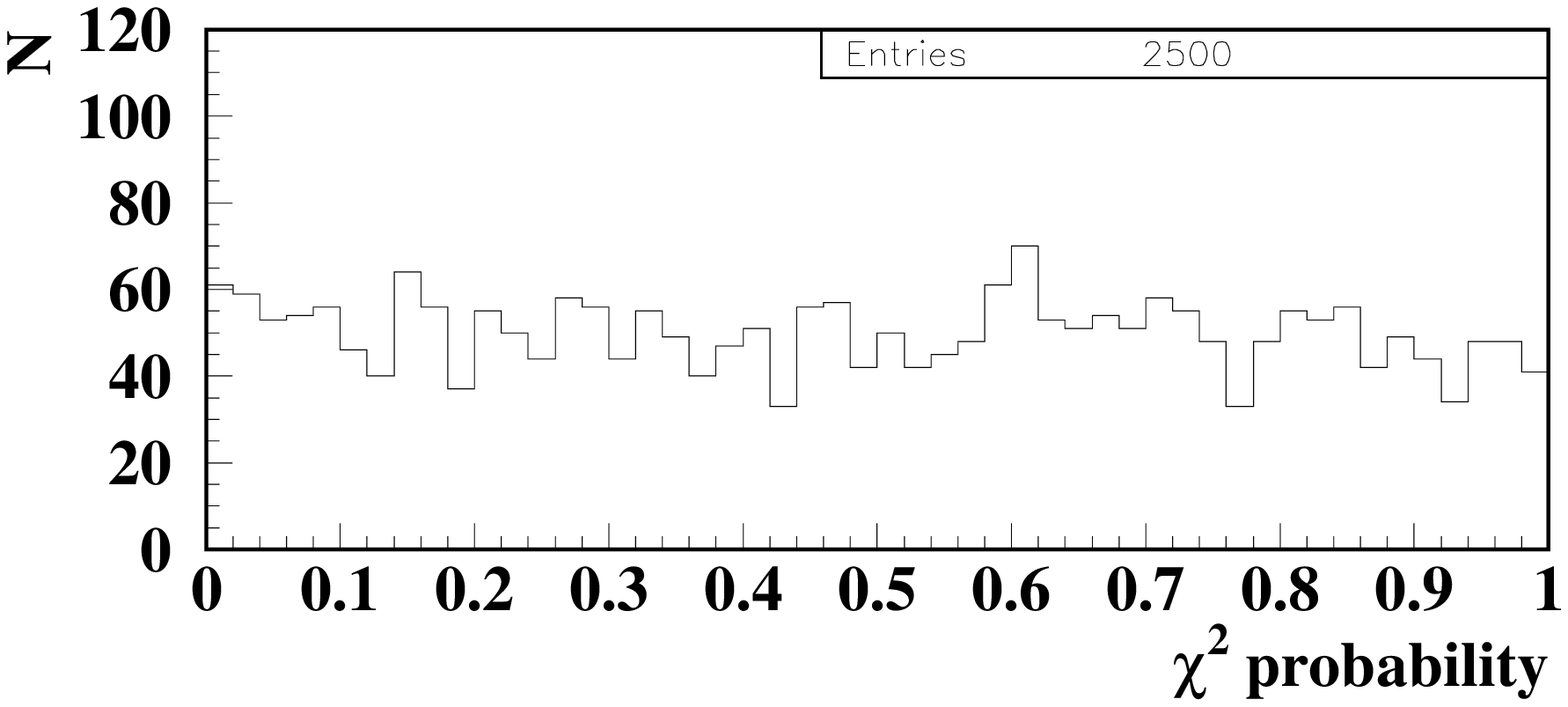,width=13cm}}}
  \put(3.5,10.7){\makebox(0,0)[t]{\large \bf (a)}}
  \put(0,0){\makebox{\epsfig{file=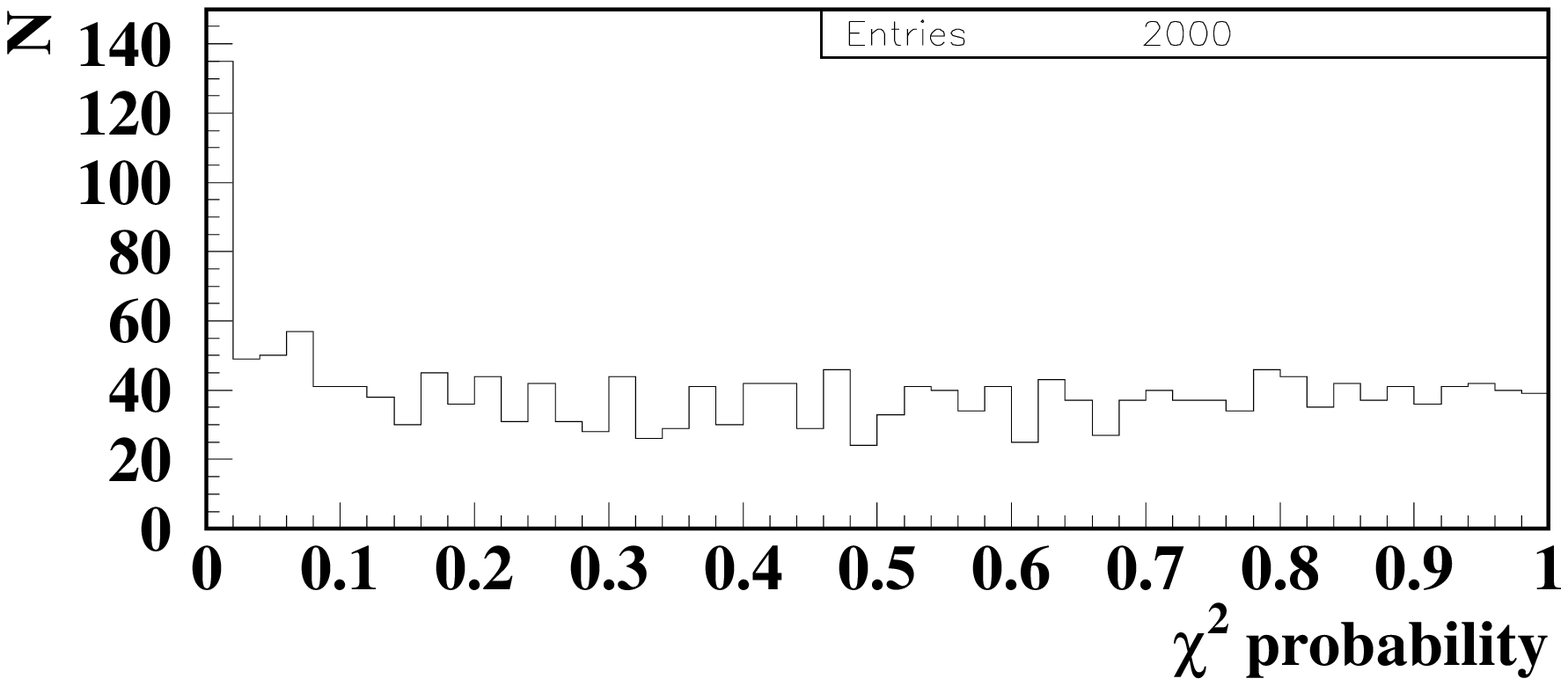,width=13cm}}}
  \put(3.5,4.7){\makebox(0,0)[t]{\large \bf (b)}}
\end{picture}
\caption{Distributions of $\chi^2$ probability (confidence level) for
  the track fit for a 10~GeV particle, (a) with Gaussian form of
  multiple scattering, (b) with Moli\`ere scattering.}
\label{fig:probms}
\end{center}
\end{figure}
\begin{figure} 
\begin{center}
\unitlength1cm
\begin{picture}(12,18)
\put(0,13.5){\makebox{\epsfig{file=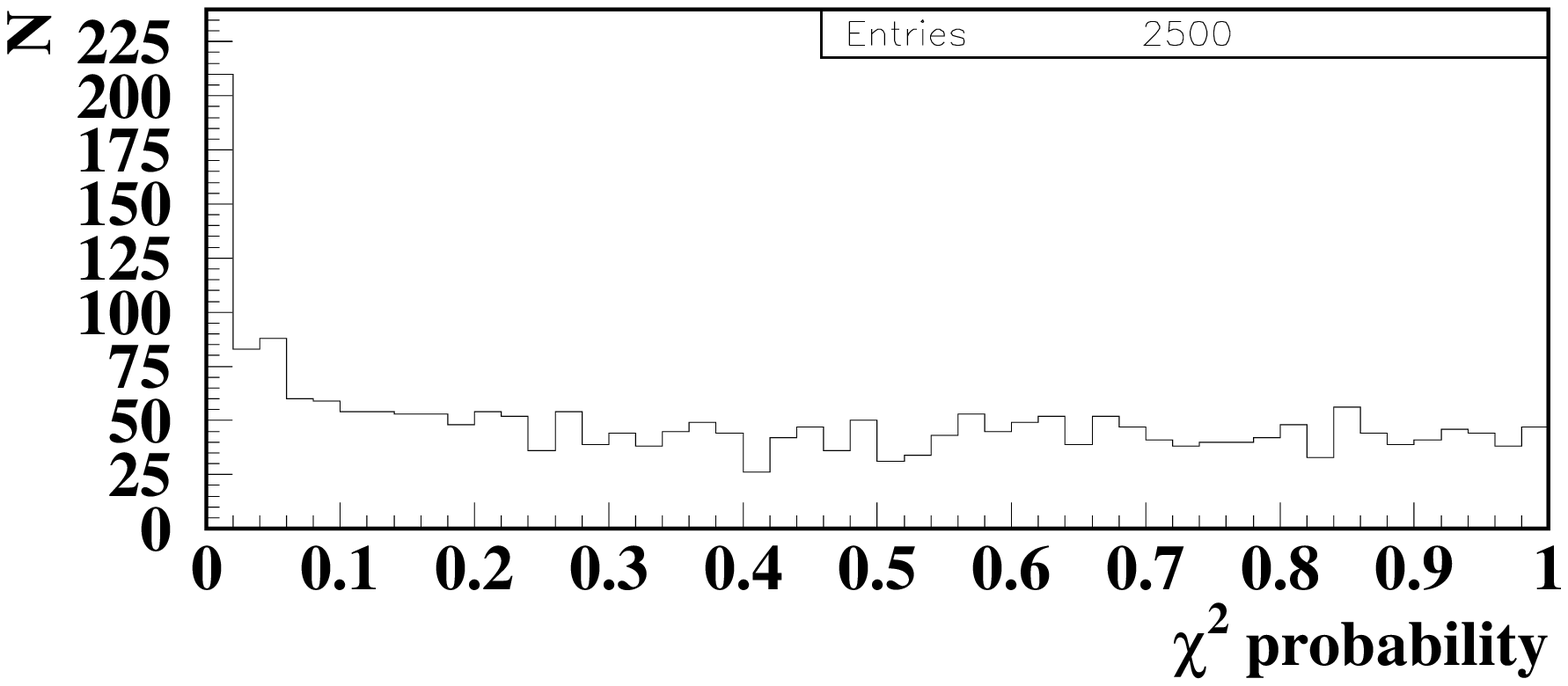,width=10cm,clip=,%
        bbllx=0pt,bblly=4pt,bburx=516pt,bbury=233pt}}}
\put(4.5,17.3){\makebox(0,0)[t]{\large \bf (a)   p=3~GeV}}
\put(0,9){\makebox{\epsfig{file=fig/prob-mol.eps,width=10cm,clip=,%
        bbllx=0pt,bblly=4pt,bburx=516pt,bbury=233pt}}}
\put(4.5,12.8){\makebox(0,0)[t]{\large \bf (b)   p=10~GeV}}
\put(0,4.5){\makebox{\epsfig{file=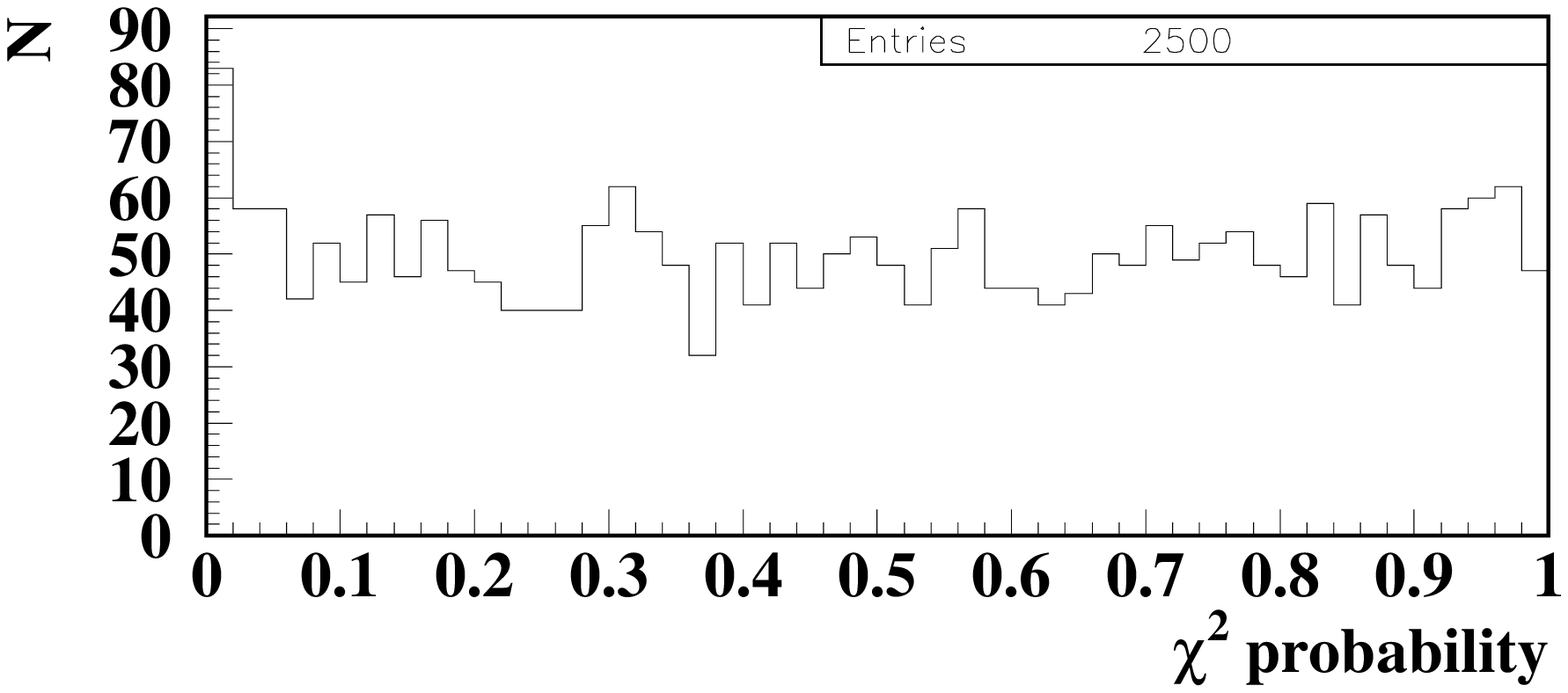,width=10cm,clip=,%
        bbllx=0pt,bblly=4pt,bburx=516pt,bbury=233pt}}}
\put(4.5,8.3){\makebox(0,0)[t]{\large \bf (c)   p=30~GeV}}
\put(0,0){\makebox{\epsfig{file=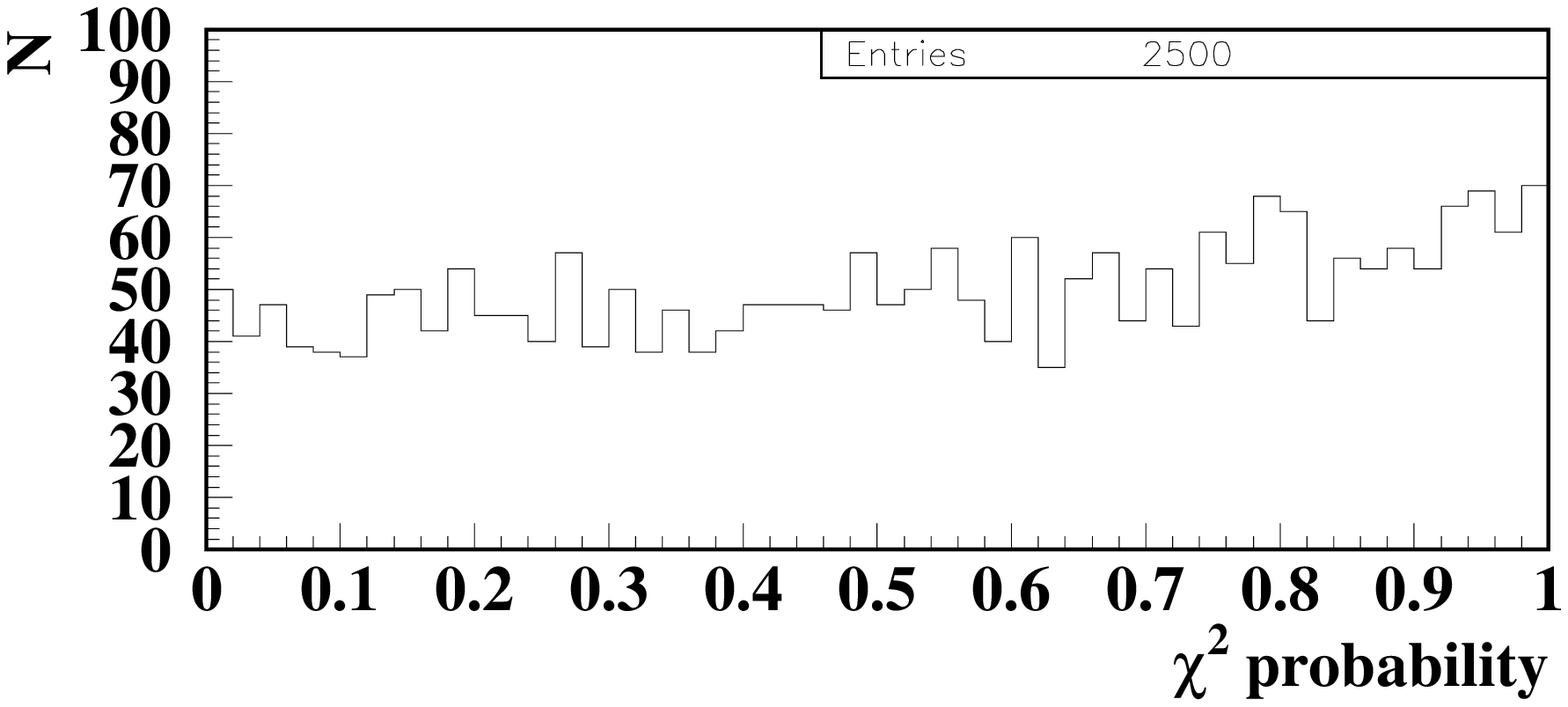,width=10cm,clip=,%
        bbllx=0pt,bblly=4pt,bburx=516pt,bbury=233pt}}}
\put(4.5,3.8){\makebox(0,0)[t]{\large \bf (d)   p=100~GeV}}
\end{picture}

\caption{Distribution of $\chi^2$ probabilities as a function of
  momentum, (a) 3.5~GeV, (b) 10~GeV, (c) 30~GeV and (d) 100~GeV.}
\label{fig:proball}
\end{center}
\end{figure}
Figure~\ref{fig:probms} compares the distribution of the $\chi^2$
probability for the Gaussian form of multiple scattering (a) and
Moli\`ere scattering (b). The peak at small probabilities in (b)
obviously does not indicate a bad behaviour of the fit, but instead
shows the inadequateness of the $\chi^2$ test with non-Gaussian random
variables.  The probability distribution for various momentum values
is displayed in fig.~\ref{fig:proball}. The increasing prominence of
the peak at low probability is clearly seen with decreasing momentum.
Small $\chi^2$ probability does not necessarily imply a bad estimation
of the parameters, hence special care is required when a $\chi^2$ cut
is to be used to eliminate improperly reconstructed tracks.

\subsection{Treatment of Ionization Energy Loss And Radiation}
\subsubsection{Ionisation energy loss}
\label{sec:ionization}
\begin{figure} 
\begin{center}
\begin{tabular}{cc}
   \unitlength1cm
   \begin{picture}(8,8)
        \put(0,0){\makebox{\epsfig{file=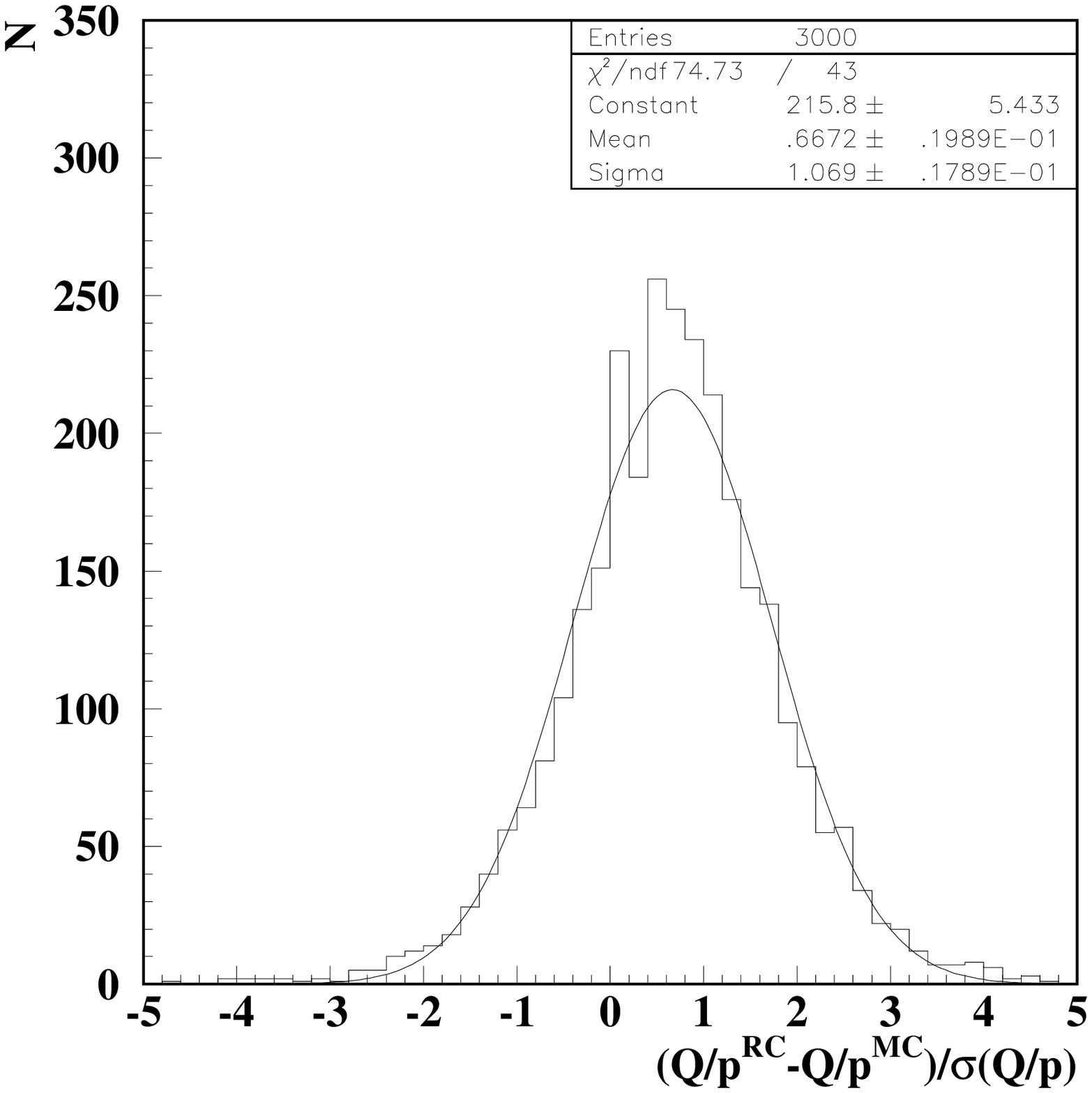,width=8cm}}}
        \put(1.3,6.5){\makebox(0,0)[l]{\large \bf (a)}}
        \put(1.3,5.7){\makebox(0,0)[l]{\large \bf $p=3~GeV$}}
        \put(1.3,5.2){\makebox(0,0)[l]{\it no correction}}
   \end{picture}
                        &
   \unitlength1cm
   \begin{picture}(8,8)
        \put(0,0){\makebox{\epsfig{file=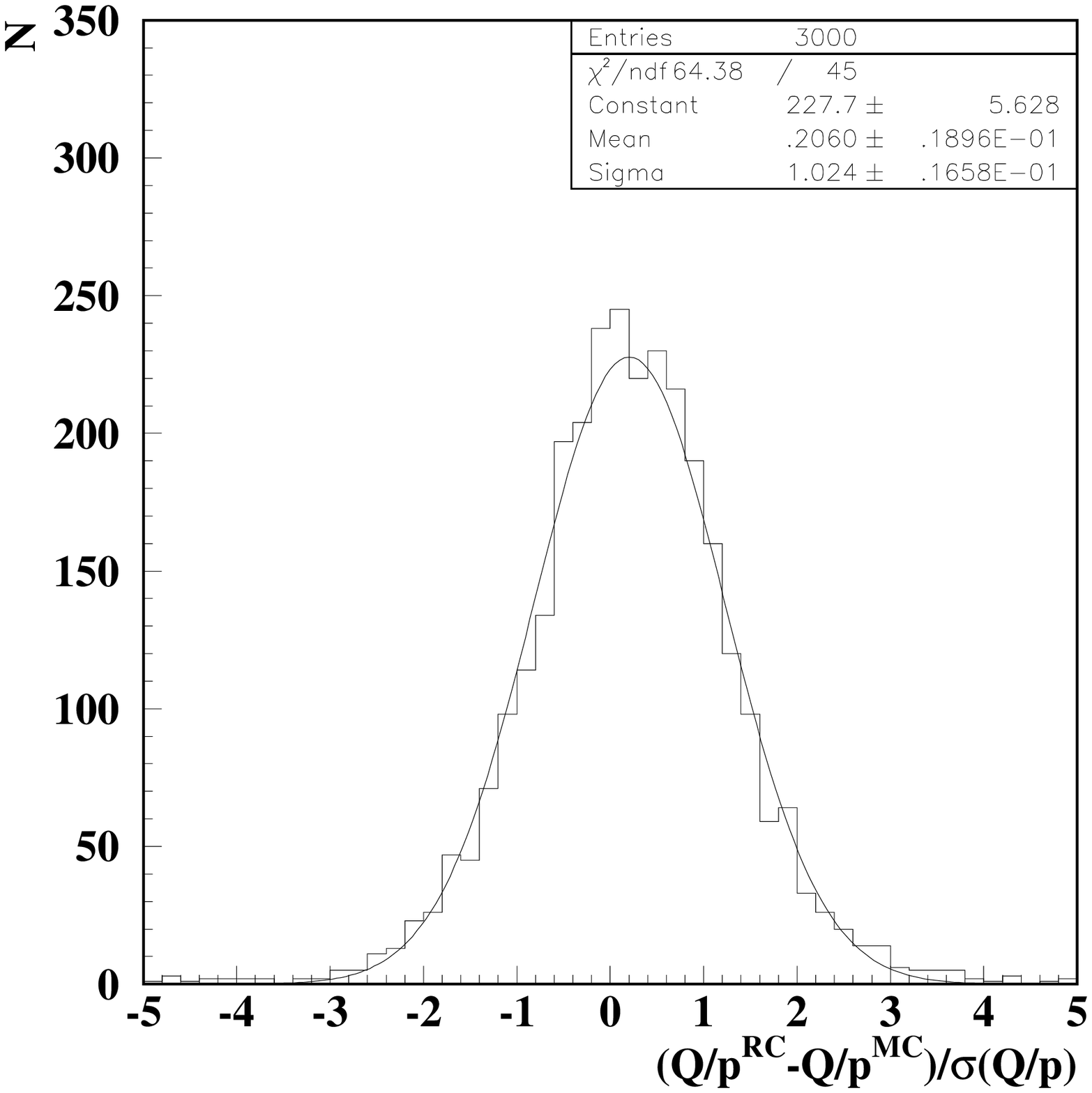,width=8cm}}}
        \put(1.3,6.5){\makebox(0,0)[l]{\large \bf (b)}}
        \put(1.3,5.7){\makebox(0,0)[l]{\large \bf $p=10~GeV$}}
        \put(1.3,5.2){\makebox(0,0)[l]{\it no correction}}
   \end{picture}
                         \\
   \unitlength1cm
   \begin{picture}(8,8)
        \put(0,0){\makebox{\epsfig{file=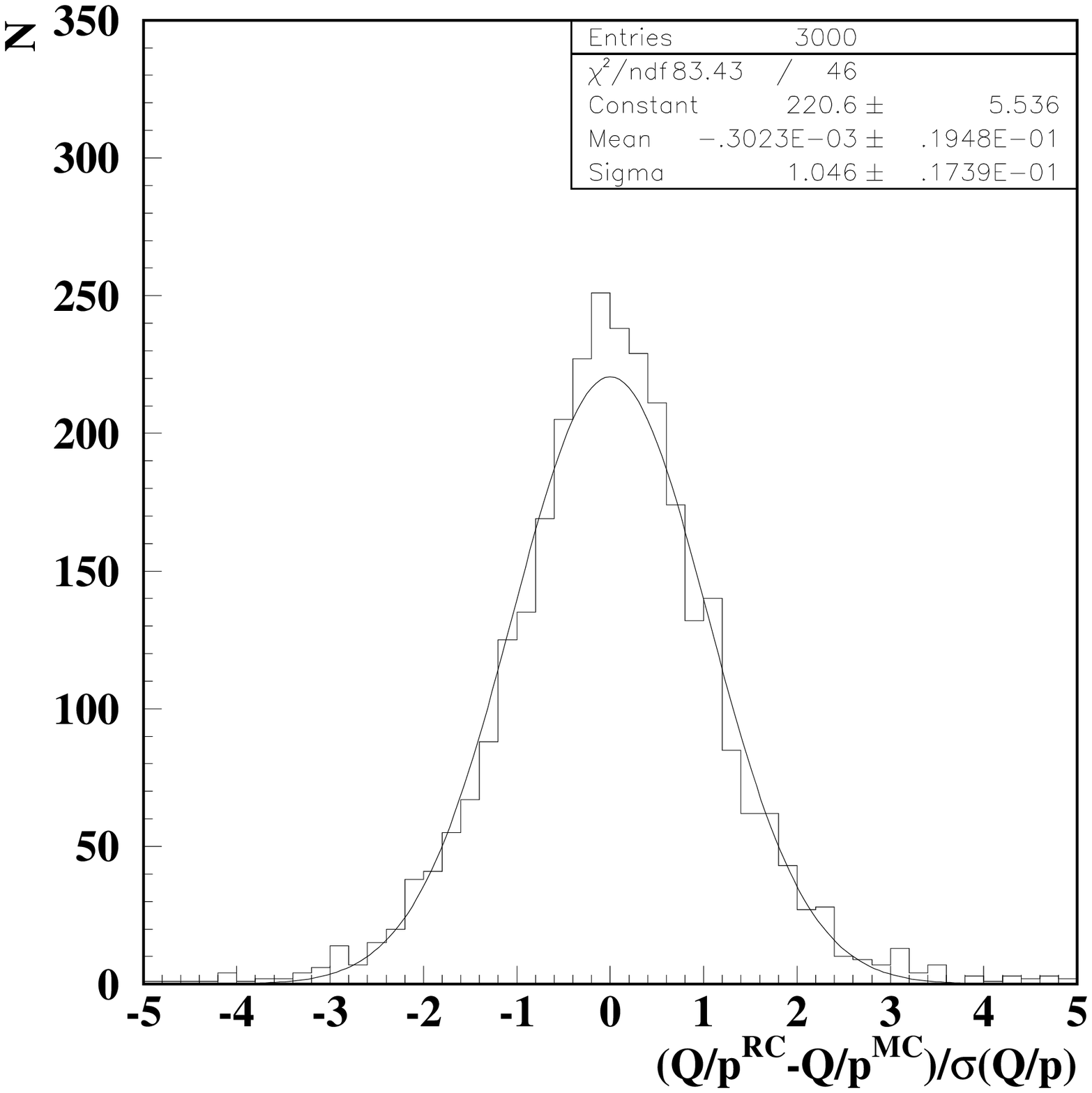,width=8cm}}}
        \put(1.3,6.5){\makebox(0,0)[l]{\large \bf (c)}}
        \put(1.3,5.7){\makebox(0,0)[l]{\large \bf $p=3~GeV$}}
        \put(1.3,5.2){\makebox(0,0)[l]{\it corrected}}
   \end{picture}
                        &  
   \unitlength1cm
   \begin{picture}(8,8)
        \put(0,0){\makebox{\epsfig{file=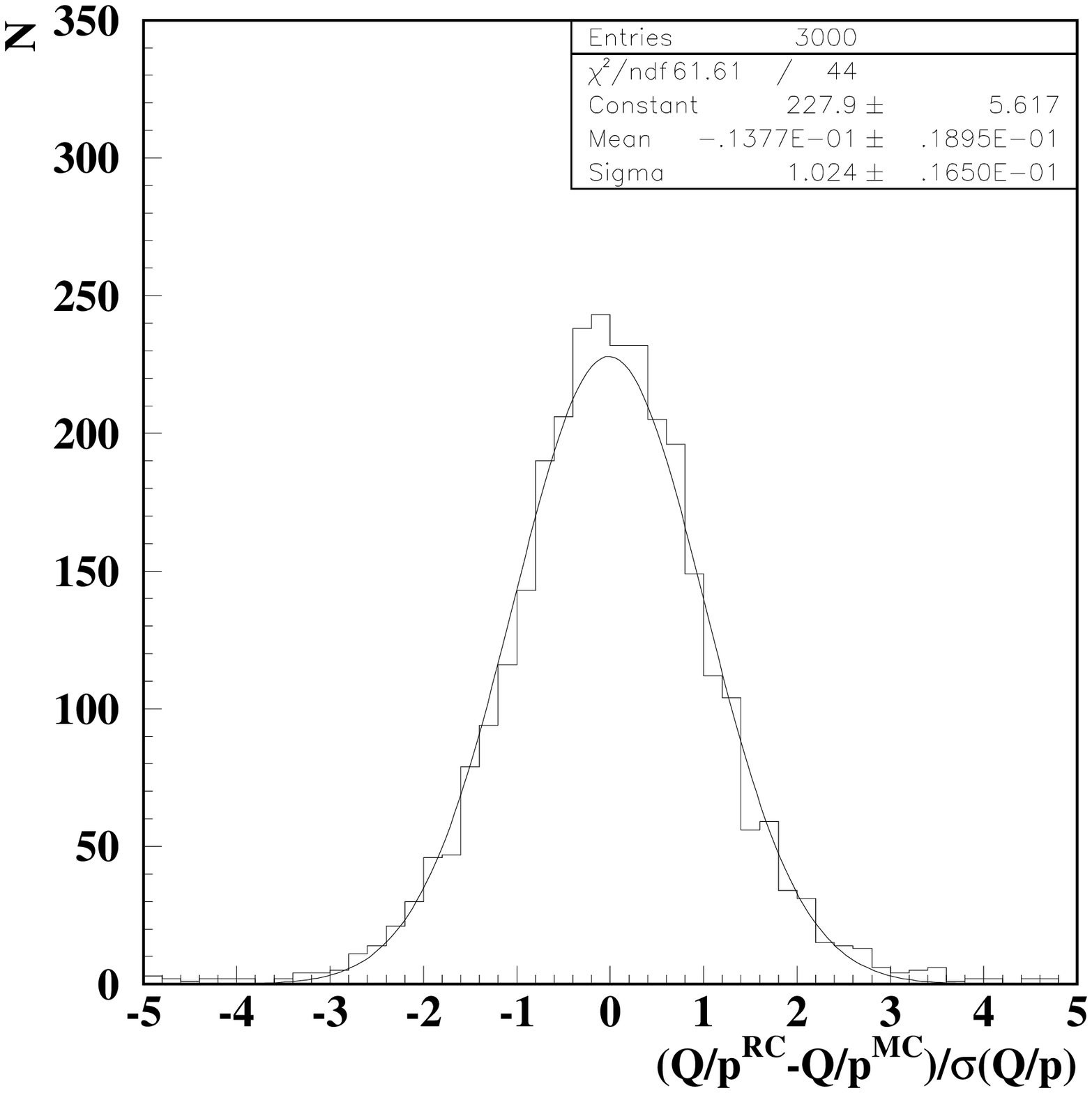,width=8cm}}}
        \put(1.3,6.5){\makebox(0,0)[l]{\large \bf (d)}}
        \put(1.3,5.7){\makebox(0,0)[l]{\large \bf $p=10~GeV$}}
        \put(1.3,5.2){\makebox(0,0)[l]{\it corrected}}
   \end{picture}
\end{tabular}
\caption{Pull distribution of the momentum parameter for $\mu^+$ 
  particles of 3~GeV (a,c) and 10~GeV (b,d). The upper pictures show
  the effect of $dE/dx$ if no correction is applied. The lower plots
  show the same when the correction is applied in the fit.}
\label{fig:eloss}
\end{center}
\end{figure}
For minimal ionizing particles in the GeV energy range, energy loss
due to ionization within the tracking system depends in good
approximation only on the amount of material that is traversed. In
this case, it is not the radiation thickness (as defined in
eq.~\ref{eq:radThickness}), but the geometrical thickness multiplied
by the mass density of the material that is relevant. Since the
energy loss depends only weakly on the energy itself in this range,
the effect will become most noticeable for low momentum particles.
This behaviour is illustrated in fig.~\ref{fig:eloss}, which shows the
normalized residual of the momentum parameter $Q/p$ for $\mu^+$
particles of 3.5 and 10~GeV with ionization energy loss simulation
turned on.  The residual distributions are shifted towards positive
values of $Q/p$, reflecting an underestimation of the energy, which is
caused by the ionization energy loss, in particular upstream of the
magnet. The visible shift corresponds to an energy loss of 12~MeV. On
the other hand, the width of the residual distributions is not
significantly increased, which in the 10~GeV case can directly be seen
by comparing with fig.~\ref{pulls10}.
 
A correction can be applied in each
filter step if the $dE/dx$ of the particle in the material is known, since
\begin{equation}
  E_{after} = E_{before} - (dE/dx) _{ion}\cdot \ell 
\end{equation}
where $\ell$ is the traversed thickness of the material. This requires
in general the knowledge of the particle mass. Since ionization energy
loss will be most notable for small particle energies where the
resolution is governed by multiple scattering, no correction to the
momentum error has been applied. The bottom part of
fig.~\ref{fig:eloss} displays the same normalized residuals with the
energy loss correction applied. The bias of the momentum estimate is
successfully eliminated by the correction.

\subsubsection{Radiative energy loss}
The corrections discussed up to now are usually sufficient for minimum
ionizing particles. For electrons\footnote{in this section the term
  {\it electron} should be interpreted to imply {\it
    positron} as well} however, the situation is more complicated
since above the {\it critical energy}, which is of the order of MeV, these
particles lose more energy through radiation of photons than through
ionization when they traverse material. This process is also of a more
notably stochastic nature than ionization energy loss, as considerable
fractions of the electron energy can be transferred to the photon.
Modern radiation-hard detectors as e.g. those under construction for the LHC
are confronted with this problem to a much higher degree than
traditional detectors, because of the significant amount of material
in the tracking system, which can easily exceed 50\% of a radiation
length.

\begin{figure} 
\begin{center}
  \epsfig{file=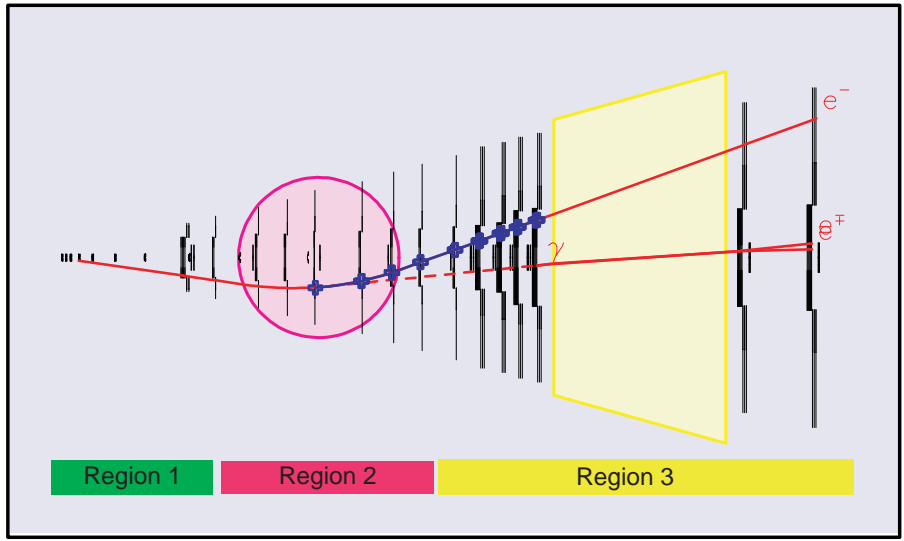,width=13cm}
  \caption{ Regions 1--3 for classifying radiative energy loss
    illustrated in the geometry of the HERA-B spectrometer. The
    simulated geometry differs in some details from the one in
    fig.~\ref{fig:herabDetector}. Also the trajectory of a simulated
    electron is shown, which radiates a photon within the magnet that
    converts into a $e^+e^-$ pair further downstream.}.
\label{fig:radrange}
\end{center}
\end{figure}
For the relevance of photon radiation on measurement of the electron,
three cases have to be distinguished regarding the range where the
radiation occurs (indicated as regions 1--3 in
fig.~\ref{fig:radrange}):
\begin{description}
\item[Region 1: between interaction point and spectrometer magnet] If
  the point of origin of the particle is not yet within the magnetic
  field -- as is typical for fixed-target setups rather than for
  collider detectors -- radiation will not change the electron
  trajectory and thus not interfere with the quality of the fit;
  however, the spectrometer will only measure the remaining momentum
  of the electron after the radiation.
\item[Region 2: within the magnetic field] In this case, the curvature
  of the trajectory changes because of the radiation, which means that
  the energy change is -- in principle -- measurable. Ignoring the
  radiation in the fit will lead to a bad description of the
  trajectory and to distortions of the parameter estimates.
\item[Region 3: beyond the magnetic field] If the electron loses
  energy downstream of the magnet, this will have no influence on the
  momentum measurement in the spectrometer. However, pair creation
  from radiated photons may lead to accompanying particles that
  can disturb pattern recognition in the downstream area.
\end{description}

The dilution due to energy loss of electrons and positrons through
emission of electromagnetic radiation can be treated by the method by
Stampfer et al.~\cite{stampfer}. According to the Bethe-Heitler
equation~\cite{betheheitler}, this energy loss is described by
\begin{equation}
  \left( \frac{dE}{dx} \right)_{rad} = \frac{E}{x_R}
\end{equation}
where $x_R$ is the radiation length of the traversed material (see
section~\ref{sec:multipleScattering}). This leads to the relation
\begin{equation}
  \left< \frac{E_{after}}{E_{before}} \right> = e^{-t}
\end{equation}
where $t$ is the traversed distance measured in radiation lengths as
defined before. For a track propagation which follows the track
opposite to its physical movement, one obtains on average
\begin{equation}
  \left( \frac{Q}{p} \right)' = \frac{Q}{p} + \Delta \left(
    \frac{Q}{p} \right) =
  \frac{Q}{p} - \frac{Q}{p}
  \frac{E_{before}-E_{after}}{E_{before}} = \frac{Q}{p} e^{-t}
\end{equation}
The contribution to the propagated covariance matrix emerges as
\begin{equation}
  \Delta {\rm cov} \left( \frac{Q}{p},\frac{Q}{p} \right) =
  \left(\frac{Q}{p}\right)^2 \left(e^{-t \frac{ln3}{ln2}} - e^{-2t}\right)
\end{equation}
This contribution can be included into the Kalman filter process noise
as introduced in eq.~\ref{eq:transport}.

\begin{figure} 
\begin{center}
\begin{tabular}{cc}

   no radiation correction & with radiation correction \\

   \unitlength1cm
   \begin{picture}(8,6)
        \put(0,0){\makebox{\epsfig{file=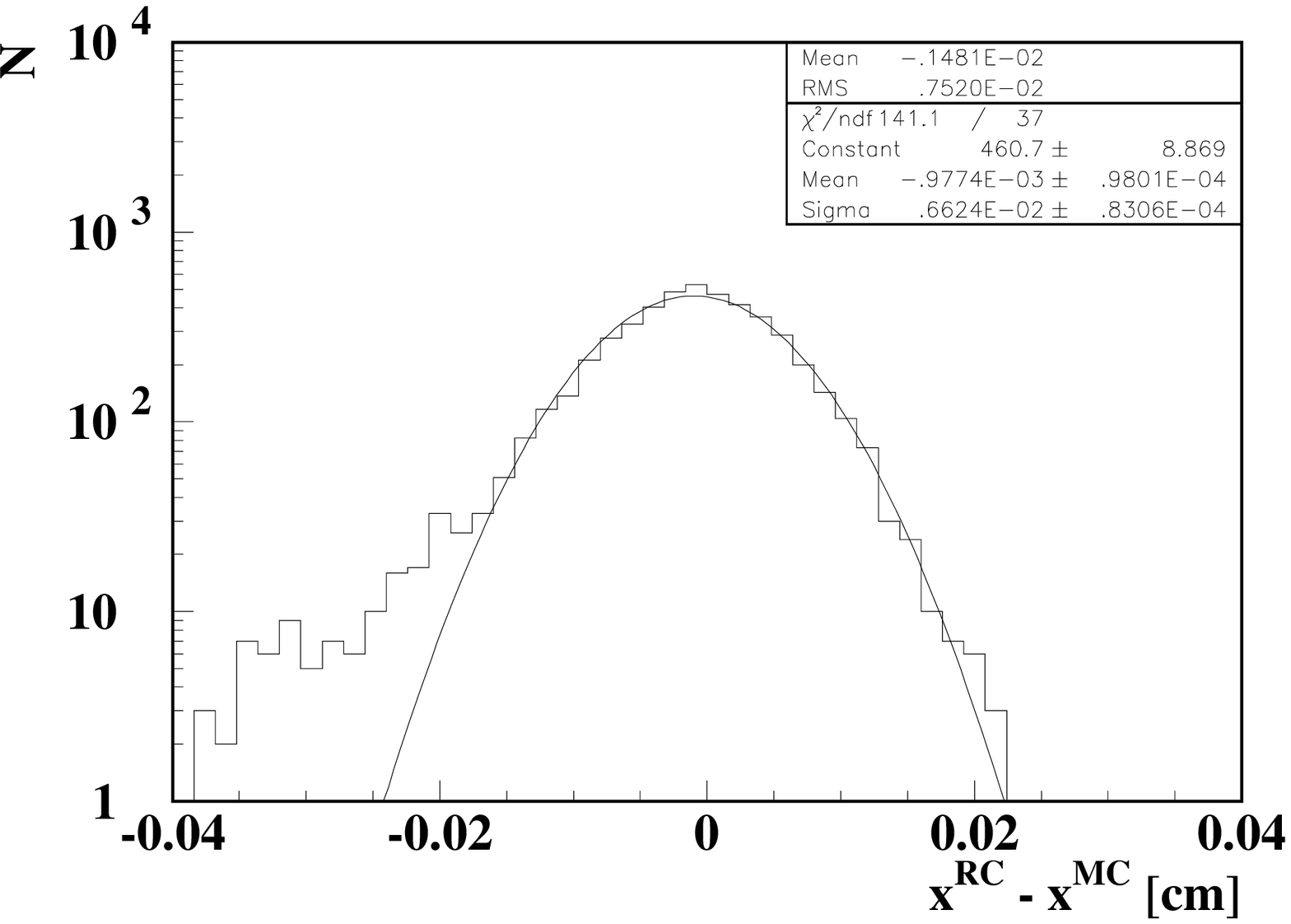,width=8cm}}}
        \put(1.5,4.9){\makebox(0,0)[t]{\bf (a)}}
        \put(2,4.0){\makebox(0,0)[t]{\huge \bf $x$}}
   \end{picture}
                        &
   \unitlength1cm
   \begin{picture}(8,6)
        \put(0,0){\makebox{\epsfig{file=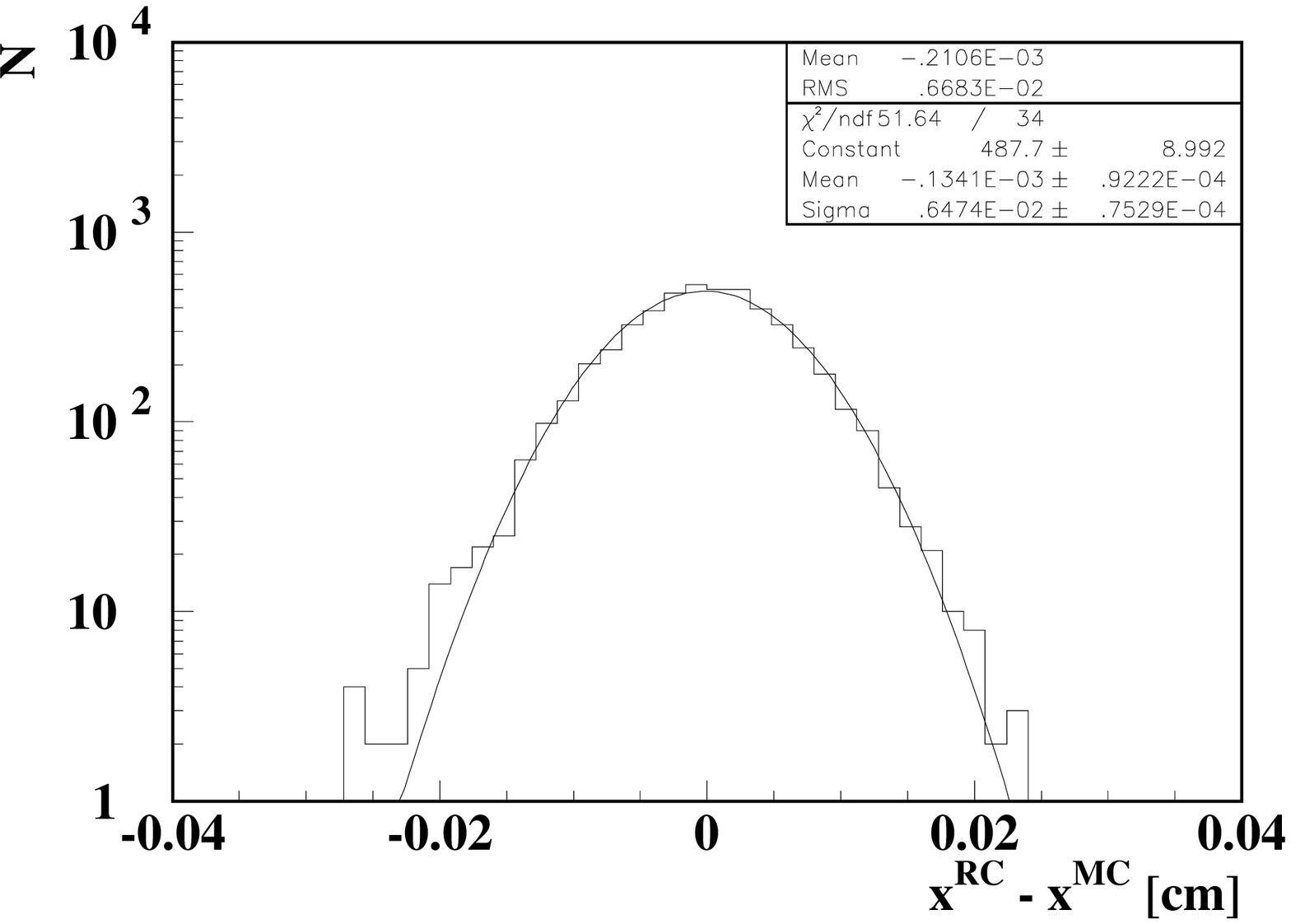,width=8cm}}}
        \put(1.5,4.9){\makebox(0,0)[t]{\bf (b)}}
        \put(2,4.0){\makebox(0,0)[t]{\huge \bf $x$}}
   \end{picture}
                         \\
   \unitlength1cm
   \begin{picture}(8,6)
        \put(0,0){\makebox{\epsfig{file=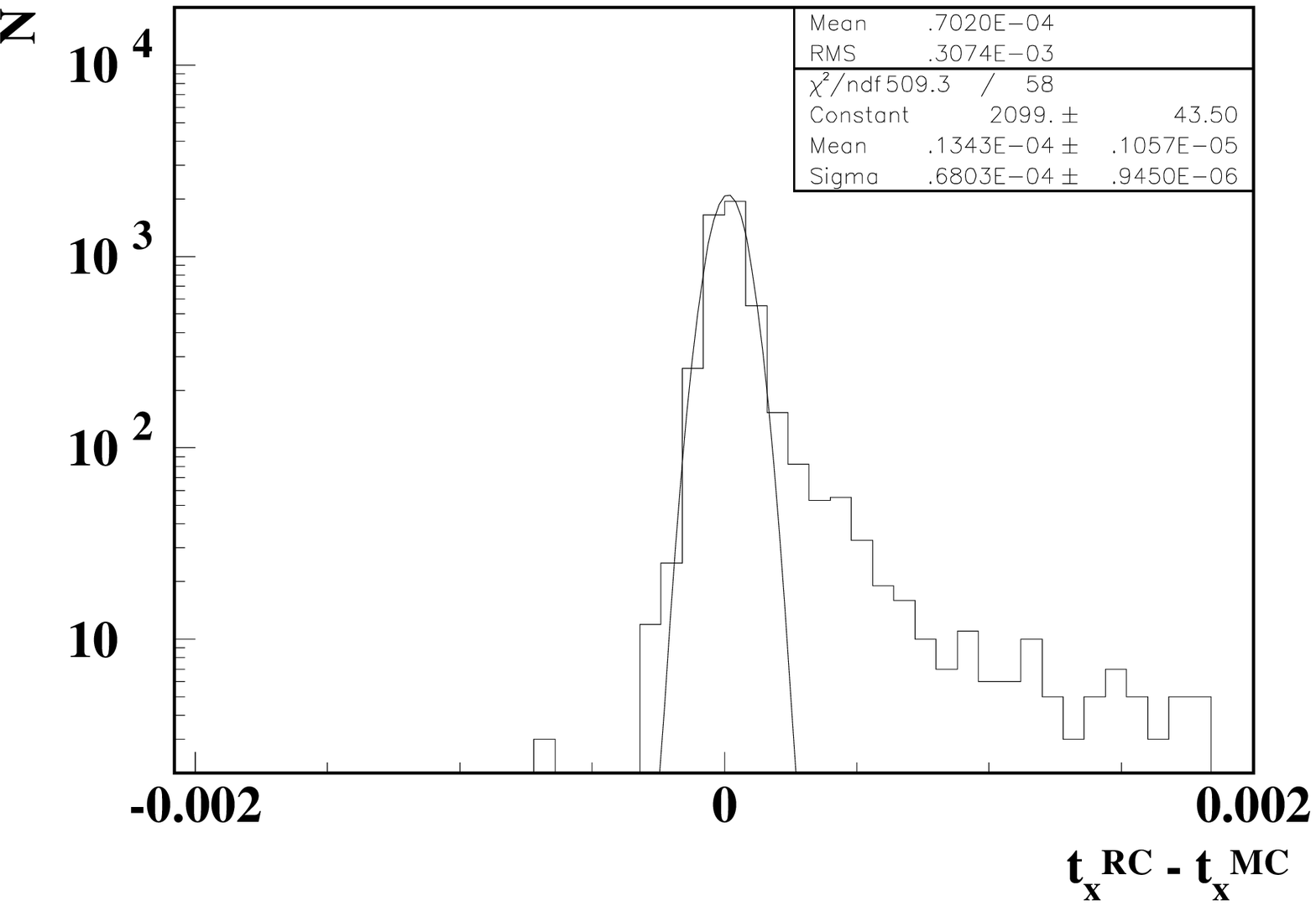,width=8cm}}}
        \put(1.5,4.9){\makebox(0,0)[t]{\bf (c)}}
        \put(2,4.0){\makebox(0,0)[t]{\huge \bf $t_x$}}
   \end{picture}
                        &  
   \unitlength1cm
   \begin{picture}(8,6)
        \put(0,0){\makebox{\epsfig{file=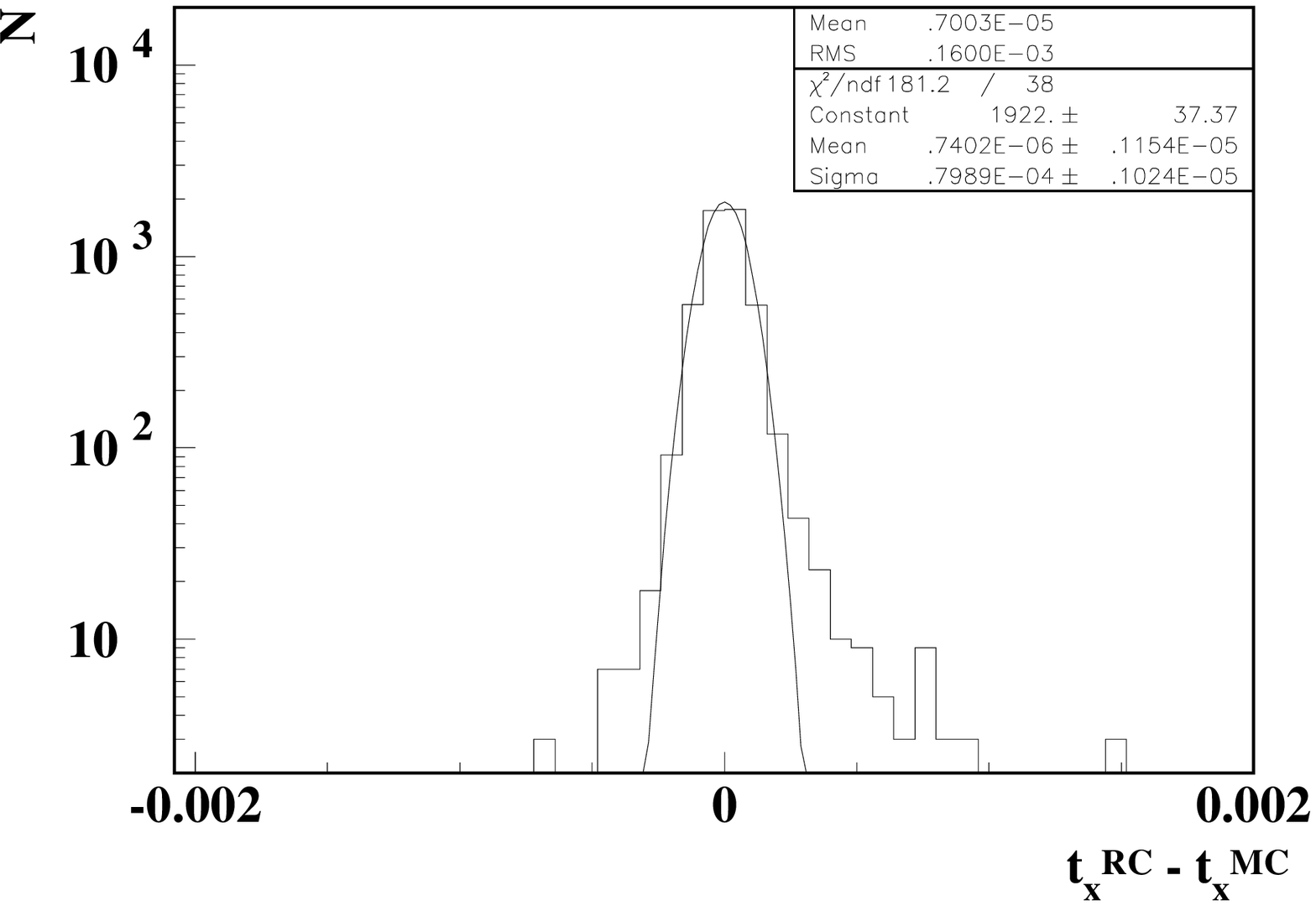,width=8cm}}}
        \put(1.5,4.9){\makebox(0,0)[t]{\bf (d)}}
        \put(2,4.0){\makebox(0,0)[t]{\huge \bf $t_x$}}
   \end{picture}
                          \\
   \unitlength1cm
   \begin{picture}(8,6)
        \put(0,0){\makebox{\epsfig{file=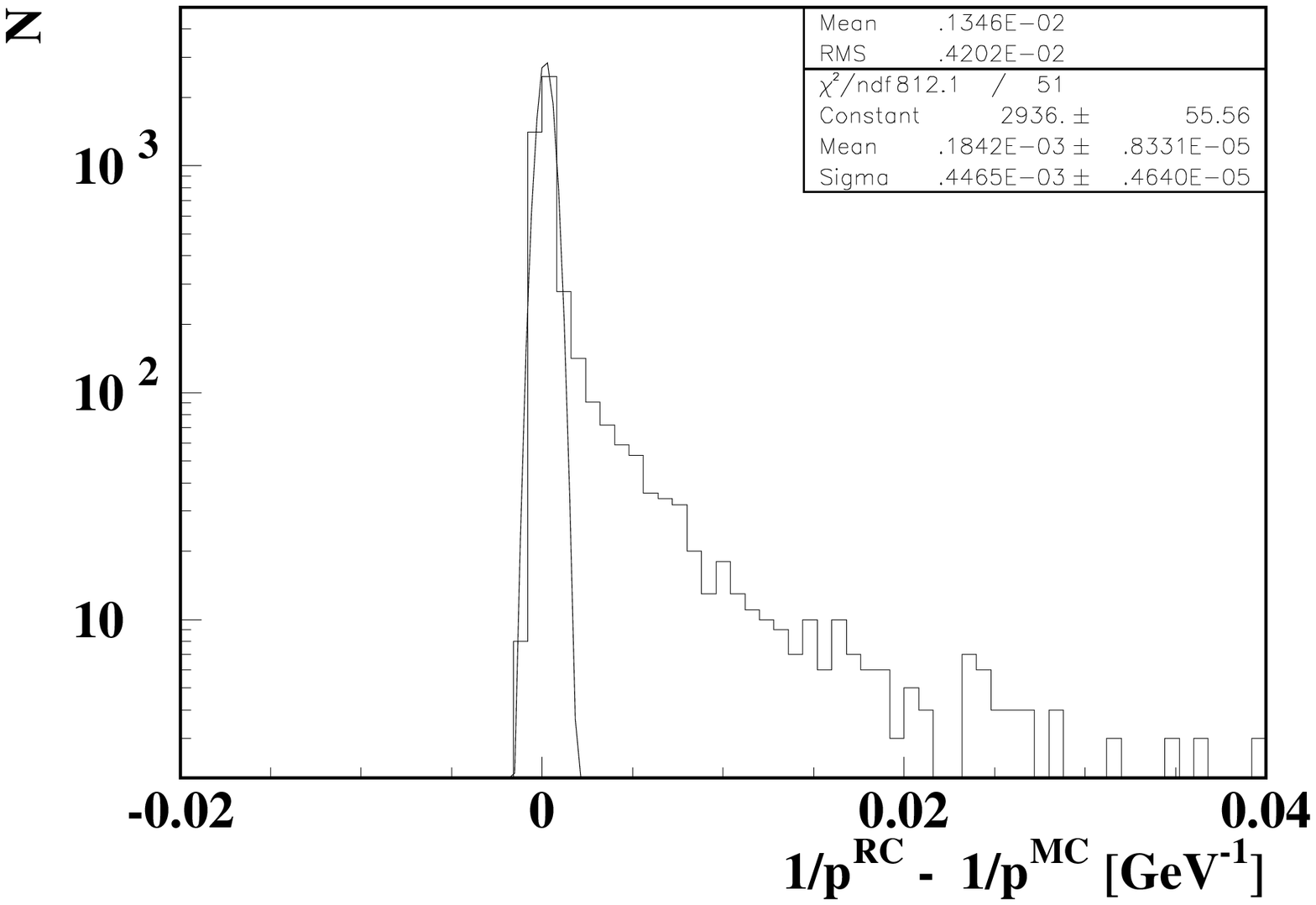,width=8cm}}}
        \put(1.5,4.9){\makebox(0,0)[t]{\bf (e)}}
        \put(2,4.0){\makebox(0,0)[t]{\huge \bf $\frac{1}{p}$}}
   \end{picture}
                          &
   \unitlength1cm
   \begin{picture}(8,6)
        \put(0,0){\makebox{\epsfig{file=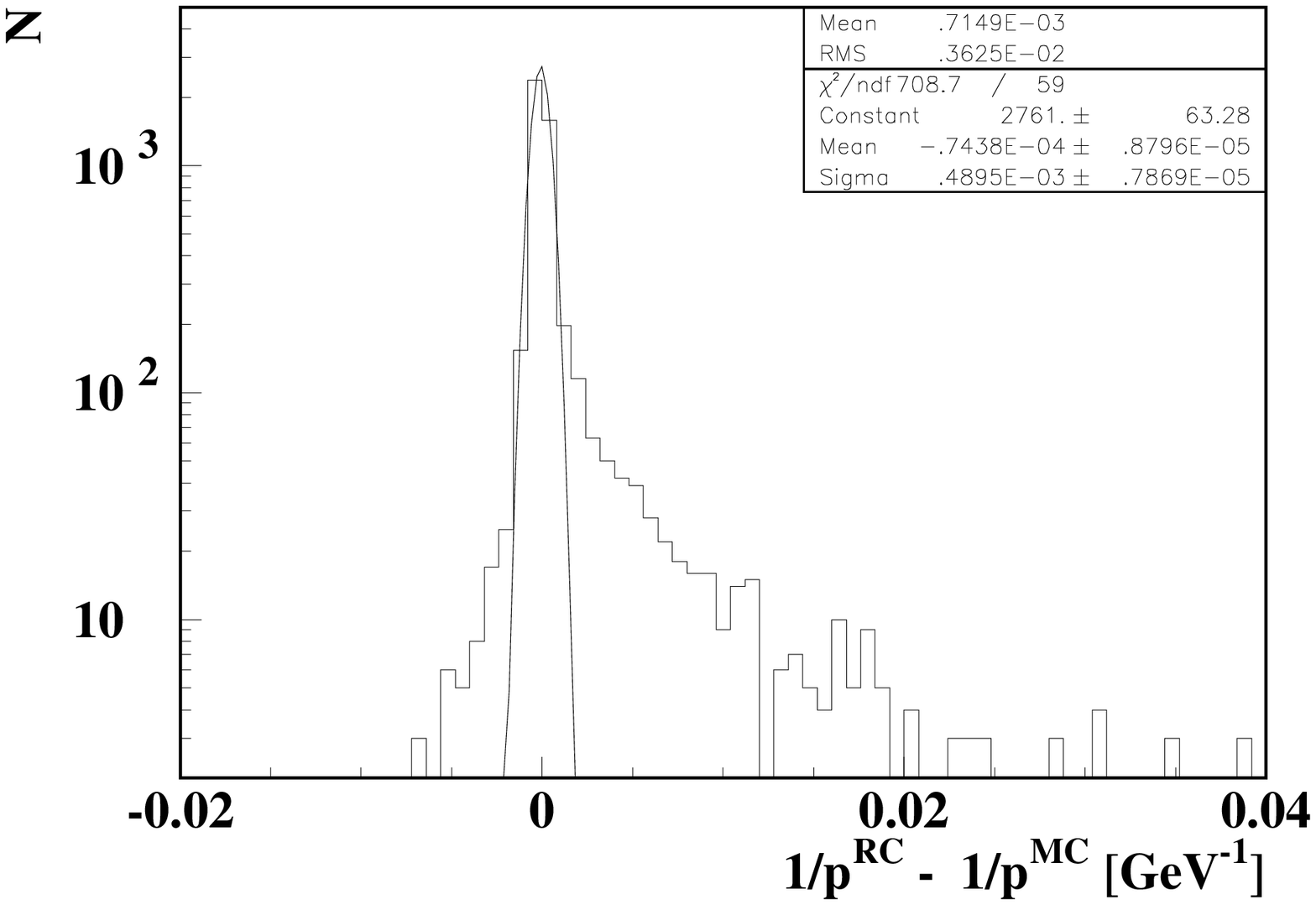,width=8cm}}}
        \put(1.5,4.9){\makebox(0,0)[t]{\bf (f)}}
        \put(2,4.0){\makebox(0,0)[t]{\huge \bf $\frac{1}{p}$}}
   \end{picture}
\end{tabular}
\caption{Distributions of parameter residuals for electrons of 100~GeV
  based on 5000 tracks, where $x$ is the impact parameter in the
  bending plane, $t_x=\tan \theta_x$ is the corresponding track slope,
  and 1/p the inverse momentum. The track fit was applied to all hits
  within the spectrometer magnet (region 2 in fig.~\ref{fig:radrange}).}
\label{radpar}
\end{center}
\end{figure}
\begin{figure} 
\begin{center}
  \epsfig{file=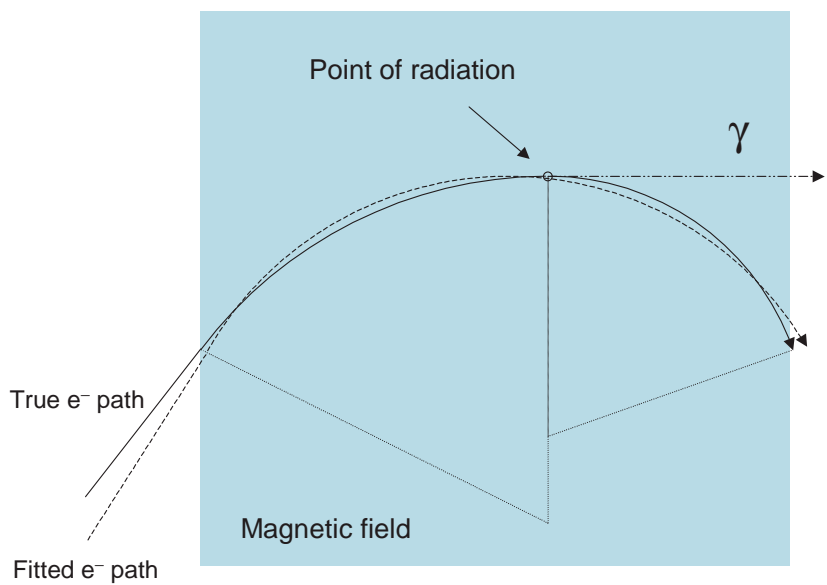,width=11cm}
  \caption{ Illustration of how radiation within the magnetic field 
    can affect the estimate of the fitted track slope. The magnetic
    field vector is pointing into the drawing plane. The electron,
    whose true path is shown by a solid line, emits a photon, which
    leads to an increase of curvature for the subsequent part of its
    trajectory within the magnet. This is illustrated by the
    curvature radii of the helices as dotted lines.  The fitted
    trajectory (dashed line) assumes a single curvature, which leads
    to an overestimation of the initial slope of the track. The
    curvatures drawn are intentionally exaggerated.}
\label{fig:radkick}
\end{center}
\end{figure}
\subsubsection{Radiation energy loss correction within the magnetic field}
Energy loss through radiation can not only interfere with the momentum
measurement, but may also affect other track parameters. This is shown
in figs.~\ref{radpar}a,c,e which display the residuals of the
parameters $x$, $t_x$ and $1/p$ for electrons produced with 100~GeV
momentum, where the fit was restricted to the magnet area (MC). Without
bremsstrahlung correction, the track slope estimate $t_x$ shows a tail
towards overestimated values, which is reflected in an underestimation
of the corresponding impact parameter, $x$. The explanation for this
effect is illustrated in fig.~\ref{fig:radkick} which for simplicity
assumes a homogeneous field: the curvature of the electron track is
abruptly increased beyond the point of radiation.  Fitting the track
with a constant momentum leads to an intermediate curvature resulting
in a shift in the measured initial track slope.

The residual distribution of the momentum parameter, $1/p$, displays a
tail towards higher values, corresponding to a mean momentum shift of
$\approx 13\%$.

\begin{figure} 
\begin{center}
\begin{tabular}{cc}

   no radiation correction & with radiation correction \\

   \unitlength1cm
   \begin{picture}(8,6)
        \put(0,0){\makebox{\epsfig{file=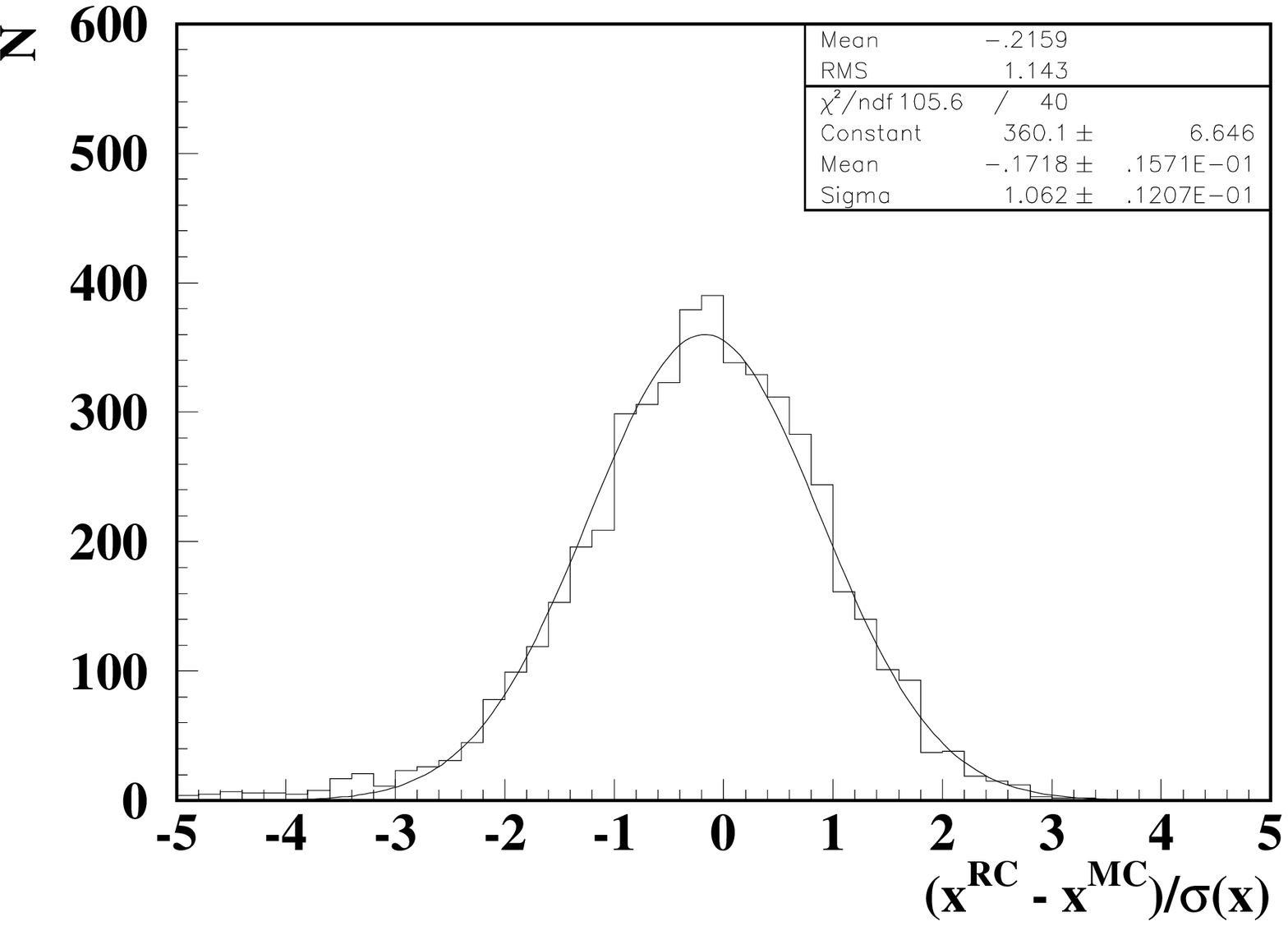,width=8cm}}}
        \put(1.5,4.9){\makebox(0,0)[t]{\bf (a)}}
        \put(2,4.0){\makebox(0,0)[t]{\huge \bf $x$}}
   \end{picture}
                        &
   \unitlength1cm
   \begin{picture}(8,6)
        \put(0,0){\makebox{\epsfig{file=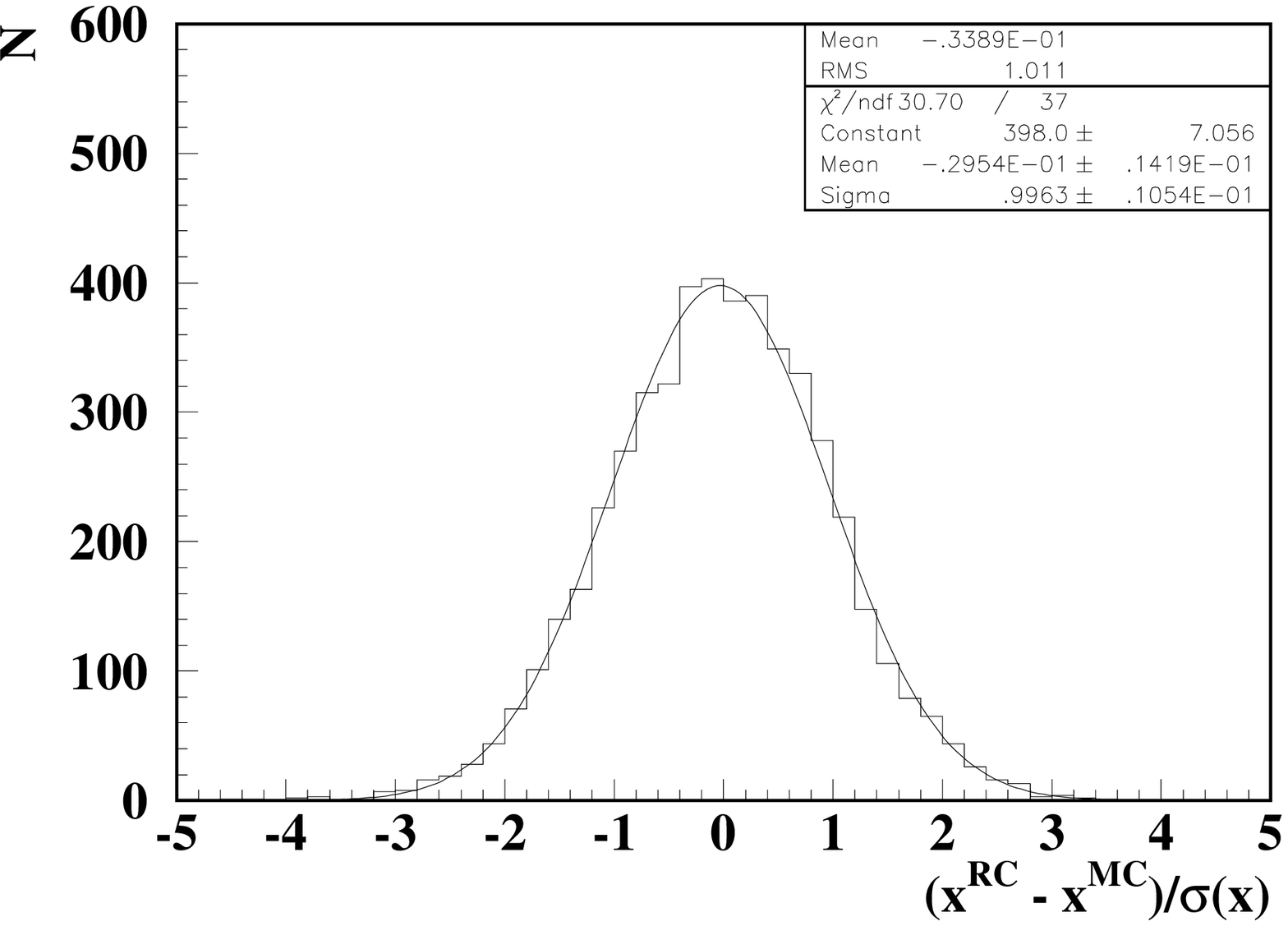,width=8cm}}}
        \put(1.5,4.9){\makebox(0,0)[t]{\bf (b)}}
        \put(2,4.0){\makebox(0,0)[t]{\huge \bf $x$}}
   \end{picture}
                         \\
   \unitlength1cm
   \begin{picture}(8,6)
        \put(0,0){\makebox{\epsfig{file=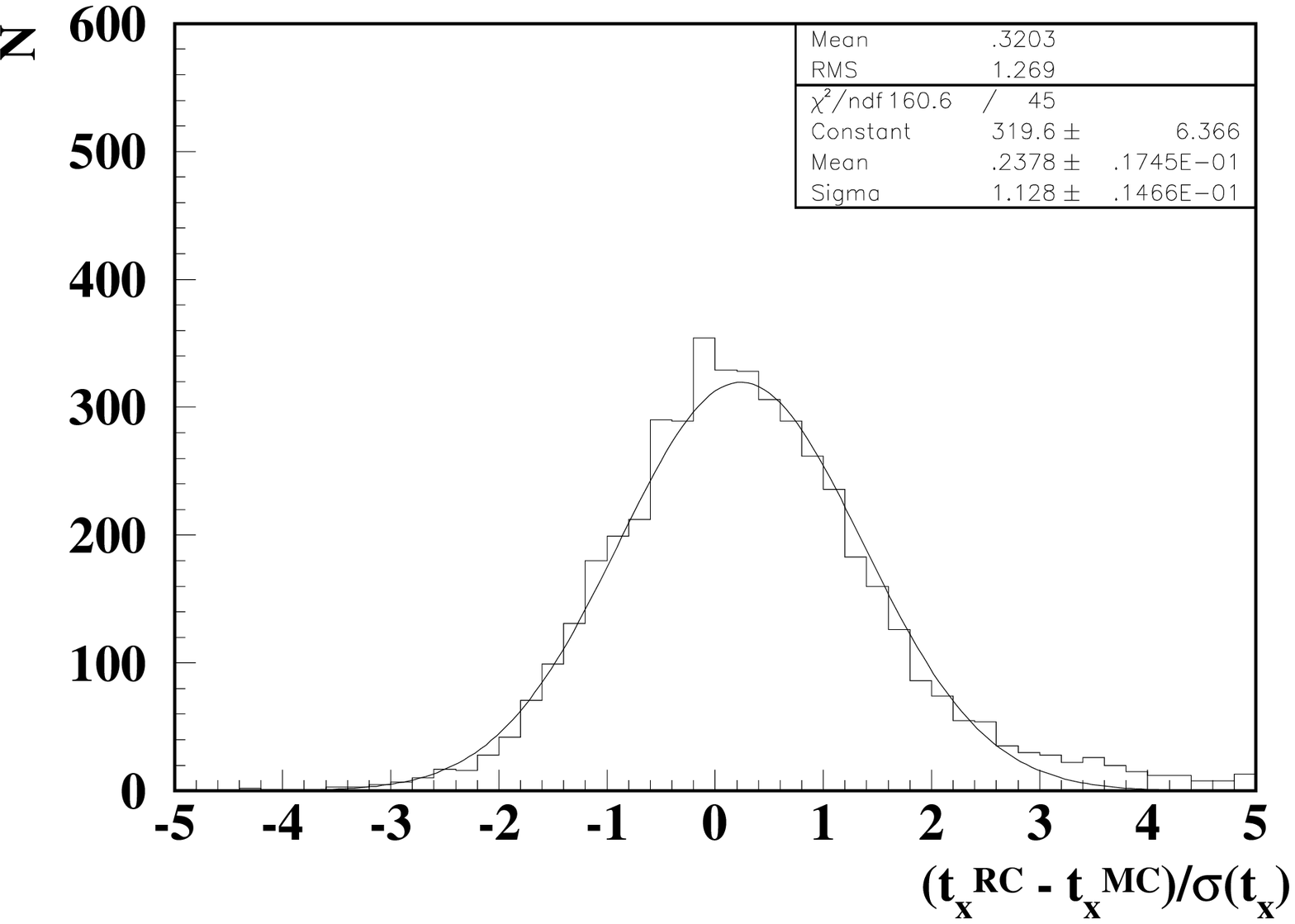,width=8cm}}}
        \put(1.5,4.9){\makebox(0,0)[t]{\bf (c)}}
        \put(2,4.0){\makebox(0,0)[t]{\huge \bf $t_x$}}
   \end{picture}
                        &  
   \unitlength1cm
   \begin{picture}(8,6)
        \put(0,0){\makebox{\epsfig{file=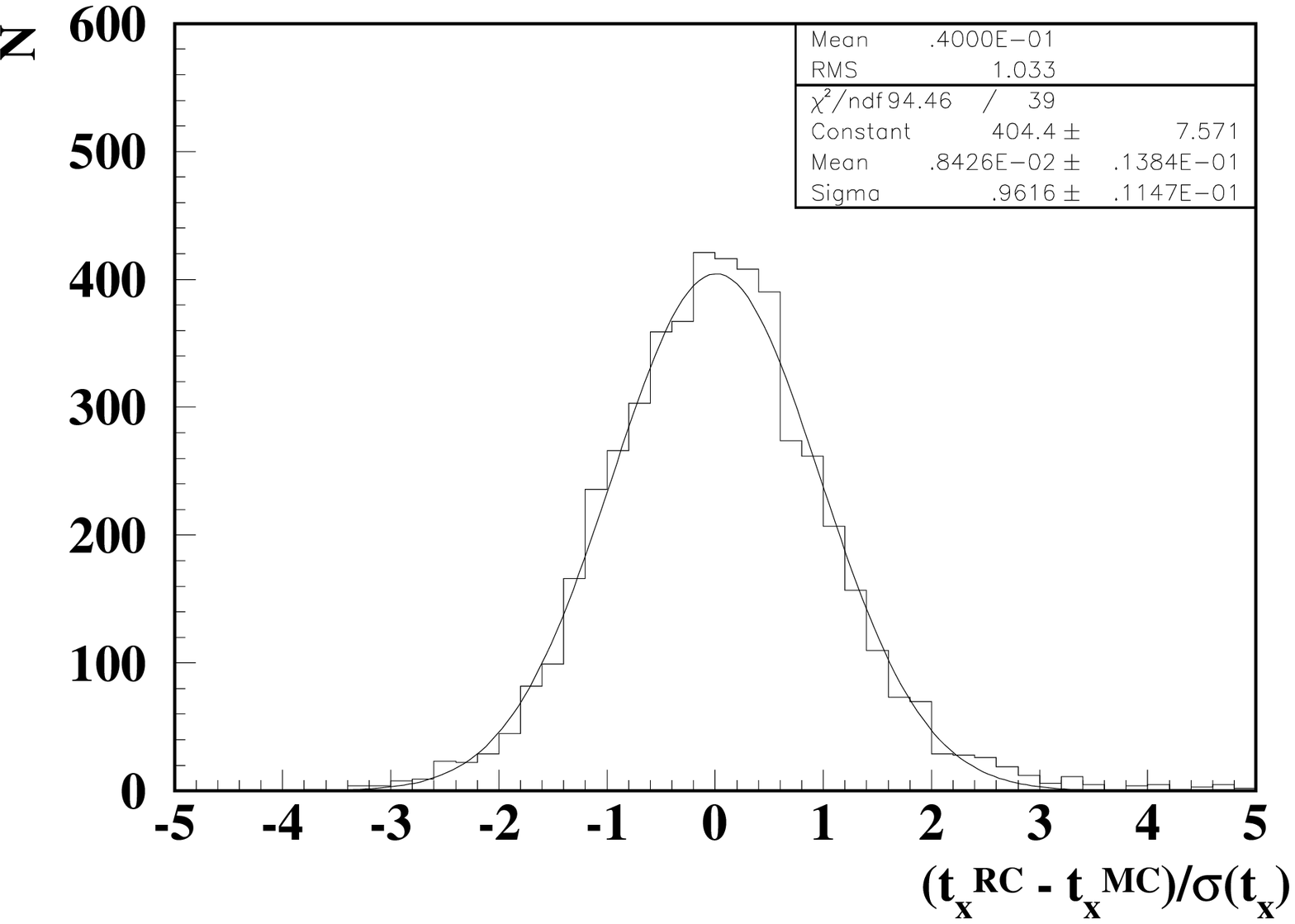,width=8cm}}}
        \put(1.5,4.9){\makebox(0,0)[t]{\bf (d)}}
        \put(2,4.0){\makebox(0,0)[t]{\huge \bf $t_x$}}
   \end{picture}
                          \\
   \unitlength1cm
   \begin{picture}(8,6)
        \put(0,0){\makebox{\epsfig{file=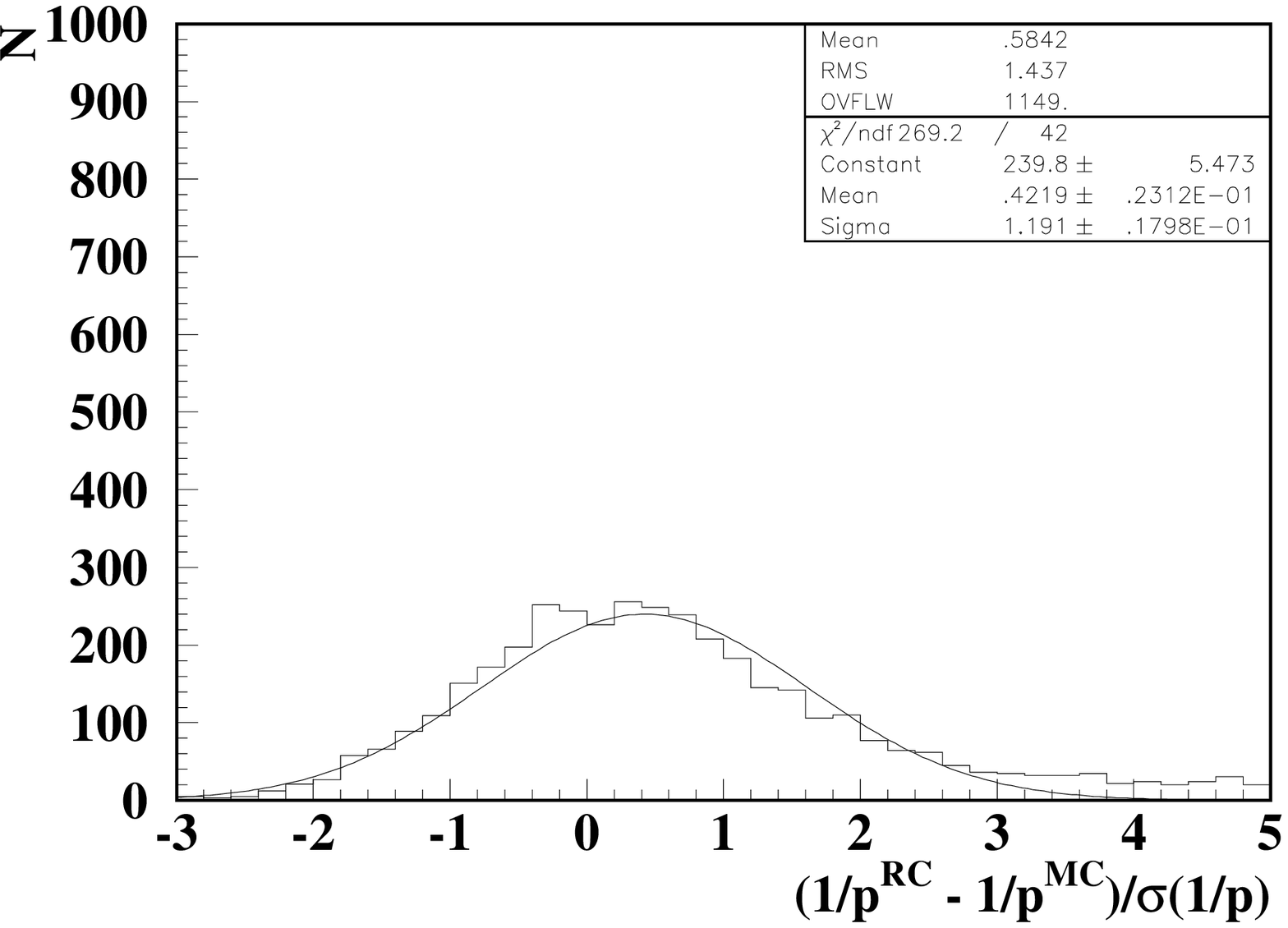,width=8cm}}}
        \put(1.5,4.9){\makebox(0,0)[t]{\bf (e)}}
        \put(2,4.0){\makebox(0,0)[t]{\huge \bf $\frac{1}{p}$}}
   \end{picture}
                          &
   \unitlength1cm
   \begin{picture}(8,6)
        \put(0,0){\makebox{\epsfig{file=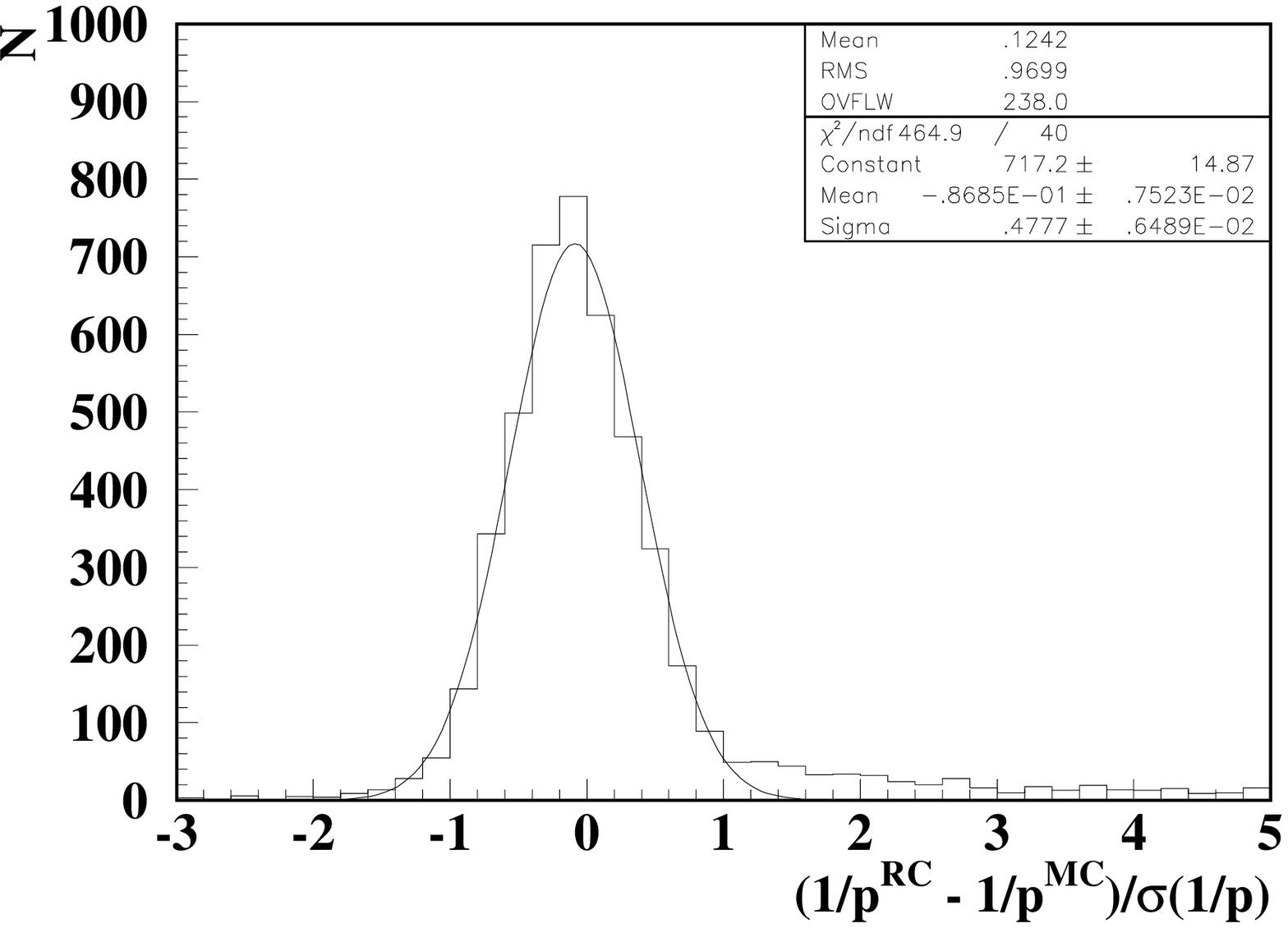,width=8cm}}}
        \put(1.5,4.9){\makebox(0,0)[t]{\bf (f)}}
        \put(2,4.0){\makebox(0,0)[t]{\huge \bf $\frac{1}{p}$}}
   \end{picture}
\end{tabular}
\caption{Distribution of normalized parameter residuals (pulls) for
  electrons of 100~GeV based on 5000 tracks, where $x$ is the impact
  parameter in the bending plane, $t_x=\tan \theta_x$ is the
  corresponding track slope, and 1/p the inverse momentum. The track
  fit was applied to all hits within the spectrometer magnet (region 2
  in fig.~\ref{fig:radrange}).}
\label{fig:radpull}
\end{center}
\end{figure}

Also the parameter errors are underestimated, which is evident from
the normalized residuals in figs.~\ref{fig:radpull}a,c,e (uncorrected
case), where the widths of the $t_x$ and $Q/p$ pull distributions
are significantly enlarged.

\begin{figure} 
\begin{center}
  \epsfig{file=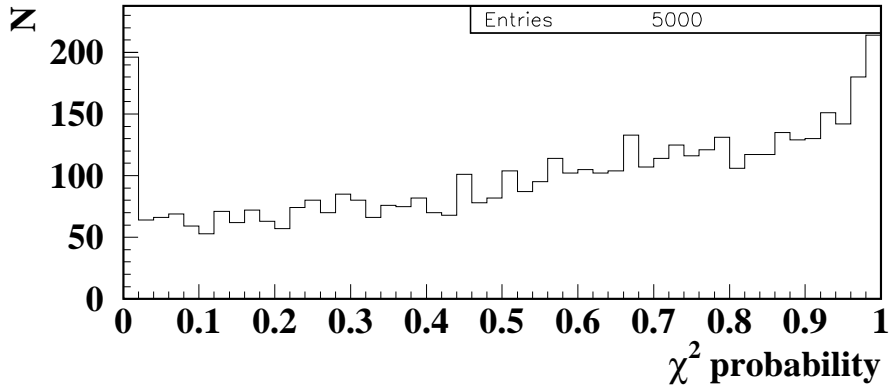,width=13cm}
\caption{Distribution of $\chi^2$ probability from the track fit of 100~GeV
  electrons in the main tracker with radiation correction. The
  non-Gaussian distribution of the radiated energy leads a non-flat
  probability distribution with a sharp peak near zero.}
\label{fig:prob-elc-main}
\end{center}
\end{figure}

Figures~\ref{radpar}b,d,f show the result with the radiation
correction applied in the fit. One can see that the tails in the
parameter estimates of $x$ and $t_x$ are far less pronounced, and the
bias in the impact parameter and track slope is considerably reduced.
Also the distortion of the mean reconstructed inverse momentum
$\delta(1/p) \approx \delta p/p^2$ is reduced from $1.3 \cdot 10^{-3}
{\rm \ GeV}^{-1}$ to $7 \cdot 10^{-4} {\rm \ GeV}^{-1}$, and the standard
deviation (RMS width) of the parameter estimates is reduced by 11\%
($x$), 48\% ($t_x$) and 14\% ($Q/p$), respectively.  Moreover, the
radiation correction brings the RMS widths of the pull distributions
close to unity (figs.~\ref{fig:radpull}b,d,f), which indicates a
reliable covariance matrix estimate.  The fit probability distribution
is shown in fig.~\ref{fig:prob-elc-main}.  It reflects a non-$\chi^2$
type distribution of the goodness-of-fit, which is expected since the
radiation of bremsstrahlung introduces a strongly non-Gaussian random
perturbation.

\begin{figure} 
\begin{center}
\begin{tabular}{cc}
  \multicolumn{2}{c}{full radiation correction}\\
  \unitlength1cm
  \begin{picture}(8,6)
    \put(0,0){\makebox{\epsfig{file=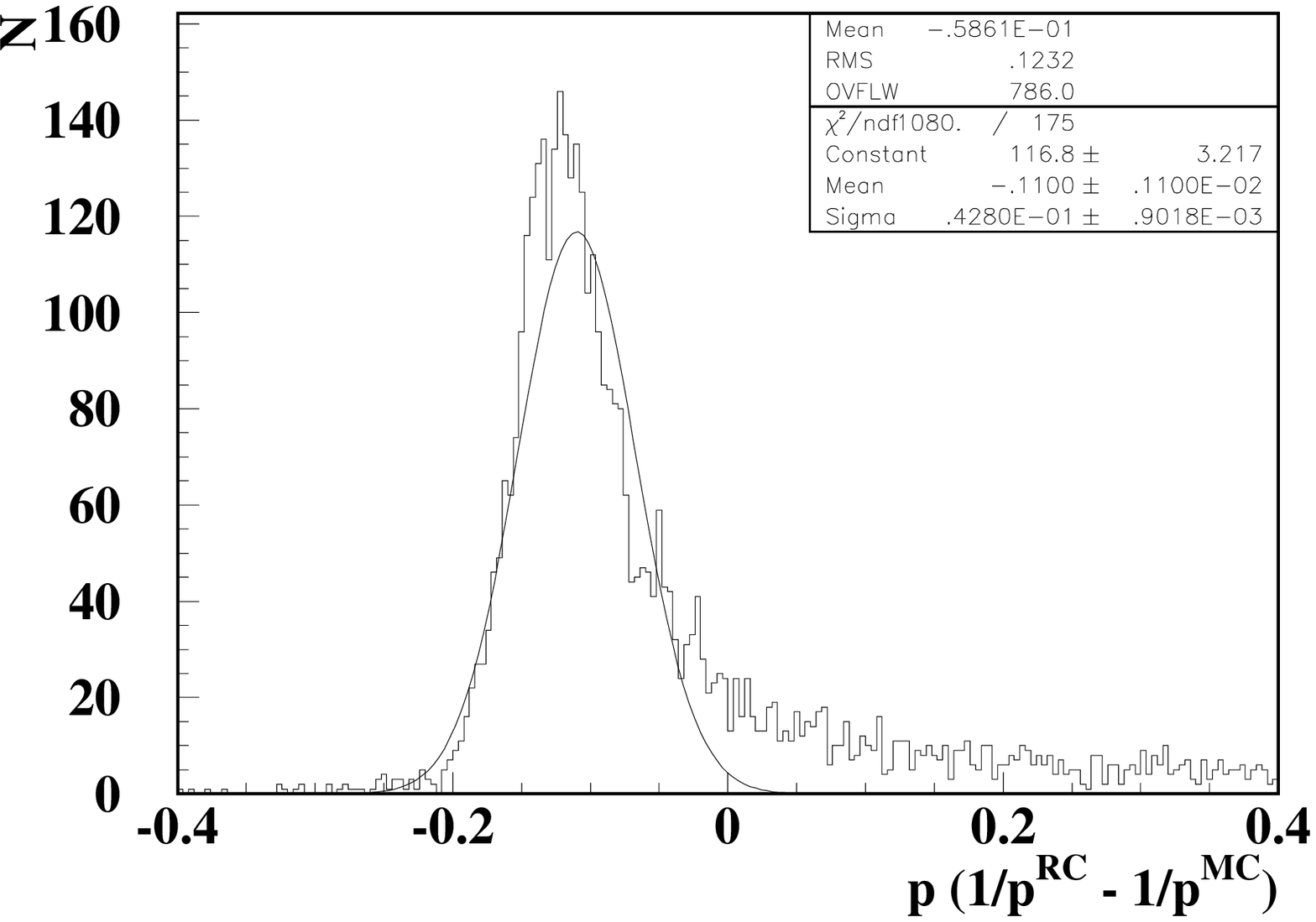,width=8cm}}}
    \put(1.5,4.9){\makebox(0,0)[t]{\bf (a)}}
    \put(2,4.0){\makebox(0,0)[t]{\huge \bf $\frac{\delta p}{p}$}}
  \end{picture}
  &
  \unitlength1cm
  \begin{picture}(8,6)
    \put(0,0){\makebox{\epsfig{file=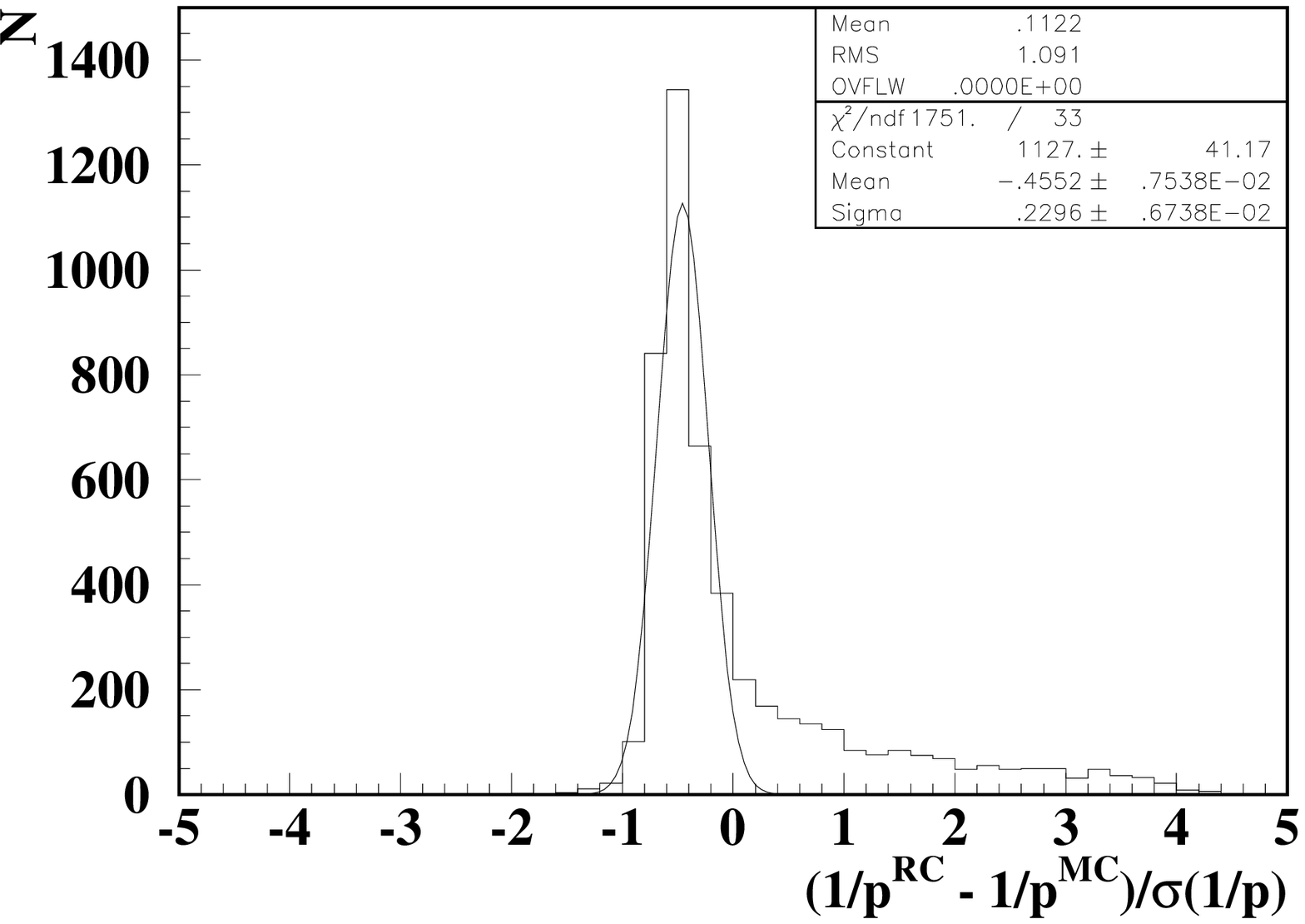,width=8cm}}}
    \put(1.5,4.9){\makebox(0,0)[t]{\bf (b)}}
    \put(2,4.0){\makebox(0,0)[t]{\huge \bf $\frac{\delta p}
        {\sigma_p}$}}
  \end{picture}
\end{tabular}
\caption{Distribution of 1/p parameter residuals multiplied by $p$
  as measure for relative momentum deviation (a), and of normalized
  1/p residuals (b) for 100~GeV electrons in the full tracker.  The
  radiation correction was applied in the whole tracking system,
  leading to the shift of the peaks described in the text.}
\label{fig:emom100}
\end{center}
\end{figure}

\begin{table}
\centering
\begin{tabular}{|c||c|c|}
\hline
Radiation correction mode & \multicolumn{2}{c|}{Fraction of fits within
momentum deviation}\\
 & $-0.1 < \delta p/p < +0.1$ & $-0.2 < \delta p/p < +0.2$ \\
\hline
none      & $0.566 \pm 0.004$  & $0.678 \pm 0.003$ \\
within magnet    & $0.635 \pm 0.003$  & $0.728 \pm 0.003$ \\
within full spectrometer      & $0.321 \pm 0.003$  & $0.786 \pm 0.002$ \\
\hline
\end{tabular}
\caption{Fraction of fits within given limits of momentum deviation,
for three variants of radiation correction}
\label{tab:edeviat}
\end{table}

The situation is different if one attempts to extend the radiation
correction to the full tracking system including regions 1 and 3 which
are outside of the magnetic field, most notably the vertex detector
whose material causes a significant energy loss for electrons. Outside
of the magnetic field, however, the trajectory shape is not modified by
radiation, which means that the fit will only apply the on-average
correction according to the traversed radiation thickness. This can
lead to bizarre results as seen in fig.~\ref{fig:emom100}, which shows
the distribution of the $1/p$ parameter residual multiplied by the
momentum itself as well as the corresponding pull distribution. The
peak has moved away from zero to negative residual values, implying
that electrons in the peak obtain an overcorrected energy value. In
fig.~\ref{fig:emom100}b, the mean value of the pull is near zero, and
the RMS width is close to one, indicating that the compensation works
correctly in the statistical sense. For intuitive plausibility,
however, it is relevant that a large fraction of measurements are in
the immediate vicinity of the quoted value. A test of this criterion
is shown in table~\ref{tab:edeviat}, which summarizes the fraction of
fits with momentum deviation of within 10\% or 20\% of the real value
for the three correction scenarios. With the 10\% criterion, the full
spectrometer correction appears worse than even in the uncorrected
case, while the correction restricted to the magnet gives the best
description in the intuitive sense. In conclusion, the magnet-based
correction appears to be provide the best compromise, though this
will in general have to be evaluated in each specific application.

\subsection{Robust Estimation}
The preceding sections have shown how intrinsically non-Gaussian
influences, as multiple scattering, or radiative energy loss of
electrons, can complicate the estimate of essential kinematic
parameters and their interpretation. A fully adequate treatment of
profoundly non-Gaussian variables is in general beyond the
capabilities of least squares estimation. Likelihood methods, on the
other hand, are in principle able to cope with random variables of any
distribution, but often cannot be used with as efficient a machinery,
in particular when it comes to computation of error matrices.

During the last years, promising concepts have been developed that
permit treatment of non-Gaussian random variables, but still allow to
use much of the powerful machinery developed with least squares
estimation. These methods are called {\it robust estimation}
techniques. One very attractive idea is based on the fact that
non-Gaussian distributions can often be approximated as superposition
of a limited number of Gaussian
distributions~\cite{kitagawa1,kitagawa2}.  For example, a distribution
resembling a Gaussian in the centre, but featuring long tails, as
is common with multiple scattering, can be approximated by a sum of a
narrow Gaussian distribution and a wide one. If one performs two
parallel least squares estimates, each based on one of the Gaussians,
the resulting parameter estimates, combined with appropriate weights,
will reflect the underlying statistics better than a single estimate
with a single Gaussian approximation. Thus, the occurence of random
variables in the tail of the distribution does not pull the estimate
as far away as it would with a traditional least square estimator, leading
to a more robust behaviour of the fit.

This is the basic idea of the {\it Gaussian Sum Filter}
(GSF)~\cite{kitagawa1,kitagawa2,fruehwirthBayes,fruehwirthGSF,
  fruehwirthNongaussian}, which uses the Kalman filter to incorporate
the individual Gaussian components. Upon each occurrence of process
noise, the distribution of which is approximated by a sum of $N$
Gaussians, the filter splits into $N$ parallel branches each of which
obtains a corresponding weight. In a detector geometry with many
scattering elements, this will lead to a repeated multiplication of
the number of linear filters to be evaluated. To avoid the explosion
of the computing effort, the number of parallel components is limited
by {\it collapsing} or {\it clustering} components of similar shape.
It has been shown that the algorithm can be designed such that the
computing effort increases linearly with the maximum number of
parallel components ($M$), and that $M \approx 6-8$ already gives good
results~\cite{fruehwirthGSF}. In a similar way, radiative energy loss
of electrons can be treated by approximating the radiated energy
distribution by superposition of several Gaussians~\cite{gsfElectron}.

\section{Event Reconstruction} After particle tracks have been
reconstructed, they form the basis for the reconstruction of the whole
event. This will ultimately include particle identification based on
$dE/dx$, time-of-flight, \v{C}erenkov or transition radiation, muon
chambers and calorimetry, as well as kinematical reconstruction of
composite particles and jets. This article will restrict itself to a
brief discussion of vertex reconstruction and kinematical constraints.

\subsection{Vertex Pattern Recognition}
\begin{figure}[htbp]
  \begin{center}
    \epsfig{file=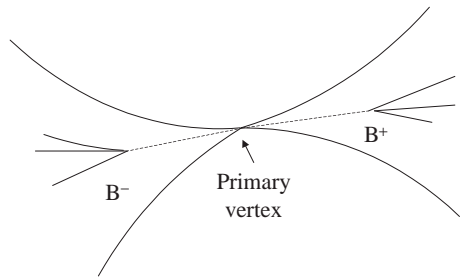,width=6cm,angle=0}
  \end{center}
  \caption{Schematic view of the event structure in an interaction of
    the type $e^+e^- \rightarrow B^+  B^- + X$}
  \label{fig:bbvertex}
\end{figure}
The vertex is an essential element of the space-time structure of an
interaction. Vertices indicate either the location where an
interaction has taken place, for example the primary interaction that
is the ultimate origin of all emerging particles, or the place where
an unstable particle has decayed. This is illustrated in
fig.~\ref{fig:bbvertex}, which schematically sketches the final state
of an interaction with associated production of two beauty mesons, as
it can occur for example at a high energy $e^+e^-$ collider. The
beauty hadrons, here a $B^+$ and a $B^-$, are produced together with
accompanying charged particles at the interaction point, travel
invisibly for some distance that is, on average, determined by their
lifetime and momentum, whereupon they decay into daughter particles.
The charged tracks coming from these decays can be used to reconstruct
the decay locations of the $B$ mesons as {\it secondary
  vertices}\footnote{We neglect here the complication that the $B$
  meson is likely to decay to a final state with a charmed particle
  which again has a non-negligible lifetime.}.  The other tracks,
together with the reconstructed $B$ mesons form the {\it primary
  vertex}, which indicates the interaction point.

\begin{figure}[htbp]
  \begin{center}
    \epsfig{file=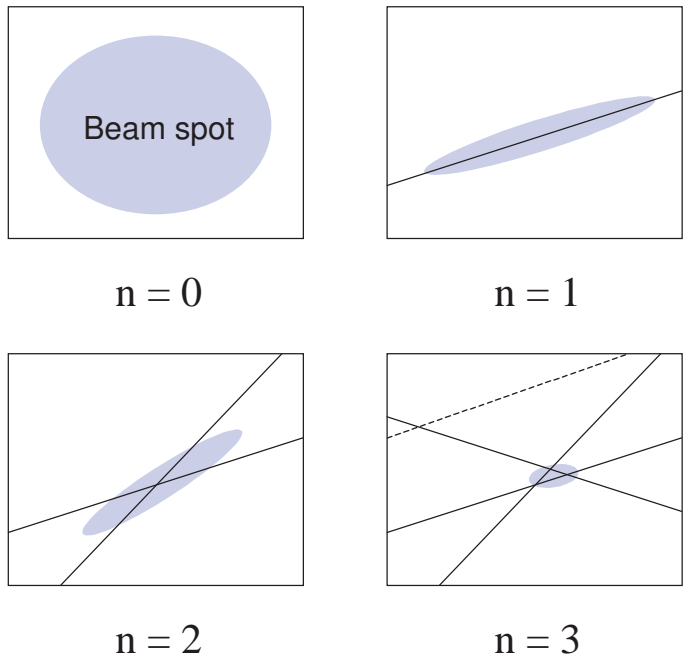,width=9cm,angle=0}
  \end{center}
  \caption{Illustration of the iterative construction of a (primary) 
    vertex, where $n$ is the number of tracks used to define the
    vertex in each step. The shaded area indicates the covariance
    ellipse of the projected vertex after each step. The dashed line
    indicates an outlier track.}
  \label{fig:vertexIter}
\end{figure}
In many practical applications, the vertex is constructed by an
iterative procedure as it is illustrated in
fig.~\ref{fig:vertexIter}. In most cases, some a-priori knowledge
about the vertex position exists, for example the shape of the beam
spot, in which interactions occur in the first place. Then a first
track is selected as a vertex seed, which already narrows down the
covariance ellipsoid in two dimensions. When a second suitable track
is added, the vertex is already closely defined in all
coordinates. This provides strong rejection power against off-vertex
particles when adding more tracks.

As in the track pattern recognition case, the danger lies in the
dependence on the starting point. It is therefore necessary to use
iterative criteria which ensure that the track forming the vertex seed
is well chosen, and even then it must be possible to scrutinize the
track ensemble of a vertex, to remove tracks that have turned out to
be off the mark, and to reconnect tracks that had been discarded at an
earlier stage of the construction. The vertex algorithm used in the
ZEUS experiment~\cite{hartnerVertex}, which internally uses the
fitting methods of~\cite{billoirQianVertex} may serve as an example:
it uses the proton beam line as a soft constraint, and then produces a
set of all track pairs that would be compatible with a common vertex
together with the beam line constraint within a suitable $\chi^2$
margin.  The track pairs are then ordered according to their degree of
compatibility with other track pairs, defined by the criterion above.
The track pair of highest compatibility forms then the first vertex
seed to be used, though also other track pairs of high compatibility
level are tried, and in the end the best set is chosen based on a
criterion of number of tracks and total $\chi^2$.  Other approaches
start by connecting all tracks to a diffuse master vertex, which is
then successively split into vertices of smaller multiplicities and
isolated tracks. A systematic investigation of different methods for
vertex reconstruction in the context of the CMS experiment can be
found in~\cite{fruehwirthVertexCMS}.

An entirely different approach is pursued in the topological vertex
finding algorithm~\cite{zvtop} developed for the vertex detector of
the SLD experiment~\cite{sldVertex}. This method assigns a {\it
  Gaussian tube} around each track extrapolation to indicate the
likelihood of an assigned vertex on a single track basis. The Gaussian
tubes of all tracks are then combined to find points with
maximum probability of a vertex. This method resembles the {\it Fuzzy
  Radon Transform} for tracks discussed in
section~\ref{sec:fuzzyRadon}. The search for maxima is then performed
by sophisticated clustering algorithms. A particularly intriguing
feature is the efficient resolution of heavy flavour cascade decays.

Direct vertex search by Hough transform is possible in cases where the
vertex location is already strongly constrained in some coordinates,
for example through the shape of a wire target~\cite{lohseVertex}.

\subsection{Vertex Fitting}
The least-squares principle can also be readily applied for vertex
fitting~\cite{saxon1,saxon2,saxon3}. The parameters of the tracks
$\vec{p}_1 \ldots \vec{p}_n$ at a given reference surface plus the
a-priori knowledge of the vertex are the input, and the calculated
vertex position together with the {\it reduced track parameters} of
each particle, which contain only directional and momentum information
at the common vertex, are the output.  A general property of vertex
fitting is the fact that, unlike track fitting, the fit is always
non-linear, since even with straight-line tracks the extrapolation to
the vertex introduces a coupling between positional and directional
parameters.

As noted earlier, already vertex pattern recognition requires
incremental, progressive fitting, with tracks added or removed one by
one. It is therefore not surprising that also for vertex fitting, the
Kalman filter is in many cases the method of
choice~\cite{luchsingerGrab}. In the vertex fitting case, the {\it
  transport} becomes trivial, and also process noise does not have an
equivalent. The filter step adds another track to the vertex and
updates the vertex position as well as the reduced track parameters.
It is very easy to remove an already filtered track from the vertex
candidate, since in the filter equations, the inverse covariance
matrix of the track acts as the {\it weight} of the track information,
and setting its sign to negative will subtract the track from the
vertex fit. We prefer not to display the Kalman filter equations for
vertex fitting here explicitly, but refer to the
literature~\cite{grotebock}.

\subsection{Kinematical Constraints}
Pattern recognition deals with merging of measured information with
a-priori knowledge. For example, in track pattern recognition the
track model enhances the measurement power of each individual hit,
while vertex assignment improves the spatial information of each
associated track. In similar fashion, a-priori knowledge can be used
in many cases in the further reconstruction of the event. A typical
example is the beam energy constraint: in $e^+e^-$ b-physics
experiments which operate at the $\Upsilon (4S)$ energy, as BaBar,
BELLE, CLEO and the earlier ARGUS, the $B$ mesons are produced in an
exclusive decay of the $\Upsilon (4S)$ resonance, and the energy of
the $B$ mesons is precisely the beam energy, which is known to a much
better precision than the $B$ meson energy reconstructed from its
measured decay particles. Imposing the beam energy constraint improves
then also the resolution of the $B$ candidate mass; this method has
been a vital tool in the investigation of exclusive $B$ decays (see
for example~\cite{beamEnergyConstraint}).

Also masses of intermediate particles in a decay chain, for example
$B^0 \rightarrow D^{*+} \pi^+ \pi^- \pi^-$, $D^{*+} \rightarrow D^0
\pi^+$, $D^0 \rightarrow K^- \pi^+$ can be used to imply kinematical
constraints. In this case, the $D^0$ is a rather stable particle whose
width is too small to resolve by direct kinematical reconstruction in
a spectrometer. Therefore, the established knowledge of the $D^0$
mass~\cite{pdg} can be imposed as a kinematical constraint. For
example, if $\vec{\alpha}$ denotes the reconstructed parameters of the
$K^-$ and $\pi^+$ particles and $V_\alpha$ their covariance matrix,
the reconstructed $D^0$ mass will be a function $M(\alpha)$ of these
parameters, and introduction of a Lagrange multiplier $\mu$ leads
to the expression
\begin{equation}
X^2 = (\vec{\alpha}_c - \vec{\alpha})^T V_\alpha^{-1}
(\vec{\alpha}_c - \vec{\alpha}) + 2 \mu (M(\vec{\alpha_c}) - m_{D^0})
\end{equation}
which has to be minimized with respect to the constrained parameters
$\vec{\alpha}_c$. If the daughter particles form a secondary vertex,
its parameters can be optimized as well. The $D^0$ mass constraint
leads in general to a considerable improvement of the $D^*$ mass peak,
which becomes much narrower than the experimental resolution.  In
comparison to the popular {\it mass difference method}, which benefits
from the correlation in the errors of the reconstructed $D$ and $D^*$
masses, this approach has the advantage that the result can be used in
turn to reconstruct more complex decay chains of angular excitations
in the $D$ systems, or of $B$ hadrons. In a next step of $B$
reconstruction, even the tabulated $D^*$ mass could be imposed as
another independent constraint.

\section{Concluding Remarks} 
The variety of pattern recognition tasks in particle physics tracking
detectors has lead to a multitude of different approaches. Several of
the global methods, as template matching or Hough
transform/histogramming play an unchallenged r\^ole in special
applications, while Hopfield networks and deformable templates
frequently appear to be either limited to favourable scenarios (e.g.
with 3D measurements and moderate occupancy), or need an excellent
initialization or combination with a track following algorithm to
become applicable at production scale.  In the case of elastic arms,
also the choice of an efficient minimization technique is essential.
Local methods of pattern recognition are still going strong, with the
Kalman filter as the mathematical backbone, and accompanied by subtle
arbitration techniques they can cope well even with high track
densities and sizable amounts of material in the tracking area. The
new generation of high energy hadron colliders, in particular the LHC
with huge track densities in piled-up events will become an important
benchmark for algorithm performance.  It can be expected that
sophisticated combination of both global and local approaches in
different passes of the procedure, matched to the particular layout of
each experiment, will become a promising path to achieving the best
performance.

The increasing abundance of material in radiation hard detectors poses
also additional challenges to track fitting. While the correction of
multiple scattering with the Kalman filter has become the accepted
general standard, Moli\`ere scattering tails require a careful
interpretation of the results. Electron energy reconstruction with
sizable radiative energy loss is a major challenge and requires very
careful treatment, and becomes a rewarding subject for robust methods
beyond least square estimation. Also vertex pattern recognition can be
expected to receive increasing attention in very complex event
topologies at LHC, where reliable tagging of heavy flavour is a
crucial prerequisite to scientific discovery.

\addcontentsline{toc}{section}{Acknowledgement}
\section*{Acknowledgement}
It is a pleasure to thank E.~Lohrmann for his valuable comments
on the manuscript.

\end{document}